\documentclass{aa}  

\usepackage{graphicx}
\usepackage{txfonts}
\makeatletter
\renewcommand*\aa@pageof{, page \thepage{} of \pageref*{LastPage}}
\makeatother

\usepackage{multirow}
\usepackage[version=4]{mhchem}
\usepackage[colorlinks=true,
            linkcolor=red,
            urlcolor=blue,
        citecolor=blue]{hyperref}
\usepackage[dvipsnames]{xcolor}
\usepackage{colortbl}
\usepackage{placeins}

\begin{document} 
\newcommand{\R}[0]{\ensuremath{R}}
\newcommand{\Rv}[1]{\ensuremath{R = #1}}
\newcommand{\SN}[0]{\ensuremath{S/N}}
\newcommand{\SNv}[1]{\ensuremath{S/N = #1}}

\newcommand{\mic}[1]{\ensuremath{#1}~\textmu m}

\newcommand{\pt}[0]{\textit{P}$-$\textit{T}}

\newcommand{\Rpl}[0]{\ensuremath{R_{\text{pl}}}}
\newcommand{\Mpl}[0]{\ensuremath{M_{\text{pl}}}}
\newcommand{\Pct}[0]{\ensuremath{P\mathrm{^{cloud}_{top}}}}
\newcommand{\Tct}[0]{\ensuremath{T\mathrm{^{cloud}_{top}}}}
\newcommand{\Teq}[0]{\ensuremath{T_\mathrm{eq}}}
\newcommand{\Pcs}[0]{\ensuremath{P\mathrm{^{cloud}_{span}}}}
\newcommand{\Rcm}[0]{\ensuremath{\bar{R}\mathrm{^{cloud}}}}
\newcommand{\Sc}[0]{\ensuremath{\sigma\mathrm{^{cloud}}}}
\newcommand{\Ab}[0]{\ensuremath{A_\mathrm{B}}}
\newcommand{\Ps}[0]{\ensuremath{P_0}}
\newcommand{\Ts}[0]{\ensuremath{T_0}}

\newcommand{\MSaOp}[0]{\ensuremath{\mathcal{M}^{\ce{H2SO4}}_{\mathrm{Op.}}}}
\newcommand{\MSaTr}[0]{\ensuremath{\mathcal{M}^{\ce{H2SO4}}_{\mathrm{Tr.}}}}
\newcommand{\MWaOp}[0]{\ensuremath{\mathcal{M}^{\ce{H2O}}_{\mathrm{Op.}}}}
\newcommand{\MCf}[0]{\ensuremath{\mathcal{M}^\mathrm{CF}}}

\newcommand{\lgrt}[1]{\ensuremath{\log_{10}(#1)}}
\newcommand{\lgrtdaj}[1]{\ensuremath{\log_{10}\left(#1\right)}}

\newcommand{\life}[0]{LIFE}
\newcommand{\lifesim}[0]{LIFE\textsc{sim}}

\newcommand{\pI}[0]{Paper~I}
\newcommand{\pII}[0]{Paper~II}
\newcommand{\pIII}[0]{Paper~III}
\newcommand{\pV}[0]{Paper~V}
\newcommand{\pIIIaV}[0]{Papers~III and~V}

\definecolor{tab:OpH2SO4}{RGB}{0, 158, 115}
\definecolor{tab:OpH2SO4t}{RGB}{191, 230, 219}
\definecolor{tab:TrH2SO4}{RGB}{0, 114, 178}
\definecolor{tab:TrH2SO4t}{RGB}{191, 219, 235}
\definecolor{tab:OpH2O}{RGB}{213, 94, 0}
\definecolor{tab:OpH2Ot}{RGB}{244, 214, 191}
\definecolor{tab:CF}{RGB}{230, 159, 0}
\definecolor{tab:CFt}{RGB}{248, 230, 191}

   \title{Large Interferometer For Exoplanets (LIFE):}
   
   \subtitle{IX. Assessing the impact of clouds on atmospheric retrievals at mid-infrared wavelengths with a Venus-twin exoplanet}

   \titlerunning{LIFE: IX. Assessing the impact of clouds on atmospheric retrievals with a Venus-twin exoplanet}

   \authorrunning{Konrad et al.}

   \author{
    B.S. Konrad \inst{1,2}
    \and
    E. Alei \inst{1,2}
    \and
    S.P. Quanz \inst{1,2,3}
    \and
    P. Mollière \inst{4}
    \and
    D. Angerhausen \inst{1,2}
    \and
    J.J. Fortney \inst{5}
    \and
    K. Hakim \inst{6,7}
    \and
    S. Jordan \inst{8}
    \and
    D. Kitzmann \inst{2,9}
    \and
    S. Rugheimer \inst{10}
    \and
    O. Shorttle \inst{8,11}
    \and
    R. Wordsworth \inst{12}
   \and
   the \textit{LIFE} Collaboration\thanks{Webpage: \url{www.life-space-mission.com}}
   }

  \institute{
    ETH Zurich, Institute for Particle Physics \& Astrophysics, Wolfgang-Pauli-Str. 27, 8093 Zurich, Switzerland\\
    e-mail: \texttt{konradb@ethz.ch}; \texttt{sascha.quanz@phys.ethz.ch}
    \and
    National Center of Competence in Research PlanetS (www.nccr-planets.ch)
    \and
    ETH Centre for Origin and Prevalence of Life, Wolfgang-Pauli-Str. 27, 8093 Zurich, Switzerland
    \and
    Max-Planck-Institut f\"ur Astronomie, Königstuhl 17, 69117 Heidelberg, Germany
    \and
    Department of Astronomy and Astrophysics, University of California, Santa Cruz, CA, USA 95064
    \and
    KU Leuven, Institute of Astronomy, Celestijnenlaan 200D, 3001 Leuven, Belgium 
    \and
    Royal Observatory of Belgium, Ringlaan 3, 1180 Brussels, Belgium
    \and
    Institute of Astronomy, University of Cambridge, CB3 0HA, UK
    \and
    University of Bern, Center for Space and Habitability, Gesellschaftsstrasse 6, 3012 Bern, Switzerland
    \and
    Department of Physics and Astronomy, York University, 4700 Keele Street, North York, Ontario 3MJ 1P3, Canada
    \and
    Department of Earth Sciences, University of Cambridge, CB2 3EQ, UK
    \and
    School of Engineering and Applied Sciences, Harvard University, Cambridge, MA 02138, USA
}

   \date{Received -; accepted -} %September 15, 1996; accepted March 16, 1997}

% \abstract{}{}{}{}{} 
% 5 {} token are mandatory
 
  \abstract
  {Terrestrial exoplanets in the habitable zone are likely a common occurrence. The long-term goal is to characterize the atmospheres of dozens of such objects. The Large Interferometer For Exoplanets (\life{}) initiative aims to develop a space-based mid-infrared (MIR) nulling interferometer to measure the thermal emission spectra of such exoplanets.}
  {We investigate how well \life{} could characterize a cloudy Venus-twin exoplanet. This allows us to: (1) test our atmospheric retrieval routine on a realistic non-Earth-like MIR emission spectrum of a known planet, (2) investigate how clouds impact retrievals, and (3) further refine the \life{} requirements derived in previous Earth-centered studies.}
  {We ran Bayesian atmospheric retrievals for simulated \life{} observations of a Venus-twin exoplanet orbiting a Sun-like star located $10$~pc from the observer. The \lifesim{} noise model accounted for all major astrophysical noise sources. We ran retrievals using different models (cloudy and cloud-free) and analyzed the performance as a function of the quality of the \life{} observation. This allowed us to determine how well the atmosphere and clouds are characterizable depending on the quality of the spectrum.}
  {At the current minimal resolution (\Rv{50}) and signal-to-noise (\SNv{10} at \mic{11.2}) requirements for \life{}, all tested models suggest a \ce{CO2}-rich atmosphere ($\geq30\%$ in mass fraction). Further, we successfully constrain the atmospheric pressure-temperature (\pt{}) structure above the cloud deck (\pt{} uncertainty $\leq\pm15$~K). However, we struggle to infer the main cloud properties. Further, the retrieved planetary radius (\Rpl{}), equilibrium temperature (\Teq{}), and Bond albedo (\Ab{}) depend on the model. Generally, a cloud-free model performs best at the current minimal quality and accurately estimates \Rpl{}, \Teq{}, and \Ab{}. If we consider higher quality spectra (especially \SNv{20}), we can infer the presence of clouds and pose first constraints on their structure.}
  {Our study shows that the minimal \R{} and \SN{} requirements for \life{} suffice to characterize the structure and composition of a Venus-like atmosphere above the cloud deck if an adequate model is chosen. Crucially, the cloud-free model is preferred by the retreival for low spectral qualities. We thus find no direct evidence for clouds at the minimal \R{} and \SN{} requirements and cannot infer the thickness of the atmosphere. Clouds are only constrainable in MIR retrievals of spectra with $\SN{}\geq20$. The model dependence of our retrieval results emphasizes the importance of developing a community-wide best-practice for atmospheric retrieval studies.
  }
  
  %This allows us to test our atmospheric retrieval routine on a realistic non-Earth-like MIR emission spectrum of a known planet. Further, Venus's cloudy atmosphere requires us to add cloud treatment to our retrieval framework and allows us to investigate how clouds impact retrievals. Finally, this study enables us to further refine the \life{} requirements derived in previous Earth-centered studies.
  
   \maketitle
%
%-------------------------------------------------------------------

\section{Introduction}

One major goal for the future of exoplanet science is to constrain the atmospheric structure and composition of a statistically significant number of terrestrial exoplanets. Special attention will be given to planets within or close to the habitable zone \citep[HZ;][]{HZ_Kasting93,HZ_kopparapu13} of the host star. Such exoplanets are expected to be common within our galaxy \citep{Petigura2013,F&M2014,DressingCharbonneau2015,Bryson_2020}, and have been detected within 20 pc of the Sun \citep[e.g.,][]{2016Natur.536..437A, 2016Natur.533..221G, 2017Natur.542..456G, Gilbert_2020}. A powerful approach to characterize exoplanets is to analyze their spectra, which contain important information about relevant properties such as the atmospheric pressure-temperature (\pt{}) structure, the chemical composition, and the possible existence of clouds and their properties. If and how well an exoplanet property can be constrained depends on the wavelength regime covered by the spectrum and the accuracy with which the spectrum is measured.

For terrestrial exoplanets orbiting their host star close to or within the HZ, detections are challenging, but possible, with current and approved future ground- and space-based observatories. However, these instruments will not be capable of obtaining detailed spectroscopic measurements for several dozens of such terrestrial exoplanets. Partially motivated by this goal, there is great interest in the community to develop a new generation of observatories. HabEx \citep{2020arXiv200106683G} and LUVOIR \citep{2017AAS...22940504P}, two flagship mission concepts that aim to directly detect and characterize HZ terrestrial exoplanets in reflected light (at ultraviolet, optical, and near-infrared or UV/O/NIR wavelengths), were evaluated in the Astro 2020 Decadal Survey in the United States \citep{NAP26141}. {As a result, the % large ($\sim$6~m primary mirror),
space-based, UV/O/NIR flagship Habitable Worlds Observatory (HWO) was recommended}. Additionally, the Voyage 2050 plan of the European Space Agency \citep[ESA;][]{ESAV2050} recommended considering a large-scale, mid-infrared (MIR), space-based mission to characterize HZ terrestrial exoplanets via their thermal emission. The Large Interferometer For Exoplanets (\life{}) initiative aims to achieve this goal using a space-based MIR nulling interferometer \citep[][]{K&QLIFE, Quanz:exoplanets_and_atmospheric_characterization, LIFE_I}.

A first step in the \life{} design phase is to derive the requirements necessary to adequately characterize the atmospheres of nearby HZ terrestrial exoplanets. This includes constraining the wavelength coverage, spectral resolution, and instrument sensitivity.
%A first step in the pursuit of \life{} is to derive the requirements. This includes constraining the wavelength coverage, spectral resolution, and instrument sensitivity required for the atmospheric characterization of a terrestrial exoplanet within or close to the HZ.
Previous studies in the \life{} series \citep[][hereafter \pIII{} and \pV{}]{konrad2021large,LIFE_V} derive first estimates for the required spectral quality. These studies use atmospheric retrievals \citep[for recent reviews on retrievals, see e.g.,][]{Madhusudhan:Atmospheric_Retrieval, deming2018, barstow_heng2020} to derive quantitative estimates for important atmospheric and planetary parameters from a simulated or observed exoplanet spectrum. Both studies focus on the characterizing Earth-like planets \citep[\pIII{} – modern Earth, \pV{} – Earth at various stages of its evolution; both assume the \lifesim{} observation noise simulator from][hereafter \pII{}]{DannertLifeSim}. % INITIAL SUGGESTION: 'One limitation of focusing on Earth-like planets to derive instrumental requirements is that any such mission will not be in the position, initially, of knowing it is an Earth being observed. It is therefore not sufficient for a mission to be able to characterize an Earth-like planet, it also needs to also be able to distinguish Earth-like from non-Earth like planets in the habitable zone.'
However, a future observatory should not only be able to characterize Earth-like exoplanets, but also discern Earth-like from non Earth-like HZ exoplanets. In addition to being Earth-centric, our previous studies do not systematically investigate the effect of clouds on exoplanet characterization. Yet, since clouds influence an exoplanet's spectrum \citep[e.g.,][]{Kitzmann2011Emission,Rugheimer2013,Vasquez2013,KomacekClouds,Feinstein2022}, a more detailed study of the impact of clouds is required to derive robust requirements for \life{}.

Earth-centered retrieval studies and retrieval studies on theoretical spectra of habitable worlds are often used to investigate the characterization performance for different quality spectra \citep[e.g.,][]{Paris2013, Brandt2014, Feng_Retrieval, Quanz:exoplanets_and_atmospheric_characterization, Leger2019, CG2020, Robinson_SS_Retrievals}. Venus – to our knowledge – has not yet been considered in a comparable retrieval study. However, terrestrial exoplanets with a Venus-like insolation could maintain habitable conditions if a surface ocean is present \citep[e.g.,][]{VenusHab?1,VenusHab?2}\footnote{Yet, the existence of large bodies of surface water on early Venus and Venus-like exoplanets is uncertain and heavily debated \citep[e.g.,][]{Kasting2021Natur.598..259K, VenusHab?3}. If Venus ever had liquid surface water, it was lost in a runaway greenhouse process \citep{Kasting1988Icar...74..472K}.}. Further, exoplanets at the inner edge of the HZ (i.e., potentially Venus-like planets) are ideal targets for \life{} \citep{LIFE_I}. Finally, in contrast to theoretical planet models, Venus and its atmosphere are a known outcome of planet formation and evolution and thus provide a realistic ground-truth for a retrieval study.

Despite Venus being Earth-like in size and mass, its atmospheric state and surface conditions are vastly different. In addition to a \ce{CO2} dominated atmosphere, the mean surface pressure reaches $93$~bar, significantly exceeding that on Earth. Further, at atmospheric pressures of 0.05~bar to 1~bar, a layer of opaque \ce{H2SO4} clouds covers the planet. This opaque cloud layer can lead to an ambiguity in characterization between thick and cloudy or tenuous and cloud-free atmospheres \citep{CloudAmb1,CloudAmb2}. Further, the detectability of \ce{H2SO4} clouds is of great interest, since it could provide constraints on the amount of liquid surface water \citep{Loftus_2019}. Finally, the high atmospheric \ce{CO2} content leads to strong atmospheric greenhouse heating, which raises the mean surface temperature to a hostile $730$~K \citep[for a recent review on Venus' atmosphere, see, e.g.,][]{Taylor2018}.

In this study, we reevaluate the \life{} requirements for wavelength range, spectral resolution (\R{}), and signal-to-noise ratio (\SN{}) from \pIII{}. We focus on the key science application of distinguishing a cloudy Venus-like planet from an Earth-like planet. To this purpose, we studied a Venus-twin exoplanet using a modified version of our retrieval framework. The opaque cloud layer in Venus' atmosphere required us to model clouds in our retrievals. This also allowed us to investigate how clouds affect exoplanet characterization with \life{}. Hence, this study improves the robustness of our instrument requirements and provides new insights into difficulties in exoplanet characterization via atmospheric retrievals.

In Sect.~\ref{sec:Methods}, we introduce the model used to simulate the Venus-twin MIR emission spectrum, our retrieval framework, and the \lifesim{} observation noise model. The retrieval results for different quality spectra are presented in Sect.~\ref{sec:results}. In Sect.~\ref{sec:discussion}, we discuss implications of our results for the \life{} requirements. Important takeaway points are summarized in Sect.~\ref{sec:conclusion}.

%--------------------------------------------------------------------
\section{Methods}\label{sec:Methods}

In Sect.~\ref{sec:venus_spectrum}, we introduce the atmosphere model used to simulate Venus' MIR thermal emission spectrum and compare our spectrum to the literature. Next, we introduce our Bayesian retrieval routine and discuss updates with respect to previous versions (Sect.~\ref{sec:ret_framework}). In Sect.~\ref{sec:ret_grid}, we introduce the noise model used to generate the input spectra for the retrievals, the different atmospheric models fitted in the retrieval, and the assumed model parameter priors.

\subsection{Cloudy Venus-twin model}\label{sec:venus_spectrum}

\begin{figure}
   \centering
   \includegraphics[width=0.47\textwidth]{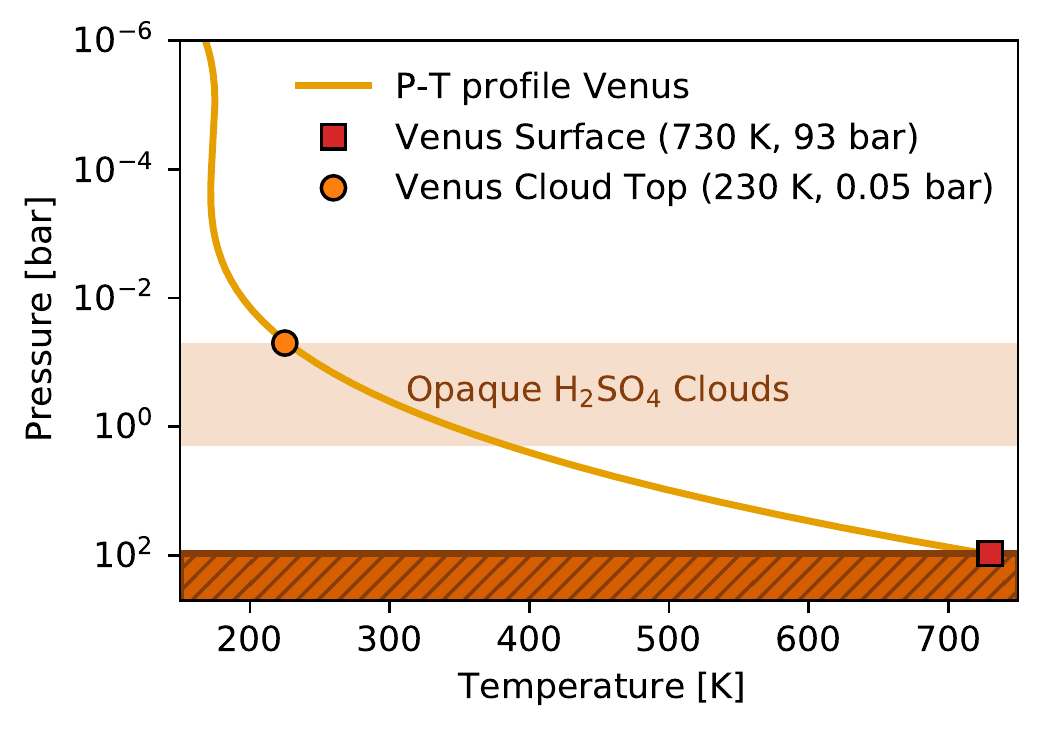}
    \caption{Schematic structure of Venus' atmospheric \pt{} profile and position of the opaque \ce{H2SO4} cloud layer.}
          \label{fig:VenusPT}%
\end{figure}

\begin{table*}
\caption{Parameters used in the different (retrieval) models, assumed true values, and prior distributions.}             % title of Table
\label{table:True_Values}      % is used to refer this table in the text
\centering                          % used for centering table
\begin{tabular}{lccc>{\centering}m{10mm}>{\centering}m{10mm}>{\centering}m{10mm}>{\centering\arraybackslash}m{10mm}}    % centered columns (4 columns)
\hline\hline                 % inserts double horizontal lines

\multirow{2}{*}{Parameter} &\multirow{2}{*}{Description} &\multirow{2}{*}{Truth}  &\multirow{2}{*}{Prior} &\multicolumn{4}{c}{Model Configuration} \\    % table heading 
\cline{5-8}
&&&&\cellcolor{tab:OpH2SO4t}\MSaOp{} &\cellcolor{tab:TrH2SO4t}\MSaTr{} &\cellcolor{tab:OpH2Ot}\MWaOp{} &\cellcolor{tab:CFt}\MCf{} \\
\hline

$a_3$           &\pt{} parameter (degree 3)  & $2.30$\tablefootmark{1}  & $\mathcal{U}(0,10)$ &\cellcolor{tab:OpH2SO4}{\textcolor{white}{$\boldsymbol{\checkmark}$}} &\cellcolor{tab:TrH2SO4}{\textcolor{white}{$\boldsymbol{\checkmark}$}} &\cellcolor{tab:OpH2O}{\textcolor{white}{$\boldsymbol{\checkmark}$}} &\cellcolor{tab:CF}{\textcolor{white}{$\boldsymbol{\checkmark}$}}\\
$a_2$           &\pt{} parameter (degree 2)  & $29.83$\tablefootmark{1}  & $\mathcal{U}(0,500)$ &\cellcolor{tab:OpH2SO4}{\textcolor{white}{$\boldsymbol{\checkmark}$}} &\cellcolor{tab:TrH2SO4}{\textcolor{white}{$\boldsymbol{\checkmark}$}} &\cellcolor{tab:OpH2O}{\textcolor{white}{$\boldsymbol{\checkmark}$}} &\cellcolor{tab:CF}{\textcolor{white}{$\boldsymbol{\checkmark}$}}\\
$a_1$           &\pt{} parameter  (degree 1) & $125.64$\tablefootmark{1} & $\mathcal{U}(0,1000)$ &\cellcolor{tab:OpH2SO4}{\textcolor{white}{$\boldsymbol{\checkmark}$}} &\cellcolor{tab:TrH2SO4}{\textcolor{white}{$\boldsymbol{\checkmark}$}} &\cellcolor{tab:OpH2O}{\textcolor{white}{$\boldsymbol{\checkmark}$}} &\cellcolor{tab:CF}{\textcolor{white}{$\boldsymbol{\checkmark}$}}\\
$a_0$           &\pt{} parameter (degree 0) & $344.94$\tablefootmark{1} & $\mathcal{U}(0,1000)$ &\cellcolor{tab:OpH2SO4}{\textcolor{white}{$\boldsymbol{\checkmark}$}} &\cellcolor{tab:TrH2SO4}{\textcolor{white}{$\boldsymbol{\checkmark}$}} &\cellcolor{tab:OpH2O}{\textcolor{white}{$\boldsymbol{\checkmark}$}} &\cellcolor{tab:CF}{\textcolor{white}{$\boldsymbol{\checkmark}$}}\\

$\lgrt{\Ps{}}$&Surface pressure $\left[\mathrm{bar}\right]$  & $1.97$\tablefootmark{2}  & $\mathcal{U}(-4,3)$ &\cellcolor{tab:OpH2SO4}{\textcolor{white}{$\boldsymbol{\times}$}} &\cellcolor{tab:TrH2SO4}{\textcolor{white}{$\boldsymbol{\checkmark}$}} &\cellcolor{tab:OpH2O}{\textcolor{white}{$\boldsymbol{\times}$}} &\cellcolor{tab:CF}{\textcolor{white}{$\boldsymbol{\checkmark}$}}\\
$\Rpl{}$&Planet radius $\left[R_\oplus\right]$& $0.95$\tablefootmark{2}  & $\mathcal{G}(0.95,0.20)$ &\cellcolor{tab:OpH2SO4}{\textcolor{white}{$\boldsymbol{\checkmark}$}} &\cellcolor{tab:TrH2SO4}{\textcolor{white}{$\boldsymbol{\checkmark}$}} &\cellcolor{tab:OpH2O}{\textcolor{white}{$\boldsymbol{\checkmark}$}} &\cellcolor{tab:CF}{\textcolor{white}{$\boldsymbol{\checkmark}$}}\\ 
$\lgrt{\Mpl{}}$&Planet mass $\left[M_\oplus\right]$ & $-0.09$\tablefootmark{2}  & $\mathcal{G}(-0.09,0.40)$ &\cellcolor{tab:OpH2SO4}{\textcolor{white}{$\boldsymbol{\checkmark}$}} &\cellcolor{tab:TrH2SO4}{\textcolor{white}{$\boldsymbol{\checkmark}$}} &\cellcolor{tab:OpH2O}{\textcolor{white}{$\boldsymbol{\checkmark}$}} &\cellcolor{tab:CF}{\textcolor{white}{$\boldsymbol{\checkmark}$}}\\

$\lgrt{\mathrm{CO_2}}$    &\ce{CO2}  mass fraction  & $-0.015$\tablefootmark{2} & $\lgrtdaj{1-10^{\mathcal{U}(-7,0)}}$ &\cellcolor{tab:OpH2SO4}{\textcolor{white}{$\boldsymbol{\checkmark}$}} &\cellcolor{tab:TrH2SO4}{\textcolor{white}{$\boldsymbol{\checkmark}$}} &\cellcolor{tab:OpH2O}{\textcolor{white}{$\boldsymbol{\checkmark}$}} &\cellcolor{tab:CF}{\textcolor{white}{$\boldsymbol{\checkmark}$}}\\
$\lgrt{\mathrm{H_2O}}$    & \ce{H2O} mass fraction  & $-4.699$\tablefootmark{2} & $\mathcal{U}(-7,0)$ &\cellcolor{tab:OpH2SO4}{\textcolor{white}{$\boldsymbol{\checkmark}$}} &\cellcolor{tab:TrH2SO4}{\textcolor{white}{$\boldsymbol{\checkmark}$}} &\cellcolor{tab:OpH2O}{\textcolor{white}{$\boldsymbol{\checkmark}$}} &\cellcolor{tab:CF}{\textcolor{white}{$\boldsymbol{\checkmark}$}}\\
$\lgrt{\mathrm{CO}}$   & \ce{CO} mass fraction    & $-4.770$\tablefootmark{2} & $\mathcal{U}(-7,0)$ &\cellcolor{tab:OpH2SO4}{\textcolor{white}{$\boldsymbol{\checkmark}$}} &\cellcolor{tab:TrH2SO4}{\textcolor{white}{$\boldsymbol{\checkmark}$}} &\cellcolor{tab:OpH2O}{\textcolor{white}{$\boldsymbol{\checkmark}$}} &\cellcolor{tab:CF}{\textcolor{white}{$\boldsymbol{\checkmark}$}}\\ 
$\lgrt{\mathrm{H_2SO_4^{cloud}}}$   & \ce{H2SO4} cloud mass fraction    & $-5.097$\tablefootmark{3} & $\mathcal{U}(-7,-1)$ &\cellcolor{tab:OpH2SO4}{\textcolor{white}{$\boldsymbol{\checkmark}$}} &\cellcolor{tab:TrH2SO4}{\textcolor{white}{$\boldsymbol{\checkmark}$}} &\cellcolor{tab:OpH2O}{\textcolor{white}{$\boldsymbol{\times}$}} &\cellcolor{tab:CF}{\textcolor{white}{$\boldsymbol{\times}$}}\\
%&(84\% \ce{H2SO4}, 26\% \ce{H2O})&&&&&&&\\
$\lgrt{\mathrm{H_2O^{cloud}}}$   & \ce{H2O} cloud mass fraction   &$-$        & $\mathcal{U}(-7,-1)$ &\cellcolor{tab:OpH2SO4}{\textcolor{white}{$\boldsymbol{\times}$}} &\cellcolor{tab:TrH2SO4}{\textcolor{white}{$\boldsymbol{\times}$}} &\cellcolor{tab:OpH2O}{\textcolor{white}{$\boldsymbol{\checkmark}$}} &\cellcolor{tab:CF}{\textcolor{white}{$\boldsymbol{\times}$}}\\
%&(100\% \ce{H2O})&&&&&&&\\
$\lgrt{\Pct{}}$    & Cloud top pressure $\left[\mathrm{bar}\right]$           & $-1.30$\tablefootmark{4} & $\mathcal{U}(-6,3)$ &\cellcolor{tab:OpH2SO4}{\textcolor{white}{$\boldsymbol{\checkmark}$}} &\cellcolor{tab:TrH2SO4}{\textcolor{white}{$\boldsymbol{\checkmark}$}} &\cellcolor{tab:OpH2O}{\textcolor{white}{$\boldsymbol{\checkmark}$}} &\cellcolor{tab:CF}{\textcolor{white}{$\boldsymbol{\times}$}}\\
$\lgrt{\Pcs{}}$    & Cloud thickness $\left[\mathrm{bar}\right]$          & $0.161$\tablefootmark{4} & $\mathcal{U}(-6,3)$ &\cellcolor{tab:OpH2SO4}{\textcolor{white}{$\boldsymbol{\checkmark}$}} &\cellcolor{tab:TrH2SO4}{\textcolor{white}{$\boldsymbol{\checkmark}$}} &\cellcolor{tab:OpH2O}{\textcolor{white}{$\boldsymbol{\checkmark}$}} &\cellcolor{tab:CF}{\textcolor{white}{$\boldsymbol{\times}$}}\\
$\lgrt{\Rcm{}}$     & Mean cloud particle radius $\left[\mathrm{cm}\right]$      & $-4.30$\tablefootmark{4} & $\mathcal{U}(-8,-3)$ &\cellcolor{tab:OpH2SO4}{\textcolor{white}{$\boldsymbol{\checkmark}$}} &\cellcolor{tab:TrH2SO4}{\textcolor{white}{$\boldsymbol{\checkmark}$}} &\cellcolor{tab:OpH2O}{\textcolor{white}{$\boldsymbol{\checkmark}$}} &\cellcolor{tab:CF}{\textcolor{white}{$\boldsymbol{\times}$}}\\
$\Sc{}$   & Log-normal particle size spread    & $1.95$\tablefootmark{4} & $\mathcal{U}(1,5)$ &\cellcolor{tab:OpH2SO4}{\textcolor{white}{$\boldsymbol{\checkmark}$}} &\cellcolor{tab:TrH2SO4}{\textcolor{white}{$\boldsymbol{\checkmark}$}} &\cellcolor{tab:OpH2O}{\textcolor{white}{$\boldsymbol{\checkmark}$}} &\cellcolor{tab:CF}{\textcolor{white}{$\boldsymbol{\times}$}}\\
\hline 
\end{tabular}
\tablefoot{{In the third column we provide the values used to generate the Venus-twin spectrum introduced in Sect.~\ref{sec:venus_spectrum}. These values provide the ground truth for all retrievals.} The fourth column lists the prior distributions used in the retrievals. With $\mathcal{U}(x,y)$, we denote a boxcar prior with lower threshold $x$ and upper threshold $y$; $\mathcal{G}(\mu,\sigma)$ indicates a Gaussian prior with mean $\mu$ and standard deviation $\sigma$. The last four columns summarize the model parameters used by the different forward models in the retrievals ({$\boldsymbol{\checkmark}$}~$=$~used, {$\boldsymbol{\times}$}~$=$~unused; see Sect.~\ref{sec:ret-models}) and the model color-coding used throughout the paper. \MSaOp{} – opaque \ce{H2SO4} clouds (84\% \ce{H2SO4}, 16\% \ce{H2O} by weight); \MSaTr{} – transparent \ce{H2SO4} clouds (84\% \ce{H2SO4}, 16\% \ce{H2O} by weight); \MWaOp{} – opaque \ce{H2O} clouds (100\% \ce{H2O}); \MCf{} – cloud-free.}

\tablebib{\tablefoottext{1}{Obtained by fitting Eq. (\ref{equ:3poly}) to the Venus \pt{} profile in Figure 1 of \cite{Mueller-Wodarg:SS_PT_Profiles};} \tablefoottext{2}{Near surface abundances from NASA's planet factsheet: \url{https://nssdc.gsfc.nasa.gov/planetary/factsheet/venusfact.html};}\tablefoottext{3}{Mean of values from \citet{OSCHLISNIOK2012940} and \citet{KRASNOPOLSKY2015327};}\tablefoottext{4}{Mean values for the mode 1 cloud particles in \citet{Esposito1983}.}}
\end{table*}

As in \pIIIaV{}, we used the 1D radiative transfer code \texttt{petitRADTRANS} \citep[][]{Molliere:petitRADTRANS, Molliere:petitRADTRANS2,LIFE_V} to model the MIR thermal emission spectrum of our Venus-twin exoplanet. \texttt{petitRADTRANS} passes a featureless black-body spectrum at the surface temperature through discrete atmospheric layers and models the interaction of each layer with the radiation. Further, it accounts for the scattering of photons by the atmosphere and the surface. This yields the MIR emission spectrum at the top of the atmosphere. Each layer is characterized by its temperature, pressure, and the opacity sources that are present. We provide a list of all model parameters along with the assumed true values in Table~\ref{table:True_Values}.

We parametrized Venus' \pt{} structure of using a polynomial. In \pIIIaV{}, this approach allowed us to minimize the number of model parameters and thus reduce the computational complexity of our retrieval. An extensive discussion, justifying this choice of \pt{} parametrization, is provided in the appendix of \pIII{}. Since, in contrast to Earth, Venus' \pt{} profile does not exhibit a temperature inversion, a third order polynomial (four parameters) is sufficient for the present study:
\begin{equation}\label{equ:3poly}
 T(P)=\sum_{i=0}^3a_iP^i.
\end{equation}
Here, $P$ is the pressure and $T$ the corresponding temperature of each atmospheric layer. The $a_i$ terms are the parameters of the polynomial \pt{} model. In Fig.~\ref{fig:VenusPT}, we show our ground-truth Venus \pt{} profile, which we obtained by fitting the polynomial model to the Venus \pt{} profile from \citet{Mueller-Wodarg:SS_PT_Profiles}. The corresponding $a_i$ values are given in Table~\ref{table:True_Values}.

To simulate the MIR emission of a Venus-twin, we accounted for different opacity sources. First, we modeled the MIR absorption and emission features of \ce{CO2}, \ce{H2O}, and \ce{CO} (see Table~\ref{table:True_Values} for the assumed mass fractions and Table~\ref{table:M-Opacities} for the used line lists, broadening coefficients, and line cutoffs). For all three gases, we assumed constant vertical abundance profiles. Second, we modeled spectral features from collision-induced absorption (CIA) by \ce{CO2}, and Rayleigh scattering by all three molecules (see Table~\ref{table:CIA,C,R-Opacities} for the used opacities). Third, we considered the opaque \ce{H2SO4} clouds (see Fig.~\ref{fig:VenusPT}), which is essential to accurately model the MIR thermal emission of Venus. We accounted for the \ce{H2SO4} clouds by adding a cloud slab to the atmosphere, which spanned multiple atmospheric layers. The cloud slab was characterized using five parameters: the pressure at the cloud-top \Pct{}, the thickness of the cloud layer \Pcs{} in bar, the mass fraction of the cloud forming substance, the mean cloud-particle radius \Rcm{}, and the standard deviation \Sc{} of the log-normal cloud-particle size distribution. The parameter \Pct{} defined the uppermost atmospheric layer that contained clouds, while the difference \Pct{}$-$\Pcs{} the corresponding lowermost layer. Throughout the cloud slab defined by these two parameters, we assumed a constant mass fraction of the cloud forming \ce{H2SO4}$-$\ce{H2O} solution (84\% \ce{H2SO4}, 16\% \ce{H2O} by weight). All other atmospheric layers were modeled to be cloud-free (mass fraction of the cloud forming substance set to zero). Our cloud model further assumed homogeneous, spherical cloud particles of variable size. %The size distribution of the cloud particles was modeled to follow a log-normal distribution defined by the mean particle radius \Rcm{} and the standard deviation \Sc{}.
We assumed both \Rcm{} and \Sc{} to be constant throughout the cloud deck. We calculated the pressure- and temperature-dependent opacities for the different cloud particle sizes from the wavelength-dependent index of refraction (see Table~\ref{table:CIA,C,R-Opacities} for sources) using Mie scattering theory. For this calculation, we relied on the software presented in \citet{Min2005}, which uses the codes from \citet{Toon:81}. Thereafter, we used the standard \texttt{petitRADTRANS} cloud modeling pathway to include clouds in the MIR Venus-twin spectrum.

\begin{table}
\caption{Molecular line opacities used to calculate MIR spectra.}             % title of Table
\label{table:M-Opacities}      % is used to refer this table in the text
\centering % used for centering table

\begin{tabular}{lccc}    % centered columns (4 columns)
\hline\hline                 % inserts double horizontal lines
Molecule    &Line List  &Pressure broadening        &Line Cutoff\\
\hline
\ce{CO2}    &HITEMP     &$\gamma_{\mathrm{air}}$    &BU69\\
\ce{H2O}    &HITEMP     &$\gamma_{\mathrm{air}}$    &HA02\\
\ce{CO}     &HITEMP     &$\gamma_{\mathrm{air}}$    &HA02\\
\hline 
\end{tabular}
\tablebib{(HITEMP) \citet{ROTHMAN20102139}; (BU69) \citet{Burch:69}; (HA02) \citet{HB02}. }
\end{table}

%Comment by Robin: I wonder if the fact that all of these coefficients are for Earth air has an effect on the results.

%We tested this as well and it only has a very minor effect on the overall MIR spectrum

\begin{table}
\caption{Continuum opacities used to calculate MIR spectra.}             % title of Table
\label{table:CIA,C,R-Opacities}      % is used to refer this table in the text
\centering % used for centering table
                      % used for centering table
\begin{tabular}{lcc}    % centered columns (4 columns)
\hline\hline                 % inserts double horizontal lines
Opacity type    &Material               &Reference\\
\hline
CIA             &\ce{CO2}-\ce{CO2}      &KA19\\
Cloud           &\ce{H2SO4} (liquid)    &PW75\\
Cloud           &\ce{H2O} (liquid)      &SE81\\
Rayleigh        &\ce{CO2}               &SU05\\
Rayleigh        &\ce{H2O}               &HA98\\
Rayleigh        &\ce{CO}                &SU05\\

\hline 
\end{tabular}
\tablefoot{The cloud opacities were calculated from the indices of refraction of the cloud species via Mie scattering theory.}
\tablebib{(KA19) \citet{KARMAN2019160}; (PW75) \citet{Palmer75H2SO4}; (SE81) \citet{siegel1981H2O}; (SU05) \citet{2005JQSRT..92..293S}; (HA98) \citet{Harvey1998}; }
\end{table}

%Comment by Robin: How large an effect does this have on the emission spectrum? Venus CO2 CIA is still pretty uncertain, at least for climate applications.

%The effect of the CO2 CIA on the MIR emission spectrum is negligible. We include it for completeness

\begin{figure}
   \centering
   \includegraphics[width=0.47\textwidth]{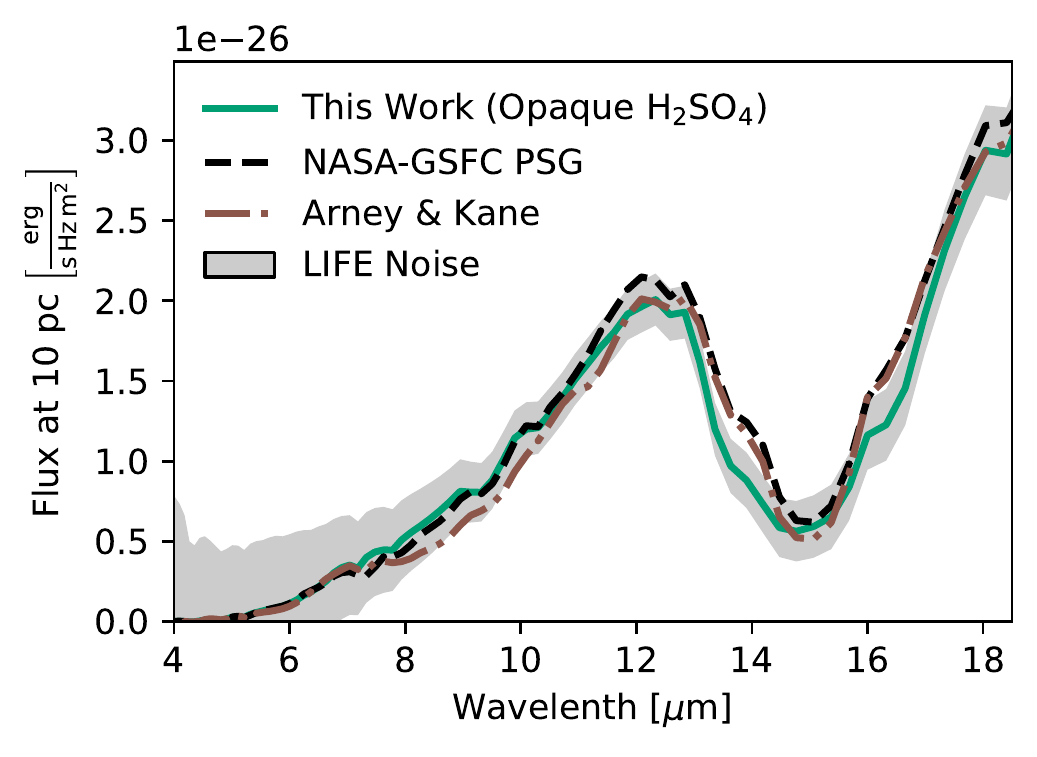}
    \caption{Comparison of our Venus-twin MIR spectrum (opaque \ce{H2SO4} clouds) to other models. We show the flux received by an observer located 10 pc from the planet. The solid line is the MIR thermal emission spectrum of our model (calculated assuming the true values listed in Table~\ref{table:True_Values}). The gray-shaded region indicates the \SNv{10} \lifesim{} uncertainty considered for our retrievals (see Sect.~\ref{sec:input_spectra}). The black-dashed line is the standard Venus emission spectrum from the NASA-GSFC Planetary Spectrum Generator (PSG) \citep[\url{https://psg.gsfc.nasa.gov};][]{Villanueva2018PSG}.%, which assumes altitude-dependent abundances \citep{EhrenreichVenusAbunds} and volcanic clouds.
    The brown-dashed-dotted line is the MIR Venus spectrum from \citet{arney2018venus}.%, which assumes altitude-dependent abundances and multi-layer \ce{H2SO4} clouds.
    }
          \label{fig:SpectraComparison}%
\end{figure}

In Fig.~\ref{fig:SpectraComparison} we compare our MIR Venus-twin spectrum, to the simulated Venus spectra from NASA's Planetary Spectrum Generator \citep[PSG\footnote{\url{https://psg.gsfc.nasa.gov}};][]{Villanueva2018PSG} and \citet{arney2018venus}. In contrast to our Venus-twin atmosphere, both models included additional atmospheric isotopes and trace gases and assumed altitude-dependent abundance profiles for all gases. Additionally, both models assumed a more complex atmospheric cloud structure (PSG: altitude-dependent volcanic clouds; \citet{arney2018venus}: altitude-dependent multilayer \ce{H2SO4} solution clouds). We observe that the general shape of all three spectra is comparable and the differences are smaller or of similar magnitude as the assumed noise level. Furthermore, our MIR Venus-twin spectrum is not missing any significant spectral absorption features, despite not taking into account various atmospheric species and isotopes. There are two possible explanations for this finding: either these species have no significant spectral lines in the MIR (e.g., \ce{O2} and \ce{N2}) or their atmospheric abundance above the opaque cloud layer is too low to cause a noticeable signature in Venus' MIR spectrum (e.g., \ce{SO2} and \ce{O3}).
% Comments:
%eleonora.alei: also differences in the continuum might be due to the fact that PSG considers refraction and pRT doesn't (but I should have the PSG folks' opinion)
%molliere: One more thing is that CO2 self-broadening is about twice as strong as CO2 air broadening, this likely plays a role, see https://arxiv.org/abs/1809.02548v1, Table 1. I have no intuition for how the refraction would / should change things.
%daniel.kitzmann: Yes, indeed. For a CO2-dominated atmosphere, the line broadening, line shapes, and CIA should be treated differently. However, given the fact that the forward model and the retrieval use the same opacities here, this shouldn't matter much. It would be more important if self-consistent atmosphere models would be computed here.
This finding justifies our approach of excluding these additional molecules from our Venus-twin model. Between \mic{6} and \mic{12}, we observe minor differences between the spectra. Since the spectrum in this wavelength range is predominantly determined by the clouds, the observed variance is likely rooted in the differences between the three cloud models. Additionally, we observe differences in the \ce{CO2} absorption feature between \mic{13} and \mic{17}. While the PSG and the \citet{arney2018venus} models yield similar results, our model deviates. This deviance is most likely evoked by differences in the assumed \pt{} profiles, but might also be partially due to differences between the line lists, pressure broadening coefficients, or line cutoffs. However, since we are interested in assessing the impact of clouds on exoplanet characterization and whether the first requirements from \pIIIaV{} are sufficient, the deviances are negligible for this study.

\subsection{Bayesian retrieval framework}\label{sec:ret_framework}

The study we present here relied on our Bayesian retrieval routine, which we first introduced in \pIII{}. Our routine relied on two subroutines. First, it applied the 1D radiative transfer code \texttt{petitRADTRANS} \citep{Molliere:petitRADTRANS, Molliere:petitRADTRANS2, LIFE_V} to calculate the theoretical MIR spectrum corresponding to a given combination of values of the forward model parameters listed in Table~\ref{table:True_Values}. Second, the routine used \texttt{pyMultiNest} \citep{Buchner:PyMultinest}, which efficiently samples the parameter space spanned by the prior probability distributions (or "priors") of the forward model parameters to determine parameter combinations that fit the simulated Venus-twin observation well. This yielded posterior probability distributions (or "posteriors") for the model parameters and estimates for the Bayesian evidence $\mathcal{Z}$ of the model. The posteriors tell us how likely different combinations of model parameter values are. The evidence $\mathcal{Z}$ measures how well the model fits the input spectrum and can be used for model selection (see Sect.~\ref{sec:ret-models}). The \texttt{pyMultiNest} package is based on the \texttt{MultiNest} \citep{Feroz:Multinest} implementation of the Nested Sampling algorithm \citep{Skilling:Nested_Sampling}. In all retrievals we performed throughout this study, we ran \texttt{pyMultiNest} using 700 live points and a sampling efficiency of 0.3 (suggested for evidence evaluation by the \texttt{MultiNest} documentation\footnote{\url{https://github.com/farhanferoz/MultiNest}}). An in depth description of our atmospheric retrieval routine can be found in \pIII{} of the \life{} series. Here, we focus on the updates and improvements implemented since \pIII{}.

In \pIII{}, we did not consider the impact surface, atmospheric, and cloud scattering processes have on the MIR thermal emission. Including scattering in the forward model, would have significantly increased the required computation time per spectrum, making the study presented in \pIII{} unfeasible. However, in both \pIIIaV{}, the surface and atmospheric scattering effects were shown to be negligible for the MIR thermal emission. %at the considered resolutions and noise levels.
Further, only cloud-free forward models were considered, which justified neglecting cloud scattering. However, for the cloudy Venus-twin retrievals we performed in this study, cloud scattering effects were no longer negligible. The limiting factor for the retrievals presented in \pIII{} was that \texttt{petitRADTRANS} calculated spectra at a predetermined resolution of \Rv{1000}. Our retrieval routine then binned down the spectra to the resolution of the input spectrum. More recent improvements to \texttt{petitRADTRANS} enabled us to compute spectra directly at the resolution of the input spectrum (see \pV{} for further information). With this update, the computation time required for one cloudy spectrum including scattering at \Rv{50} dropped from $\approx$~20.0~seconds to just $\approx$~1.5~seconds. This reduction in the computation time per spectrum enabled us to run retrievals accounting for scattering effects, making the present study feasible. {We validate the updated retrieval routine in Appendix~\ref{sec:validation}.}

\subsection{Retrieval setup}\label{sec:ret_grid}

In Sect.~\ref{sec:input_spectra}, we discuss the Venus-twin input spectrum and the noise model used in the retrievals. Thereafter, we introduce four atmospheric models we used as forward models during the retrievals (Sect.~\ref{sec:ret-models}). Lastly, the prior distributions assumed in our retrievals are motivated in Sect.~\ref{sec:input_priors}.

\subsubsection{Input spectra and noise terms}\label{sec:input_spectra}

We generated the Venus-twin input spectra for our retrieval study using \texttt{petitRADTRANS} and the cloudy Venus-twin model introduced in Sect.~\ref{sec:venus_spectrum}.
%For the validation retrieval discussed in Sect.~\ref{sec:validation}, we studied a \mic{3-20}, \Rv{1000}, and \SNv{50} Venus-twin spectrum. For the \life{} retrieval studies presented in Sect.~\ref{sec:results}, we first studied a spectrum that matched the minimal requirements determined in \pIIIaV{} (\mic{4-18.5}, \Rv{50}, and \SNv{10}). The goal was to understand if the minimal design requirements from \pIII{} are sufficient to characterize a Venus-twin exoplanet. In a second step, we explored whether the characterization of a Venus-twin with \life{} is improved significantly when considering spectra of higher quality. To this purpose, we ran retrievals for a grid of input spectra. The grid covered all possible combinations of the following spectral parameters: Wavelength ranges – \mic{3-20}, and \mic{4-18.5}; spectral resolution (\R{}) – $50$, and $100$; signal-to-noise ratio (\SN{}) – $10$, $15$, and $20$.
We defined the resolution \R{} of a spectrum as $\lambda/\Delta\lambda$, where $\lambda$ was the wavelength at the center of a wavelength bin and $\Delta\lambda$ was the bin width. {Further, we used noise models to estimate the wavelength dependent \SN{}. We defined the \SN{} of the input spectrum as the \SN{} at the \mic{11.2} reference bin, since this bin did not coincide with strong absorption features from the considered atmospheric species.}% We then scaled the \SN{} for all other wavelength bins according to the assumed noise model.

For all \life{} retrievals in Sect.~\ref{sec:results}, we used the \lifesim{} noise model introduced in \pII{}. \lifesim{} provides estimates for the wavelength-dependent \SN{} expected for observations with \life{} by accounting for noise contributions from the photon noise of the planet's emission, stellar leakage, and local- as well as exozodiacal dust emission. We hence implicitly assumed that a large future space mission like \life{} will not be dominated by instrumental noise terms (Dannert et al., in prep.). {Possible consequences of this assumption are mentioned in Sect.~\ref{sec:limitations}}. For our study, we assumed a Venus-twin exoplanet orbiting a G2V Star on an $0.72$~AU orbit at a distance of $10$~pc from the observer. We further set the exozodiacal dust emission to be three times the level of the local zodiacal light. This value corresponds to the median exozodi level found for Sun-like stars in the HOSTS survey \citep{ertel2020}.

In all retrievals, we interpreted the noise as uncertainty to the simulated spectral points and assumed that the noise does not impact the predicted flux values. As discussed in \citet{Feng_Retrieval} and the Appendix of \pIII{}, randomizing the individual spectral data points according to the \SN{} would simulate more accurate observational instances. However, a retrieval study based on a single noise instance will result in biased estimates for the retrieval's characterization performance due to the random placement of the few spectral points. An ideal retrieval study should thus consider multiple ($\geq10$) different noise realizations of each input spectrum and evaluate the instrument performance by considering the average retrieved parameter posterior. However, the vast number of different retrievals ({12 retrievals for each of the four models introduced in Sect.~\ref{sec:ret-models} resulting in a total of 48 retrievals}) we executed for this study and the average computation time per retrieval ($\sim$~1~day on 20~CPUs) made such a study computationally unfeasible ($\geq5$~months of total cluster time). In addition, in the Appendix of \pIII{}, we motivated that by retrieving the unrandomized input spectra we obtain reliable estimates for the average expected retrieval performance.

\subsubsection{Atmospheric forward models in the retrievals}\label{sec:ret-models}

To test our retrieval framework's sensitivity for Venus' clouds, we analyzed how our routine performed for different atmospheric forward models. This approach enabled us to test if the \life{} design requirements from \pIIIaV{} are sufficient to infer the presence of clouds and to accurately characterize the clouds in the atmosphere of a Venus twin. Additionally, this approach provided us with important new insights into the biases that arise when assuming an incorrect atmospheric model in a retrieval study. We considered four atmospheric forward models (see Table~\ref{table:True_Values} for the parameter configuration of each model):
\begin{enumerate}
    \item \textit{Opaque \ce{H2SO4} clouds} – (14 parameters; \MSaOp{}): As is true for Venus, we assumed that opaque \ce{H2SO4} clouds blocked the contributions from the lower atmospheric layers and surface to the outgoing MIR emission spectrum. By fixing the surface pressure $P_0$ to an arbitrary $10^4$ bar, we forced the retrieval to add an opaque cloud layer to the atmosphere.
    
    \item \textit{Transparent \ce{H2SO4} clouds} – (15 parameters; \MSaTr{}): In contrast to the opaque model, we assumed that contributions from the lower atmosphere are not fully blocked. Therefore, we tried to retrieve for the surface pressure $P_0$.
    
    \item \textit{Opaque \ce{H2O} clouds} – (14 parameters; \MWaOp{}): Similar to the opaque \ce{H2SO4} model, we assumed an opaque cloud layer to be present and fixed the surface pressure $P_0$ to $10^4$ bar. However, we assumed pure \ce{H2O} clouds to determine if we could identify the correct cloud species in retrievals.
    
    \item \textit{Cloud-free} – (ten parameters; \MCf{}): We assumed no clouds to be present in the atmosphere. With this model, we investigated whether the presence of clouds in an atmosphere can be inferred at the considered input qualities.
\end{enumerate}
We used Bayesian model selection to determine which model performed best as a function of the quality of the input spectrum. If we run two retrievals that assume different atmospheric models $\mathcal{A}$ and $\mathcal{B}$, both retrieval results will be characterized by their evidences $\ln\left(\mathcal{Z_A}\right)$ and $\ln\left(\mathcal{Z_B}\right)$. We can use the evidences to identify the better fitting model via the Bayes factor $K$:
\begin{equation}\label{eq:BayesFactor}
    \lgrt{K}=\frac{\ln\left(\mathcal{Z}_{\mathcal{A}}\right)-\ln\left(\mathcal{Z}_{\mathcal{B}}\right)}{\ln\left(10\right)}.
\end{equation}
The Jeffreys scale \citep[][see Table~\ref{Table:Jeffrey}]{Jeffreys:Theory_of_prob} provides a possible interpretation for the value of the Bayes factor $K$.

\begin{table}
\caption{Jeffreys scale \citep{Jeffreys:Theory_of_prob}.}% title of Table
\label{Table:Jeffrey}      % is used to refer this table in the text
\centering                          % used for centering table
\begin{tabular}{l c c}        % centered columns (4 columns)
\hline\hline                 % inserts double horizontal lines
$\lgrt{K}$ &Probability &Strength of Evidence\\    % table heading 
\hline 
   $<0$     &$<0.5$         &Support for $\mathcal{B}$\\
   $0-0.5$  &$0.5-0.75$     &Very weak support for $\mathcal{A}$\\
   $0.5-1$  &$0.75-0.91$    &Substantial support for $\mathcal{A}$\\
   $1-2$    &$0.91-0.99$    &Strong support for $\mathcal{A}$\\
   $>2$     &$>0.99$        &Decisive support for $\mathcal{A}$\\ 
\hline 
\end{tabular}
\tablefoot{Scale for interpreting the values of the Bayes' factor $\lgrt{K}=\left(\ln\left(\mathcal{Z}_{\mathcal{A}}\right)-\ln\left(\mathcal{Z}_{\mathcal{B}}\right)\right)/\ln\left(10\right)$. The scale is symmetrical, i.e., negative values of $\lgrt{K}$ correspond to very weak, substantial, strong, or decisive support for model $\mathcal{B}$.}
\end{table}

\subsubsection{Prior distributions}\label{sec:input_priors}

In Table~\ref{table:True_Values}, we provide a summary of the assumed prior distributions, which define the range of parameter space sampled by \texttt{pyMultiNest}. For the \pt{} parameters $a_i$ and the surface pressure $P_0$, we chose broad uniform priors such that the corresponding \pt{} profiles covered a wide range of atmospheric structures. For the abundances of the atmospheric species and the cloud parameters, we assumed broad and uniform priors that spanned large regions of parameter space. In contrast to \pIII{}, we assumed narrower abundance priors for \ce{H2O} and \ce{CO}, %The previous lower abundance limit of $10^{-15}$ was excessive, given that the assumed true abundances of all considered gases were much higher.
since the lowest abundance detectable by our retrieval routine for the quality of input spectra considered is approximately $10^{-7}$ in mass fraction (cf. \pIII{}). For \ce{CO2}, we used a prior that covers the full abundance range, but samples high abundances more densely. This prior allows us to better estimate the \ce{CO2} abundance and better identify a potential upper limit.

As in \pIIIaV{}, we chose a Gaussian prior for \Rpl{}. For the mean, we assumed Venus' true radius, for the standard deviation of 20\% of the true value. This choice for \Rpl{} was motivated by findings presented in \pII{}, which demonstrated that the detection of a planet during \life{}'s search phase would yield such constraints for \Rpl{} (for a terrestrial planet around the HZ, we expect a radius estimate $R_\mathrm{est}$ for the true radius $R_\mathrm{true}$ with $R_\mathrm{est}/R_\mathrm{true}=0.97\pm0.18$). We then used the \Rpl{} prior to derive a Gaussian prior for the planet mass \lgrt{\Mpl{}} using the statistical mass-radius relation \texttt{Forecaster}\footnote{\url{https://github.com/chenjj2/forecaster}} \citep{Kipping:Forecaster}.

%--------------------------------------------------------------------
\section{Retrieval results}\label{sec:results}

Here, we present the retrieval results for the Venus-twin mock observations with \life{} for the different assumed forward models (see Sect.~\ref{sec:ret-models} and Table~\ref{table:True_Values}). In Sect.~\ref{sec:LIFEsim_papIII_ret_res}, we discuss the results obtained for a Venus-twin spectrum at the minimal \life{} requirements determined in \pIII{} (\mic{4-18.5}, \Rv{50}, and \SNv{10}). Thereafter, in Sect.~\ref{sec:high_LIFEsim_res}, we discuss if and how the retrieval's characterization performance is improved when considering higher quality spectra. To this purpose, we ran retrievals for various spectra of different wavelength coverage (\mic{4-18.5}, \mic{3-20}), \R{} ($50$, $100$), and \SN{} ($10$, $15$, $20$).

\subsection{Results for the current minimal \life{} requirements}\label{sec:LIFEsim_papIII_ret_res}

\begin{figure*}
\centering
\includegraphics[width=0.975\textwidth]{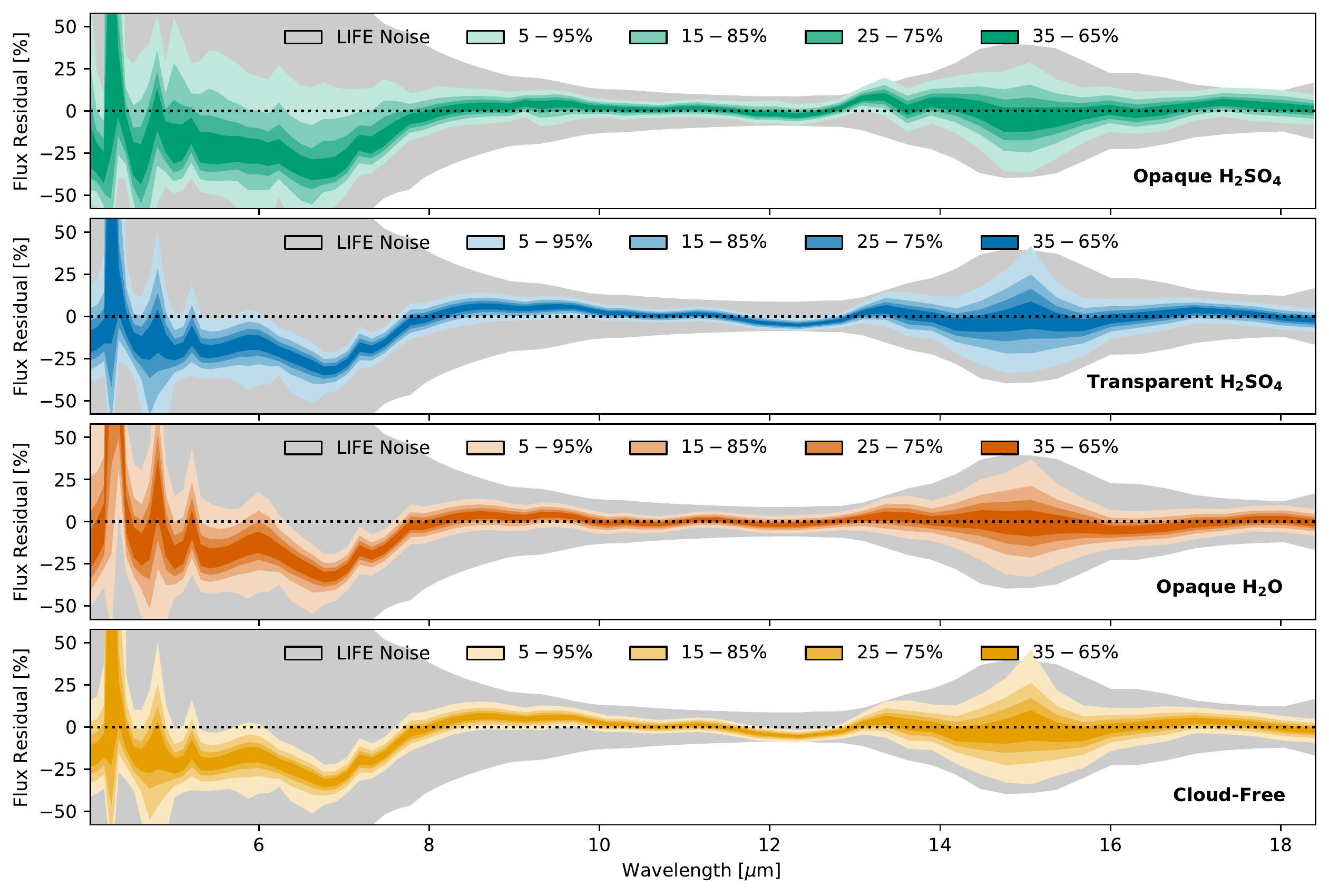}
\caption{Residuals of the spectra corresponding to the posteriors retrieved for the \mic{4-18.5}, \Rv{50}, \lifesim{} \SNv{10} Venus-twin spectrum (opaque \ce{H2SO4} clouds). Color-shaded areas represent different residual quantiles. The gray area marks the 1$\sigma$ \lifesim{} noise level. Each panel contains the results for a different forward model (see Sect.~\ref{sec:ret-models}). From top to bottom: Opaque \ce{H2SO4} clouds, transparent \ce{H2SO4} clouds, opaque \ce{H2O} clouds, and cloud-free.}
\label{fig:LIFEsim_spec_resid}
\end{figure*}

\begin{figure*}
\centering
\includegraphics[width=0.98\textwidth]{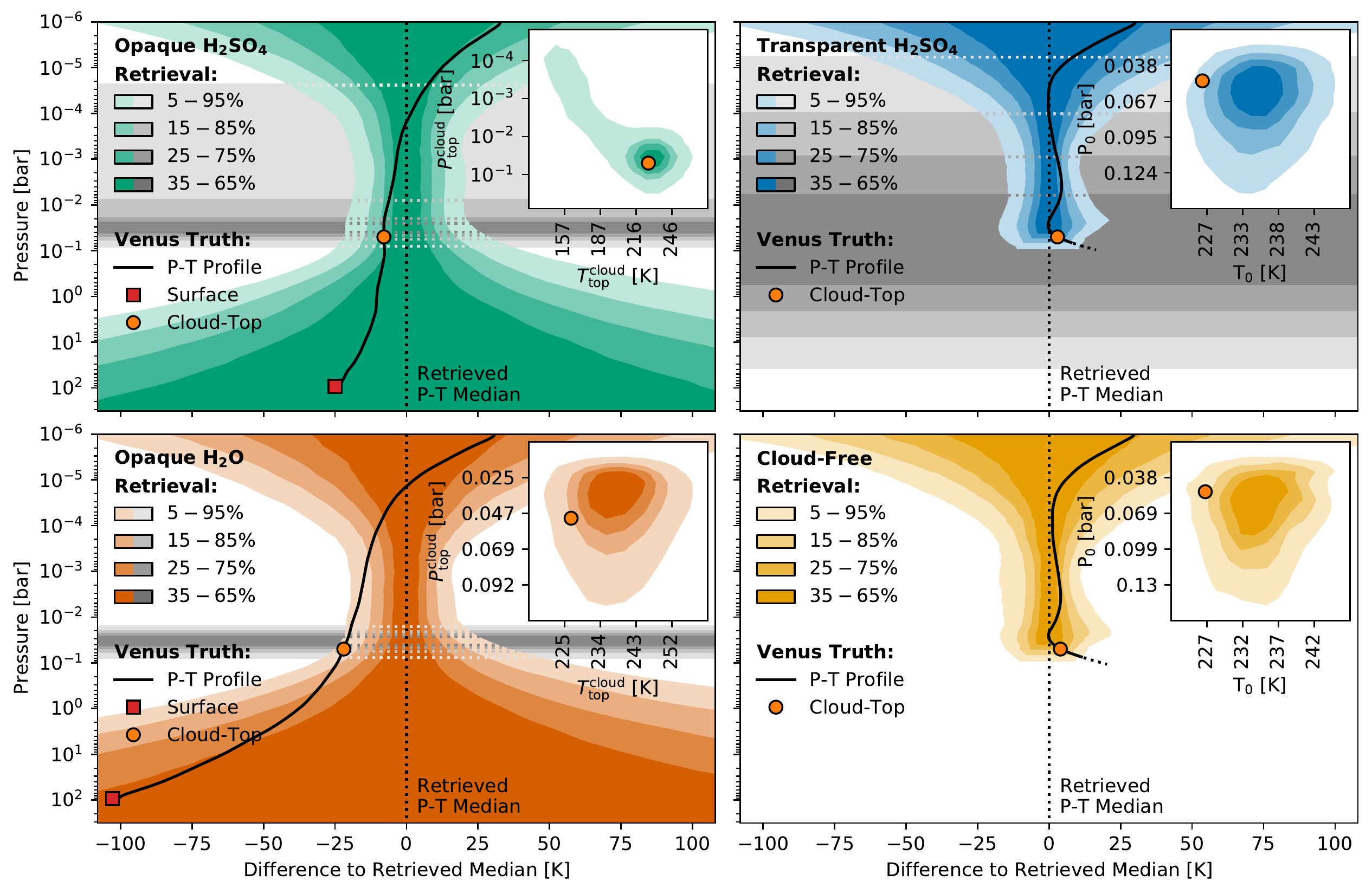}
\caption{Residuals of the \pt{} profiles retrieved for the \mic{4-18.5}, \Rv{50}, \lifesim{} \SNv{10} Venus-twin spectrum (opaque \ce{H2SO4} clouds) relative to the retrieved median \pt{} profile. Color-shaded areas indicate percentiles of the \pt{} residuals. If the model includes clouds, the gray shaded regions indicate percentiles of the retrieved cloud-top pressure. The solid black line, the orange circular marker, and the red square marker represent the true profile relative to the retrieved median (for the transparent \ce{H2SO4} cloud and cloud-free models, we cannot plot the true surface and the \pt{} profile at pressures higher than retrieved surface pressure). In the top right, we plot the 2D \Ps{}-\Ts{} posterior (if retrieved; otherwise \Pct{}-\Tct{}). Each panel summarizes the result for one of the four different forward models (see Sect.~\ref{sec:ret-models}). From top-left to bottom-right: Opaque \ce{H2SO4} clouds, transparent \ce{H2SO4} clouds, opaque \ce{H2O} clouds, and cloud-free.}
\label{fig:LIFEsim_pt_resid}
\end{figure*}

\begin{figure*}
\includegraphics[width=0.96\textwidth]{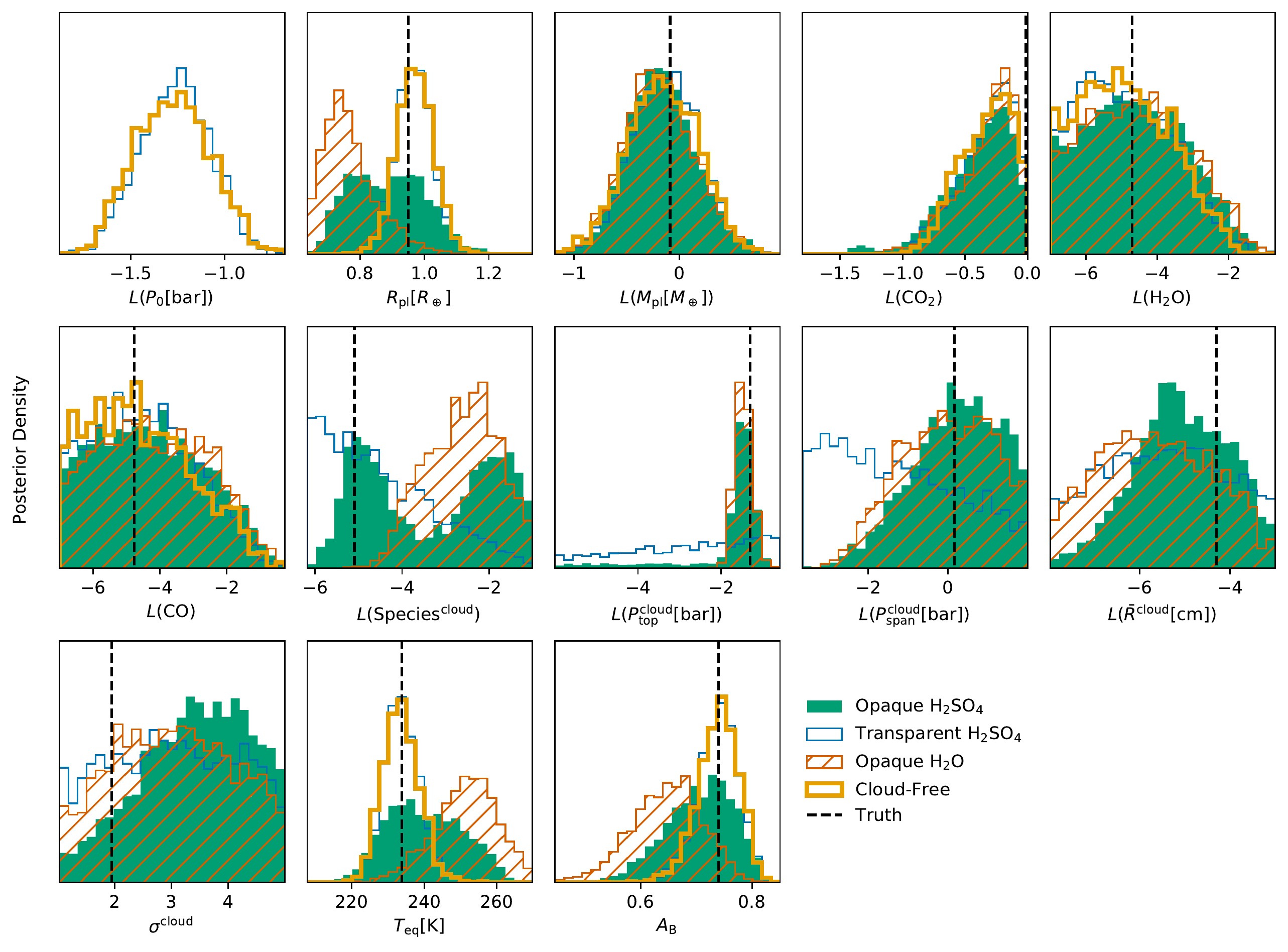}
\caption{Posteriors retrieved for the \mic{4-18.5}, \Rv{50}, \lifesim{} \SNv{10} Venus-twin spectrum (opaque \ce{H2SO4} clouds) for different forward models (see Sect.~\ref{sec:ret-models}). Here, $L(\cdot)$ abbreviates $\lgrt{\cdot}$. We include the planetary equilibrium temperature \Teq{} and the Bond albedo \Ab{}, which we derived from the posteriors (see Appendix~\ref{sec:albedo_calc}). Dashed black lines indicate the true values (see Table~\ref{table:True_Values}). For $\mathrm{Species^{cloud}}$, the true value is the \ce{H2SO4} mass fraction in the Venus-twin model. Solid green area $-$ opaque \ce{H2SO4} clouds; thin-blue outlined area $-$ transparent \ce{H2SO4} clouds; orange-hatched area $-$ opaque \ce{H2O} clouds; thick-yellow outlined area $-$ cloud-free.}
\label{fig:LIFEsim_post}
\end{figure*}

We present the retrieval results for the different forward models at the minimal \life{} design requirements (\mic{4-18.5}, \Rv{50}, \SNv{10}) in Figs.~\ref{fig:LIFEsim_spec_resid} (spectrum residuals), \ref{fig:LIFEsim_pt_resid} (\pt{} profile residuals), and \ref{fig:LIFEsim_post} (posteriors). The full corner plots, the absolute retrieved \pt{} profiles and spectra, the wavelength- and pressure-dependent contribution to the emission spectrum, and tables with the retrieved values can be found in Appendix~\ref{app:corners}.

\subsubsection{Fit to the Venus-twin spectrum}\label{sec:ret_spec_resid}

The spectrum residuals in Fig.~\ref{fig:LIFEsim_spec_resid} show that the fit of all four forward models to the Venus-twin spectrum lies well within the \lifesim{} noise level. This indicates that all models can reproduce the Venus-twin input with sufficient accuracy. Above \mic{8}, the retrieved quantile envelopes of all forward models are similar, roughly centered on the truth, and smaller than the \lifesim{} noise. Below \mic{8}, the quantile envelopes become larger and deviate more from the truth as the \lifesim{} noise level increases. {We discuss the origin of these deviations in Sect.~\ref{sec:ret_param_post}.} The spread of the quantiles is largest for the true model (opaque \ce{H2SO4} clouds). For the other models, the spread is smaller, but the residual deviates from the input. This indicates that the three models cannot reproduce the input accurately. However, due to the large \lifesim{} errors in this wavelength range, these deviations will not affect the retrieval performance significantly.

\subsubsection{Retrieved P–T structure}

The \pt{} profile residuals in Fig.~\ref{fig:LIFEsim_pt_resid} show that the fit of all four models is best above Venus' cloud layer (roughly between $10^{-1}$~bar and $10^{-4}$~bar). The retrieved means for the cloud-free and \ce{H2SO4} cloud forward models lie within $\pm15$~K of the truth. For the opaque \ce{H2O} cloud model, the deviations from the truth are larger (within $\pm25$~K). The uncertainty on the retrieved \pt{} structure in this pressure range is roughly $\pm15$~K for all four forward models. When considering higher or lower pressures, the deviations from the truth increase and the uncertainties grow. This behavior is due to a lack of significant spectral features from the high and low pressure atmospheric layers in the Venus-twin MIR spectrum (see emission contribution plots in Appendix~\ref{app:corners}). The constraints on the percentile envelopes for these layers stem from extrapolation of the \pt{} model (nonphysical polynomial model). Thus, we cannot trust the \pt{} predictions for these atmospheric layers and cannot estimate Venus' surface pressure \Ps{} and temperature \Ts{} accurately.

For the cloud-free and the transparent \ce{H2SO4} cloud model, the retrieved \Ps{} roughly corresponds to the position of the cloud-top in Venus' atmosphere and the retrieved \Ts{} slightly overestimates the cloud-top temperature (by roughly 10~K for both models). Additionally, for the transparent \ce{H2SO4} cloud model, the large spread in the retrieved cloud-top pressure and temperature indicates that these two parameters are no longer well constrained. In contrast, \Pct{} is accurately retrieved and \Tct{} is slightly overestimated (by roughly 25~K for \ce{H2O} and 10~K for \ce{H2SO4} clouds) in the retrievals that assume opaque cloud forward models. Thus, we find accurate estimates for the position of the cloud-top in all retrievals.

\subsubsection{Retrieved parameter posteriors}\label{sec:ret_param_post}

Lastly, we consider the retrieved posteriors displayed in Fig.~\ref{fig:LIFEsim_post}. The figure further includes distributions for the planetary equilibrium temperature \Teq{} and Bond albedo \Ab{}, which we derived from the posteriors using the method outlined in Appendix~\ref{sec:albedo_calc}.

First, we consider the results for the planetary surface pressure \Ps{}. We see, that if \Ps{} is a model parameter (transparent \ce{H2SO4} clouds or cloud-free models, see Table \ref{table:True_Values}), the posterior is strongly constrained. However, the retrieved value does not correspond to Venus' true surface pressure, but coincides with the cloud-top pressure ($\lgrt{\Pct{}[\mathrm{bar}]}=-1.3$). This is in agreement with the findings for the \pt{} profiles we outlined in the previous section. The forward models that assumed an opaque cloud layer did not retrieve for \Ps{} and thus yielded no estimates.

The \Mpl{} posterior is roughly Gaussian in log space ($\mu = -0.18$, $\sigma = 0.33$) for all forward models and is not strongly constrained over the assumed Gaussian prior ($\mu = -0.09$, $\sigma = 0.4$). This failure to further constrain \Mpl{} was also observed in \pIIIaV{} and is due to the well known degeneracy between the planet mass (surface gravity) and the abundances of the atmospheric trace gases \citep[see also, e.g.,][]{Molliere:Gravity_Abundance_Degeneracy, Feng_Retrieval, Madhusudhan:Atmospheric_Retrieval, Quanz:exoplanets_and_atmospheric_characterization}. 

For \Rpl{}, the retrieved posterior strongly depends on the forward model. For the transparent \ce{H2SO4} and the cloud-free model, the \Rpl{} posterior is roughly centered on the truth, approximately Gaussian ($\mu = 0.97\,\mathrm{R}_\oplus$, $\sigma = 0.05 \,\mathrm{R}_\oplus$), and significantly constrained over the Gaussian prior ($\mu = 0.95\,\mathrm{R}_\oplus$, $\sigma = 0.20 \,\mathrm{R}_\oplus$). In contrast, the \Rpl{} posteriors for the forward models assuming opaque clouds are broader, non-Gaussian, and not centered on the truth. When assuming opaque \ce{H2O} clouds, the posterior is shifted relative to the true value and roughly Gaussian (slightly asymmetric, with a tail to larger radii). The retrieved median strongly underestimates the planet radius by $0.2\,\mathrm{R}_\oplus$. For the opaque \ce{H2SO4} cloud forward model, the resulting \Rpl{} posterior is significantly broader. We observe two separate peaks, one of which is centered on the truth. The other is shifted to the left and underestimates \Rpl{} by approximately $0.18 \,\mathrm{R}_\oplus$. 

Since the \Teq{} and \Ab{} distributions are derived from the \Rpl{} posterior (see Appendix~\ref{sec:albedo_calc}), they inherit the forward model dependence of \Rpl{}. Thus, retrievals using the transparent \ce{H2SO4} or the cloud-free forward model result in accurate, Gaussian-shaped estimates for \Teq{} ($\mu=233$~K, $\sigma=5$~K) and \Ab{} ($\mu=0.74$, $\sigma=0.04$). For the forward models that assume an opaque cloud layer, the underestimation of \Rpl{} results in an overestimation of \Teq{}. A higher \Teq{} can only occur if the planet retains more of the incident stellar radiation, which manifests itself in a lower Bond albedo \Ab{}. As a result, we overestimate \Teq{} by roughly 20~K and underestimate \Ab{} by approximately $0.1$ for the opaque \ce{H2O} cloud model. For the opaque \ce{H2SO4} clouds, the \Teq{} posterior is similar to the \Rpl{} posterior. It is non-Gaussian in shape, exhibits a peak at the true \Teq{} value, and extends significantly toward higher \Teq{} values. For the \Ab{} distribution, the peak coincides with the truth, but the distribution shows a significant tail toward lower \Ab{} values.

We retrieve high atmospheric \ce{CO2} abundances ($\geq30\%$ in mass fraction) for all forward models. Thus, we can easily differentiate between a Venus-like, \ce{CO2} dominated atmosphere and an Earth-like atmosphere with lower \ce{CO2} abundances. In contrast, \ce{H2O} and \ce{CO} are not detected at the considered input quality, since the signatures in Venus' spectrum {lie below \mic{8} and are therefore not significant compared to the high \lifesim{} noise level. For \ce{H2O}, the drop in the posterior at high abundances rules out abundances $\gtrsim 10^{-3}$. This limit on \ce{H2O} and the unconstrained \ce{CO} abundance cause the drop in the spectrum residual below \mic{8} observed in Sect.~\ref{sec:ret_spec_resid}. Both posteriors extend to abundances significantly above the truth. Thus, on average, the spectra corresponding to the retrieved parameters have stronger \ce{H2O} and \ce{CO} absorption features than the true Venus-twin spectrum, which leads to the observed drop in the residual below \mic{8}.}

Last, we consider the cloud parameter posteriors. The cloud-top pressure (\Pct{}) posterior for both opaque cloud forward models provides a good approximation of the truth and is well described by a Gaussian ($\mu=-1.5$, $\sigma=0.2$). Furthermore, we manage to retrieve a value for the minimal possible cloud thickness (\Pcs{}; $\lgrt{\Pcs{} [\mathrm{bar}]} \gtrsim-1.5\pm0.8$). Interestingly, even with the opaque \ce{H2O} forward model, which assumes a wrong cloud composition, we obtain accurate estimates for the position of the cloud deck in the atmosphere. In contrast, for the transparent \ce{H2SO4} cloud model, we do not manage to significantly constrain either the cloud-top position or the cloud thickness. The posteriors are flat and unconstrained with respect to the assumed priors. Similarly, also the cloud particle mass fraction ($\mathrm{Species^{cloud}}$) is unconstrained for the transparent \ce{H2SO4} forward model.

The $\mathrm{Species^{cloud}}$ posterior for the opaque \ce{H2SO4} forward model is strongly bimodal. The lower of the two peaks is centered on the true \ce{H2SO4} mass fraction, while the other overestimates the abundance by roughly $3$~dex. A more in-depth analysis of the posterior distributions (see corner plots in Appendix~\ref{app:corners}) reveals a strong correlation between the retrieved cloud species abundance and \Rpl{}. Interestingly, an overestimated cloud particle abundance is linked to an underestimated \Rpl{} and thus also correlated with the \Teq{} and \Ab{} posteriors. For the opaque \ce{H2O} forward model, the retrieved median abundance lies roughly $2.5$~dex above the true \ce{H2SO4} cloud particle abundance.

Finally, the parameters describing the cloud particle size, \Rcm{} and \Sc{}, are not well constrained for any of the forward models. In an in-depth analysis of the posteriors for the two opaque cloud models (see corner plots in Appendix~\ref{app:corners}), we find a degeneracy between these two parameters. This indicates that a smaller \Rcm{} can be compensated with a larger \Sc{} for the considered spectral quality. For the transparent \ce{H2SO4} model, we observe no degeneracy between the two parameters. The lack of constraints on all cloud parameters for the transparent \ce{H2SO4} model is caused by the addition of the surface pressure \Ps{} to the retrieval. The retrieval sets the surface at the cloud-top, which alleviates the need to model an opaque cloud layer.

\subsection{Retrieval results for higher quality spectra}\label{sec:high_LIFEsim_res}

\begin{figure*}
   \centering
    \includegraphics[width=0.0192756\textwidth]{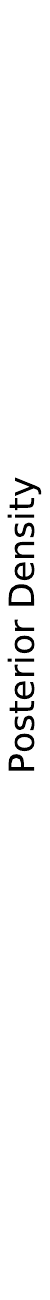}
    \includegraphics[width=0.24\textwidth]{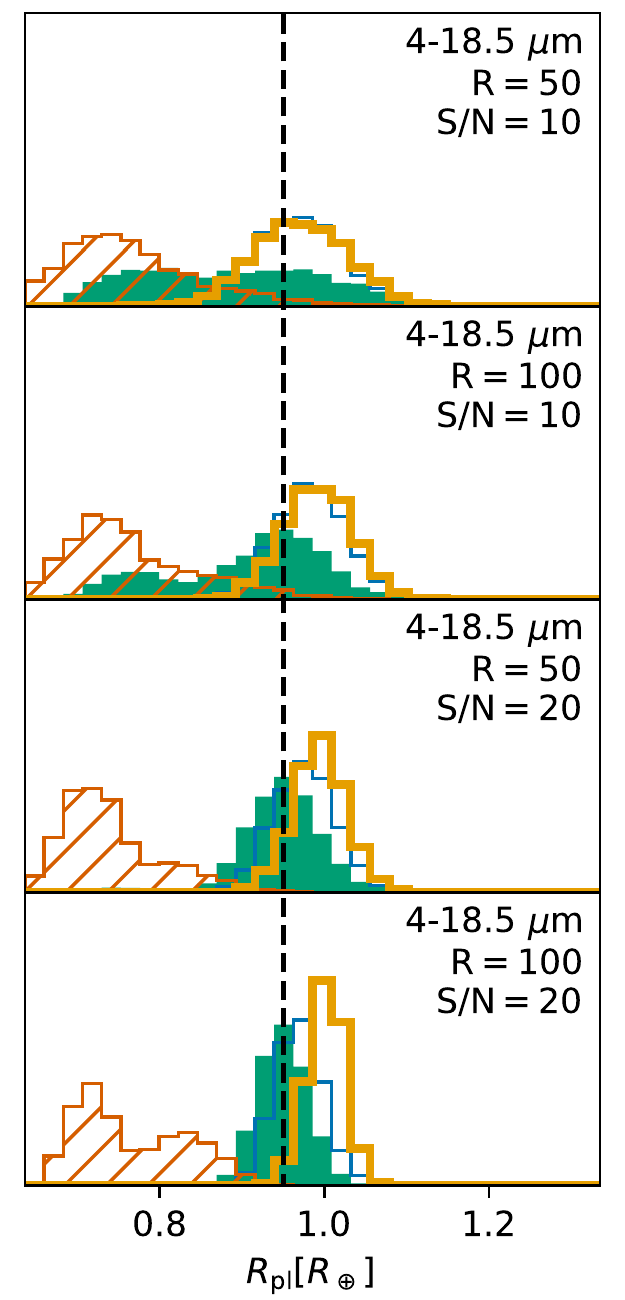}
    \includegraphics[width=0.24\textwidth]{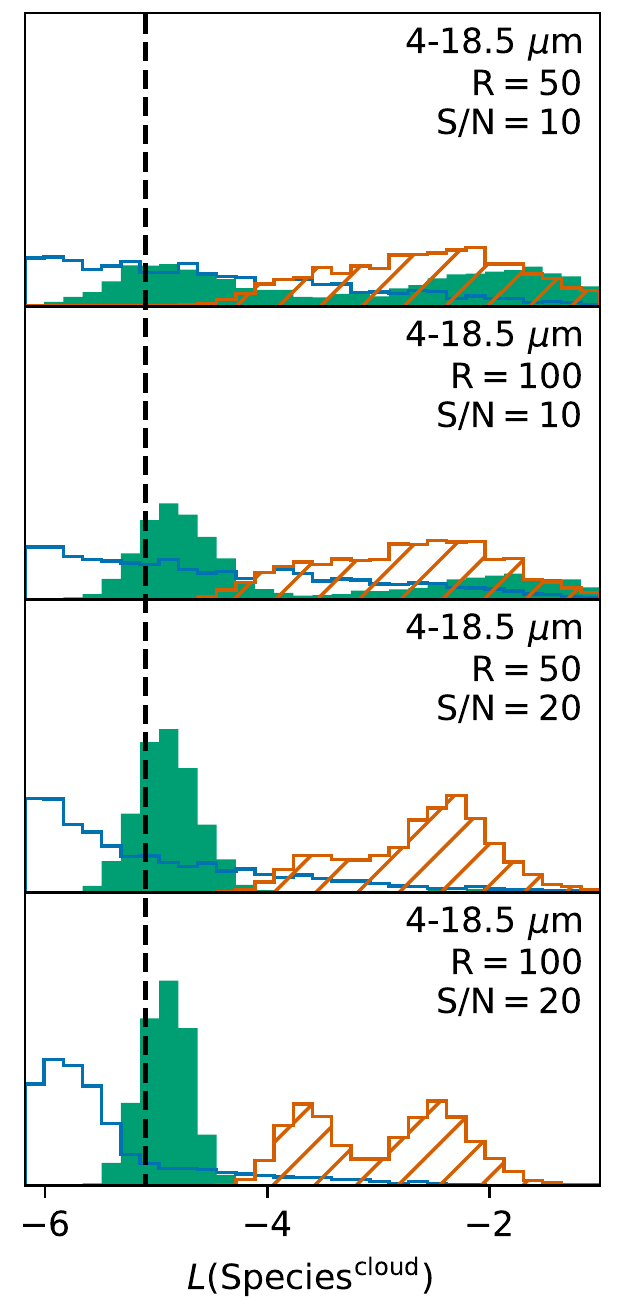}
    \includegraphics[width=0.24\textwidth]{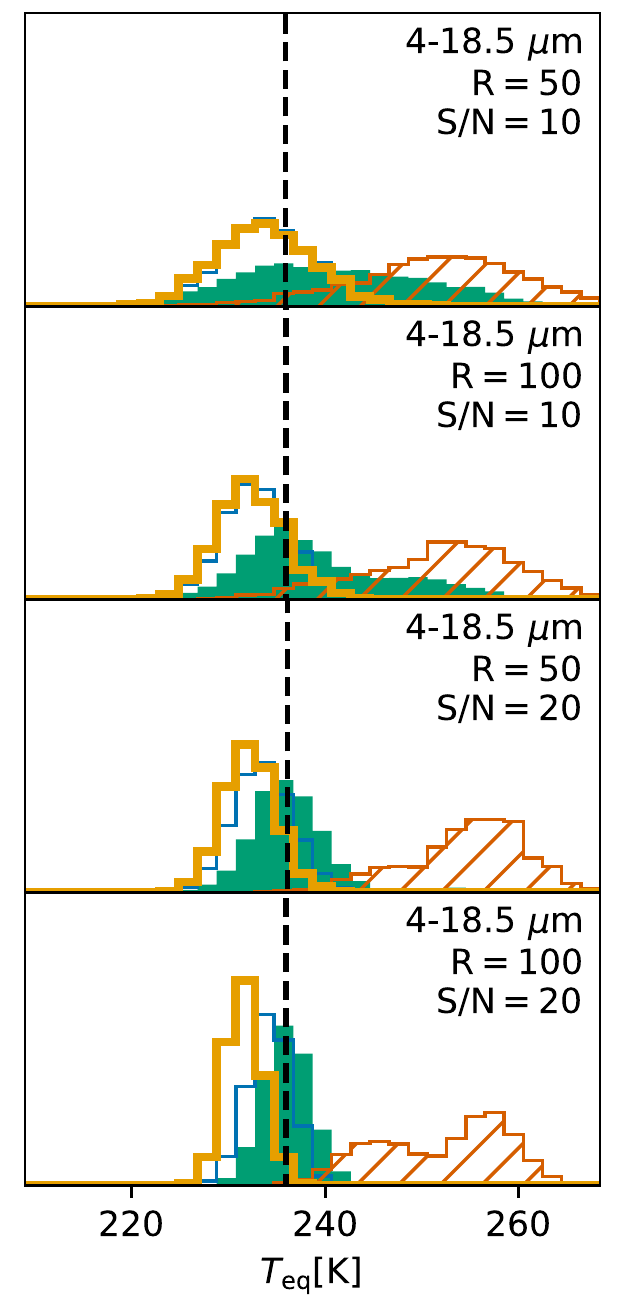}
    \includegraphics[width=0.24\textwidth]{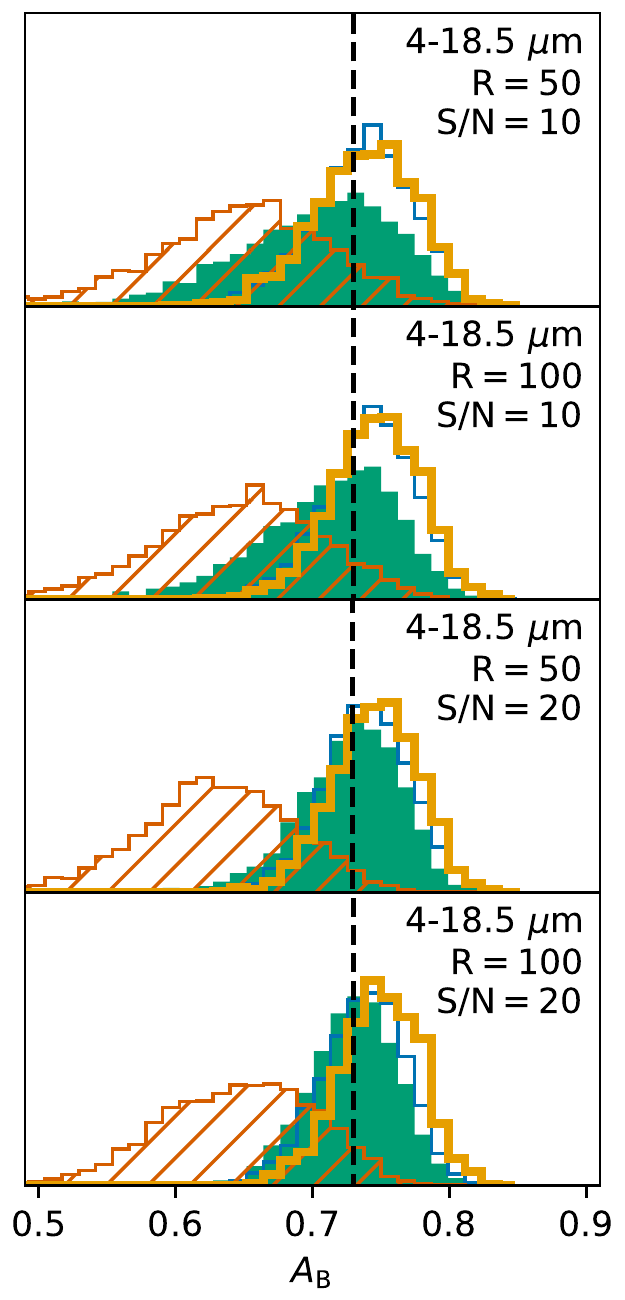}
    \includegraphics[width=0.74\textwidth]{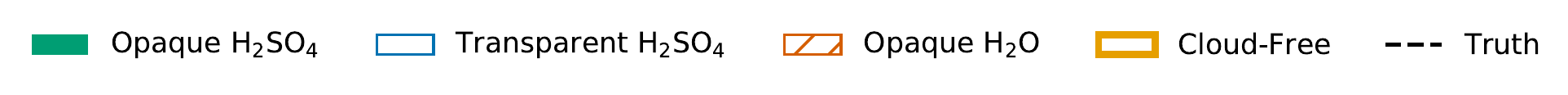}
    \caption{Model parameter posteriors for increased quality \mic{4-18.5} Venus-twin spectra (opaque \ce{H2SO4} clouds). Here, $L(\cdot)$ abbreviates $\lgrt{\cdot}$. Dashed black lines indicate the true values (see Table~\ref{table:True_Values}). For $\mathrm{Species^{cloud}}$, the true value is the \ce{H2SO4} mass fraction in the Venus-twin model. Solid green area $-$ opaque \ce{H2SO4} clouds; thin-blue outlined area $-$ transparent \ce{H2SO4} clouds; orange-hatched area $-$ opaque \ce{H2O} clouds; thick-yellow outlined area $-$ cloud-free. Columns (left to right) show the results for the planet radius \Rpl{}, cloud species abundance $\mathrm{Species^{cloud}}$, equilibrium temperature \Teq{}, Bond albedo \Ab{} (\Teq{} and \Ab{} were calculated following Appendix~\ref{sec:albedo_calc}; for the posteriors of all model parameters see Appendix~\ref{app:extended_retrievals}). Rows (top to bottom) represent different input qualities: \Rv{50}, \SNv{10}; \Rv{100}, \SNv{10}; \Rv{50}, \SNv{20}; \Rv{100}, \SNv{20}.}
\label{fig:Better_input_post_4-185}
\end{figure*}

We now investigate how the retrieval results change if we consider higher quality Venus-twin spectra. For most model parameters, both increases in \R{} and \SN{} do not significantly change the retrieval results. Generally, most parameters are better constrained as we move to higher quality spectra, but the general shape of the posterior distributions remains unchanged. Further, increasing the wavelength coverage from \mic{4-18.5} to \mic{3-20} does not significantly impact the results either. In Fig.~\ref{fig:Better_input_post_4-185}, we focus on the parameter posteriors that significantly change when increasing the quality of the \lifesim{} input spectrum (\Rpl{}, abundance of cloud species, \Teq{}, and \Ab{}). We plot the results for the \mic{4-18.5} input spectra and exclude the intermediate \SNv{15} retrievals to increase readability. We provide the retrieved posteriors for all remaining model parameters and the retrieval results for the \SNv{15} and \mic{3-20} input spectra in Appendix~\ref{app:extended_retrievals}. 

For the opaque \ce{H2SO4} cloud model, the correlated bimodal nature of the \Rpl{} and the \ce{H2SO4} abundance posteriors (see Sect.~\ref{sec:LIFEsim_papIII_ret_res}) diminishes strongly as we consider higher quality spectra. While neither of the two modes in these posteriors is preferred for the \Rv{50}, \SNv{10} input spectrum (both peaks are equally high), the modes centered on the true values are clearly preferred for the higher quality spectra. For the \SNv{20} input spectra, the bimodalities disappear completely and both \Rpl{} and the \ce{H2SO4} abundance are accurately determined.

For retrievals with the opaque \ce{H2O} cloud model, \Rpl{} is underestimated by roughly $0.2\,R_\oplus$ independent of the quality of the spectrum. Similarly, the retrieved median cloud particle abundance lies roughly 3~dex above the true \ce{H2SO4} abundance, irrespective of the \R{} and \SN{} of the spectrum. However, for the \SNv{20} inputs, bimodalities in the \Rpl{} and the \ce{H2O} cloud species posteriors emerge. These bimodalities are likely also present at lower \SN{}, but are not observable due to the larger uncertainties on the individual posterior modes.

For the transparent \ce{H2SO4} and the cloud-free model, the constraint on \Rpl{} is increased for higher \R{} and \SN{} spectra. However, while the posterior is centered on the true value for the \Rv{50}, \SNv{10} retrieval, \Rpl{} is overestimated in retrievals of higher quality spectra. This bias is stronger for the cloud-free model ($\approx0.05\,R_\oplus$) and is likely due to differences between the model assumed to generate the Venus-twin input spectrum and the forward model. It is not observable at \Rv{50}, \SNv{10}, due to the larger uncertainties on the posterior. Finally, for the transparent \ce{H2SO4} cloud forward model, the retrieved cloud species abundance converges toward abundances below the true \ce{H2SO4} abundances as we consider higher quality \lifesim{} spectra.

Lastly, we consider the \Teq{} and \Ab{} distributions. When assuming the opaque \ce{H2SO4} model, the tail toward high \Teq{} observed at \Rv{50}, \SNv{10} diminishes analogously to the \Rpl{} bimodality as we consider higher quality inputs. Consequentially, the tail toward low \Ab{} decreases. Especially for \SNv{20} spectra, the \Ab{} and \Teq{} distributions provide accurate Gaussian estimates that are centered on the truth. In contrast, systematic offsets from the truth emerge for \Teq{} for the transparent \ce{H2SO4} and the cloud-free forward models as we consider higher quality spectra. These offsets are linked to the offsets in the \Rpl{} posteriors discussed above. Albeit less prominent, the systematic shifts also appear in the \Ab{} distributions. They are less noticeable since the offset in \Teq{} is small compared to the uncertainties on the other parameters used to calculate \Ab{} (see Appendix~\ref{sec:albedo_calc}). Finally, for the opaque \ce{H2O} cloud model, the \Teq{} and \Ab{} estimates are not improved significantly as we move to higher \R{} and \SN{}.

In summary, we observe significant changes in the posteriors of some parameters when considering higher quality spectra. While the transparent \ce{H2SO4} and the cloud-free forward models perform well at \Rv{50}, \SNv{20}, biases emerge for higher resolutions. This indicates that for high quality spectra, these models are likely not sufficient. In contrast, results for the opaque \ce{H2SO4} forward model are further refined with the input quality increase. While the estimates for many parameters are weak and biased at \Rv{50}, \SNv{10}, they improve significantly as we consider higher quality spectra (especially for \SNv{20}). Finally, the results for the wrong opaque \ce{H2O} cloud model do not improve significantly when considering higher quality spectra.

%--------------------------------------------------------------------
\section{Discussion}\label{sec:discussion}

After summarizing the main results from our retrieval analysis in Sect.~\ref{sec:results}, we discuss how well one can characterize a Venus-twin exoplanet from simulated \life{} MIR observations of different quality. %As a first step, we compare our Venus-twin results to the findings from the Earth-centered studies presented in \pIIIaV{} (Sect.~\ref{sec:compIIIaV}). Subsequently, 
In Sect.~\ref{sec:model_selection}, we compare the performance of the different forward models to see whether we can find evidence for clouds by analyzing Venus' MIR thermal emission spectrum. We further discuss potential alternative pathways for cloud inference. Thereafter, in Sect.~\ref{sec:limitations}, we address the limitations of our approach and motivate potential future studies.

\subsection{Forward model selection and interpretation}\label{sec:model_selection}

In Sect.~\ref{sec:results}, we find that the retrieved \pt{} profile shape and the posterior distributions of the atmospheric gases (\ce{CO2}, \ce{H2O}, and \ce{CO}) exhibit only minor variations with the forward model and input quality. In contrast, the posteriors for \Rpl{} and the cloud parameters, as well as the inferred distributions for \Teq{} and \Ab{} depend significantly on the forward model and the input quality. This shows that incorrect model assumptions or an inadequate level of model complexity can result in incorrect exoplanetary characterization. These dependencies of the posteriors are problematic, because the true atmospheric structure and composition is unknown for an observed exoplanet. Consequentially, we will not be able to verify if the parameter values we retrieve assuming a forward model characterize the observed exoplanet correctly. Therefore, we require a method of determining an adequate forward model for a given exoplanet spectrum (Sect.~\ref{sec:Life_Baye_model}). Furthermore, once an adequate forward model is determined, we need to understand how to link the obtained retrieval results to the conditions present on the exoplanet (Sect.~\ref{sec:life_model_interpretation}).

\subsubsection{Forward model selection via the Bayes factor}\label{sec:Life_Baye_model}
\begin{table}

\caption{Model comparison via the Bayes factor $K$.}             % title of Table
\label{table:BayeLIFEsim}      % is used to refer this table in the text
\centering                          % used for centering table
\begin{tabular}{lcc}    % centered columns (4 columns)
\hline\hline                 % inserts double horizontal lines

Compared models        &\lgrt{K}        &Preferred model\\    % table heading 
\hline
\MSaOp{} versus \MSaTr{}   &$-0.4\pm0.1$   &\MSaTr{}\\ 
\MSaOp{} versus \MWaOp{}   &$0.1\pm0.1$    &Either\\
\MSaOp{} versus \MCf{}     &$-1.0\pm0.1$   &\MCf{}\\
\MSaTr{} versus \MWaOp{}   &$0.5\pm0.1$    &\MSaTr{}\\
\MSaTr{} versus \MCf{}     &$-0.6\pm0.1$   &\MCf{}\\
\MWaOp{} versus \MCf{}     &$-1.1\pm0.1$   &\MCf{}\\
\hline 
\end{tabular}
\tablefoot{Performance comparison of forward models for retrievals of the \mic{4-18.5}, \Rv{50}, \SNv{10} Venus-twin (opaque \ce{H2SO4} clouds) \lifesim{} spectrum. We calculate \lgrt{K} for pairs of models with Eq.~\ref{eq:BayesFactor} and interpret its value via the Jeffreys scale \citep[see Table~\ref{Table:Jeffrey},][]{Jeffreys:Theory_of_prob}.}
\end{table}

\begin{figure*}
    \centering
    \includegraphics[width=0.93\textwidth]{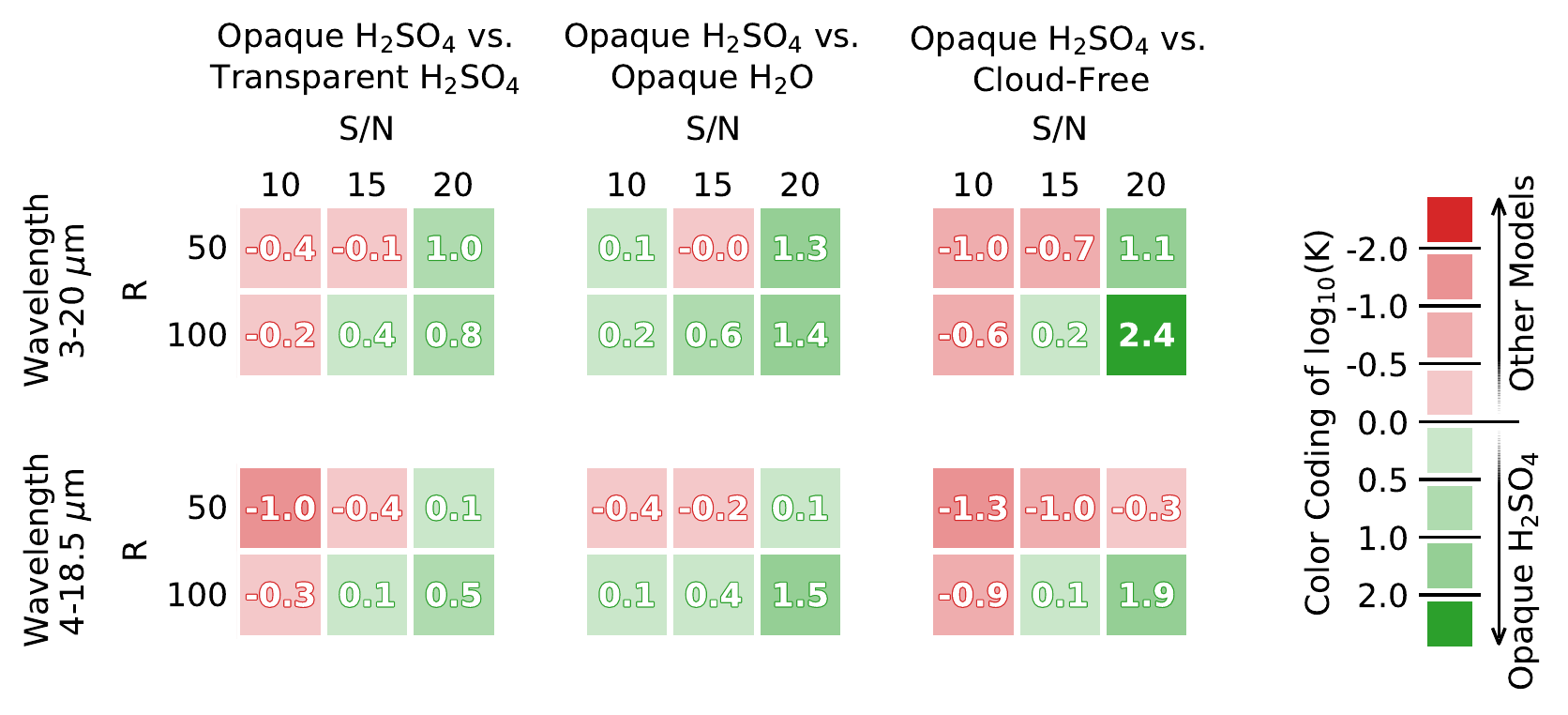}
    \caption{Comparison of the retrieval performance of the true opaque \ce{H2SO4} cloud model with the other models via the Bayes factor $K$. We use the Jeffreys scale as metric for comparison \citep[see Table~\ref{Table:Jeffrey},][]{Jeffreys:Theory_of_prob} and consider different wavelength coverages, \R{}, and \SN{} of the simulated Venus-twin \life{} observations. Green squares (positive \lgrt{K} values) indicate that the correct model (opaque \ce{H2SO4} clouds) is preferred, red squares (negative \lgrt{K} values) mark preference for the incorrect model. The intensity of coloring represents the preference strength.}
    \label{fig:ModelComparison}%
\end{figure*}

As outlined in Sect.~\ref{sec:input_spectra}, the Bayes' factor $K$ allows us to compare the retrieval performance of different forward models for a given exoplanet spectrum. Importantly, the Bayes' factor $K$ does not tell us if a model is correct or not. It provides a metric that measures which model out of a set of models is best suited to explain an observed exoplanet spectrum. Here, we investigate if clouds can be detected and characterized in retrievals for different quality input spectra. We do this by comparing the retrieval performance of the correct opaque \ce{H2SO4} cloud forward model to the other tested forward models using the Bayes' factor $K$.

First, we compare the performance of different forward models at the minimal \life{} requirements from \pIII{} (\mic{4-18.5}, \Rv{50}, and \SNv{10}). We list the Bayes factor $K$ for all combinations of forward models in Table~\ref{table:BayeLIFEsim}. No forward model we tested is decisively ruled out or preferred ($\left|K\right|\leq2$ for all comparisons). We only find slight performance differences between the forward models. Overall, the cloud-free model performs best (always preferred; $\left|K\right|\geq0.6$), while the opaque cloud models perform worst (never preferred, despite the opaque \ce{H2SO4} cloud model being the true model). Importantly, the cloud-free model also uses fewer parameters than the cloudy scenarios (ten versus 14–15 parameters). The finding that the cloud-free forward model yields the best retrieval performance indicates that the additional parameters required to model clouds are not justifiable.

For the Venus twin, this suggests that MIR retrievals at the current minimal \life{} requirements are not sufficient to find evidence for atmospheric clouds. Crucially, this does not rule out clouds. It merely indicates, that the considered spectrum is satisfactorily described by the cloud-free model. This agrees well with the findings for cloud-free retrievals on cloudy input spectra presented in \pV{}.

Second, we investigate if our findings change for higher quality \lifesim{} spectra. In Fig.~\ref{fig:ModelComparison}, we compare the retrieval performance of the opaque \ce{H2SO4} cloud model (model used to produce the input spectra) to the other models via the Bayes factor $K$. Green fields mark positive \lgrt{K} values and thus preference for the true opaque \ce{H2SO4} cloud model. Red fields (negative \lgrt{K}) indicate a preference for the incorrect model.

For spectra with $\SN{}\leq20$, we observe that both the cloud-free and the transparent \ce{H2SO4} models perform better than (or comparable to) the opaque \ce{H2SO4} cloud model. Further, the cloud-free model outperforms the transparent \ce{H2SO4} model, as we see from the lower \lgrt{K} values. The two models assuming opaque clouds perform equally well. As before, these findings suggest that no evidence for clouds in Venus' atmosphere can be found via retrieval studies of spectra with $\SN{}<20$.

For \SNv{20} spectra, a preference for the opaque \ce{H2SO4} forward model emerges. This preference is generally stronger for spectra with larger wavelength coverage and higher \R{}, since they contain more information. This suggests that an \SN{} of at least 20 is required to infer the presence of clouds in a retrieval study on the MIR thermal emission spectrum of a Venus-twin exoplanet.

Crucially, the \SN{}-dependent model preferences agree with our findings for the posteriors (see Sect.~\ref{sec:results}). For $\SN{}<20$ spectra, the preferred cloud-free model yields good estimates for Venus' atmospheric structure and composition above the cloud-top. The planetary parameters \Rpl{}, \Teq{}, and \Ab{} are correctly retrieved. Similarly, the transparent \ce{H2SO4} cloud model (second-best performance) approximates the aforementioned parameters well. The retrieved \Ps{} corresponds to the cloud-top, which alleviates the need to model an opaque cloud layer. Thus, the cloud parameters are unconstrained. In contrast, the opaque cloud retrievals (lowest preference) yield weak and biased estimates for \Rpl{}, \Teq{}, \Ab{}, and the cloud parameters. This suggests, that for spectra with $\SN{}<20$, the spectral information content is not sufficient to constrain these additional parameters. Hence, the cloud-free model with fewer parameters is preferred.

For \SNv{20} spectra, we notice significant changes in both the posteriors and the model preference. For the transparent \ce{H2SO4} and the cloud-free model, the estimates for \Rpl{}, \Teq{}, and \Ab{} are offset from the true value and thus yield biased estimates. For the opaque \ce{H2O} cloud model, we see no significant improvements in the posteriors over the $\SN{}<20$ retrievals. For the opaque \ce{H2SO4} cloud model, which performs best on the \SNv{20} spectra, the biases we find for $\SN{}<20$ spectra are no longer present. The posteriors for cloud and planet parameters are unbiased and provide good estimates. This suggests that at \SNv{20}, the information content of the input spectrum is sufficient to justify the additional cloud parameters. These observations for the posteriors agree well with the shift in forward model preference from the cloud-free ($\SN{}<20$) to the opaque \ce{H2SO4} cloud (\SNv{20}) forward model.

In conclusion, we find that for low quality \lifesim{} Venus-twin spectra ($\SN{}<20$) cloud presence is not inferrable via MIR retrievals. For these inputs, the cloud-free model yields accurate estimates for fundamental planetary and atmospheric parameters. The accuracy of the constraints on these parameters are in accordance with the findings presented in \pIIIaV{}. For the \SNv{20} spectra (especially if \Rv{100}), we manage to find weak evidence for clouds in the atmosphere of the Venus twin and to constrain the cloud properties. We emphasize that our findings are based on the assumption of a Venus twin. However, the conclusion that clouds are hard to infer and constrain via retrievals of low quality MIR thermal emission spectra is likely generalizable to arbitrary terrestrial exoplanets. Further testing of this important result is foreseen for the future.

\subsubsection{Interpretation of model selection results}\label{sec:life_model_interpretation}

Inferring cloud presence for terrestrial exoplanets via MIR thermal emission retrievals is challenging. At the minimal \life{} specifications, we find the cloud-free model to perform best, and thus no direct evidence for a cloud deck. There are two simple interpretations of the retrieval's preference for the cloud-free forward model. In the first interpretation, the retrieved surface pressure \Ps{} is incorrect. The true \Ps{} is larger and not retrieved correctly, since the atmospheric high pressure layers are optically thick and thus leave no signatures in the spectrum. For our study, this would suggest that the emission spectrum contains no information about the exoplanet's lower atmosphere ($\gtrsim0.05$~bar) and surface conditions. In the second interpretation, the retrieved \Ps{} corresponds to the truth and the surface contributes to the emission spectrum. In this case, the exoplanet would possess a thin atmosphere ($\approx0.05$~bar). In both cases, the exoplanet is characterized by a high bond albedo ($\Ab{}=0.74\pm0.4$).

By considering the \Ab{} of Solar System objects, we attempt to link the retrieved \Ab{} to planet properties. In Table~\ref{Table:SS_Ab}, we list the \Ab{} of selected objects along with the main source of the MIR continuum emission. If an object has a low \Ab{} ($\lesssim0.3$), the continuum emission typically originates from the planet's surface (rocky objects with no or an optically thin atmosphere). For high \Ab{} objects ($\gtrsim0.5$), the MIR continuum emission stems from either clouds (e.g., Venus) or predominantly ice and frost covered surfaces (e.g., Europa). Thus, our retrieval results suggest an exoplanet with either a cloudy atmosphere (first scenario) or an icy surface (second scenario). In the second scenario, the retrieved \Ps{} ($\approx0.05$~bar) and the corresponding temperature ($\Ts{}=235\pm 4$~K) would allow for a water ice surface\footnote{This possibility of strongly underestimating the surface temperature of a cloudy exoplanet has already been discussed for habitable Earth-like exoplanets \citep[e.g.,][]{Kitzmann2011Emission}.}. Such "snowball" states have occurred on Earth \citep{SnowballKirschvink,SnowballHoffman,HoffmanSnowball2} and are also conceivable toward the inner edge of the HZ \citep{Wordsworth_snowballs_inner_Hz,Graham2021}. In a snowball state, the majority of the incident stellar radiation is reflected due to the high albedo of the planet's surface, which leads to low surface temperatures. Even for high incident radiation from the host star and large concentrations of atmospheric greenhouse gases, the planet surface can remain in a stable frozen state \citep{Budyko1969,Sellers1969}.

Yet, the icy scenario appears improbable given the exoplanet-star separation ($0.72$~AU) and the high retrieved levels of the strong greenhouse gas \ce{CO2} ($\geq30$\% in mass fraction). In addition, the low retrieved surface pressure ($\approx0.05$~bar) seems unlikely for an evolved Venus-sized exoplanet \citep[e.g.,][]{2020NatSR..1010907O}. Lastly, the long-term stability of such a planet is uncertain and depends on various factors such as the rate of volcanic outgassing \citep[e.g.,][]{pierrehumbert_2010}. However, the icy scenario cannot be ruled out solely via low quality MIR observations ($\SN{}<20$). Increasing the spectrum's \SN{} to at least 20 allows us to infer cloud presence, yet still not robustly. Furthermore, an increase in \SN{} would require significantly more observation time (for \SNv{20} roughly four times longer than for \SNv{10}). Thus, alternative cloud inference pathways are desirable to resolve this ambiguity in interpretation.

% Kaustubh: if needed, can add Hamano et al. (2013) - https://ui.adsabs.harvard.edu/abs/2013Natur.497..607H/exportcitation
%{Comment by Sascha: So, you are suggesting that a planet with a high, icy albedo and a surface temperature of 230 K can also feature a (admittedly thin) CO2 atmosphere at 0.72 AU from the Sun. I somehow find it hard to believe that this works, but I cannot prove it to you… Is this something worth double-checking with some climate modeling colleagues? Or do the papers you cite provide enough information?}

%{Comment by Robin: Snowball states inside the inner edge of the habitable zone are conceivable if the planet's atmosphere is dominated by gases like N2. But if the atmosphere is CO2-rich, they will not be stable unless albedo is extremely high, because of the CO2 greenhouse effect. Long-term stability would depend on things like the rate of volcanic outgassing and possibly condensation of CO2 on the surface.}

\begin{table}
\caption{Bond albedos \Ab{} of selected Solar System objects.}% title of Table
\label{Table:SS_Ab}      % is used to refer this table in the text
\centering                          % used for centering table
\begin{tabular}{l c c c}        % centered columns (4 columns)
\hline\hline                 % inserts double horizontal lines
Object      &\Ab{}      &MIR Continuum Emission   &Reference\\    % table heading 
\hline 
Mercury     &0.08       &Rocky surface          &1\\
Venus       &0.76       &Clouds                 &2\\
Earth       &0.30       &Surface and clouds     &3\\
Moon        &0.14       &Rocky surface          &4\\
Mars        &0.24       &Rocky surface          &5\\
Jupiter     &0.53       &Clouds                 &6\\
Europa      &0.55       &Icy surface            &7\\
Saturn      &0.34       &Clouds                 &8\\
Tethys      &0.67       &Icy surface            &9\\
Enceladus   &0.81       &Icy surface            &9\\
\hline 
\end{tabular}
\tablefoot{The third column specifies the main source of the MIR continuum emission (for clear atmospheres the continuum originates from the surface, for opaque cloudy atmospheres it stems from the cloud deck).}
\tablebib{(1)~\citet{AbMercury}; (2)~\citet{AbVenus}; (3)~\citet{AbEarth}; (4)~\citet{AbMoon}; (5)~\citet{AbMars}; (6)~\citet{AbJupiter}; (7)~\citet{AbEuropa}; (8)~\citet{AbSaturn}; (9)~\citet{AbSatsat}}
\end{table}

\begin{figure*}
   \centering
   \includegraphics[width=0.99\textwidth]{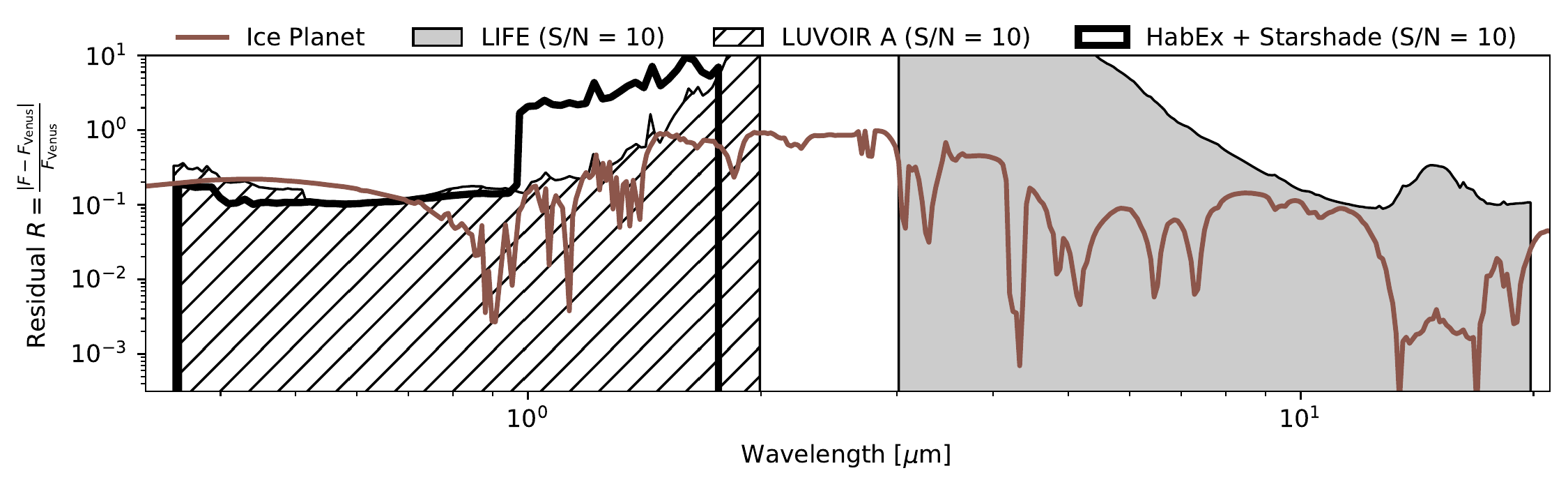}
    \caption{Flux difference between a cloudy Venus-twin exoplanet (opaque \ce{H2SO4} clouds) and an icy world with a thin \ce{CO2}-dominated atmosphere (see Appendix~\ref{app:abs_flux_comp} for a comparison of the absolute fluxes). In brown, we plot the flux residual of the icy world with respect to the Venus twin as a function of wavelength. The gray shaded region indicates the 1$\sigma$ \lifesim{} noise level at \SNv{10}. The thin, hatched, and black area represents the expected wavelength-dependent noise for the LUVOIR A mission concept \citep{2017AAS...22940504P}; the thick, solid black line represents the expected noise for the HabEx + Starshade mission concept \citep{2020arXiv200106683G}.}
          \label{fig:Clouds_VS_Ice}
\end{figure*}

A potential remedy to the aforementioned ambiguity, is to leverage 1D or 3D (photo-)chemistry and climate models (e.g., \texttt{Atmos}\footnote{\url{https://github.com/VirtualPlanetaryLaboratory/atmos}}, \texttt{ROCKE-3D}\footnote{\url{https://www.giss.nasa.gov/projects/astrobio/}}, or \texttt{PlaSim}\footnote{\url{https://www.mi.uni-hamburg.de/en/arbeitsgruppen/theoretische-meteorologie/modelle/plasim.html}}). Such a simulative approach can help us identify and rule out nonphysical planetary states. For example, for the icy exoplanet scenario motivated above, there are two crucial questions that could be studied via climate simulations. First, one has to investigate if an icy surface together with the retrieved atmospheric structure and composition describes a physically possible and stable state. Since our retrieval framework does not model the atmospheric physics (e.g., convection, photochemistry), not all points in the posterior distribution result in stable atmospheres. In the case of the icy exoplanet, the atmosphere described by the retrieved posteriors might enter a rapid runaway greenhouse phase, which would lead to melting and subsequent evaporation of the surface \ce{H2O}. Under these circumstances, the icy scenario would be highly unlikely, and thus the cloudy scenario preferred. Studies similar to \citet{Boukrouche2021}, \citet{Chaverot2022}, or \citet{Graham2022} could help us better understand the stability of the icy scenario. If we find that the icy state to be realistic, the second question to tackle is if it can be reached, given the exoplanets proximity to the host star and the high atmospheric \ce{CO2} abundance. To answer this question, studies similar to \citet{Wordsworth_snowballs_inner_Hz} or \citet{Graham2021}, that investigate a wide range of different atmospheres, could provide answers. Such studies are an example of potential future synergies between atmospheric retrievals and the theoretical modeling of atmospheres. However, recent intercomparison efforts show that both retrievals \citep[e.g., ][]{Barstov2020retrievals} and atmospheric modeling \citep[e.g., ][]{Sergeev2022} depend on the model choice. Community efforts, such as the CUISINES Working Group\footnote{\url{https://nexss.info/cuisines/}}, that benchmark, compare, and validate different models, are vital to the studies proposed above.

An alternative approach is to not only consider a terrestrial exoplanet's MIR emission, but also the stellar light it reflects in the UV/O/NIR. While we do not expect the scenarios to differ noticeably in MIR emission, the different reflective properties of clouds and ice will significantly impact the UV/O/NIR reflected light spectrum. To test this approach, we simulated spectra between \mic{0.3} and \mic{21} using \texttt{petitRADTRANS}. For the first scenario, we used our cloudy Venus-twin exoplanet (opaque \ce{H2SO4} clouds; see Sect.~\ref{sec:venus_spectrum} and Table~\ref{table:True_Values}). For the second scenario, we modeled an icy surface at a pressure of 0.05~bar (\Ps{} from cloud-free retrievals) and a cloud-free atmosphere with the same composition and \pt{} profile as for the first scenario. We used data from the ECOSTRESS Spectral Library\footnote{\url{https://speclib.jpl.nasa.gov}} \citep{ECOSTRESS1,ECOSTRESS2} to model the wavelength-dependent reflectivity of the icy surface. An 80\% frost- and 20\% ice-covered surface yielded a total reflectance of 0.75 in the UV/O/NIR ($\approx$~retrieved \Ab{}). In Fig.~\ref{fig:Clouds_VS_Ice}, we plot the residual of the ice-planet spectrum relative to the cloudy Venus-twin spectrum and the \lifesim{} noise (\SNv{10}). We provide an analogous plot comparing the absolute fluxes in Appendix~\ref{app:abs_flux_comp}. In the UV/O/NIR, we show the noise expected for observations of the same planets with LUVOIR A \citep{2017AAS...22940504P} or HabEx + Starshade \citep{2020arXiv200106683G} \citep[\SNv{10} at \mic{0.6}, calculated with the NASA-GSFC Planetary Spectrum Generator\footnote{\url{https://psg.gsfc.nasa.gov}};][]{Villanueva2018PSG}. While the residual lies below the \lifesim{} noise level in the MIR, it is significantly larger than both the expected LUVOIR and HabEx noise in the UV/O/NIR below \mic{0.7}. This indicates that the reflected light spectrum is more suitable to differentiate between the icy and cloudy scenario than the MIR thermal emission spectrum. It further exemplifies the complementarity of UV/O/NIR reflected light and MIR thermal emission observations and highlights the importance of following both strategies. Retrieval studies on combined UV/O/NIR and MIR spectra are foreseen for the future.

%Comment Paul: Didn't I give you the self-consistent experimental version of pRT Björn (including the surface treatment)? You could feed in the retrieved abundance / cloud profiles, surface albedos, pressures, etc. and see what you get.

\subsection{Limitations and future work}\label{sec:limitations}

The study we present here provides us with first estimates for how well a Venus-like exoplanet could be characterized by \life{}. Further, we obtain insights into how atmospheric clouds can complicate retrieval studies and the interpretation of their output. As we are making several assumptions in our approach, our findings cannot readily be generalized to arbitrary science cases. Here, we discuss these limitations in detail.

%Sean: You are correct to not over-reach with your conclusions, but I do think that you have demonstrated these results in a generalisable way that shouldn't be understated. 
%The links between Radius and cloud abundance, and between equilibrium temperature and bond albedo, leading to bimodal posteriors with two degenerate interpretations, have no direct specificity to being an exact Venus-twin right? precisely because the high P regions are unobservable.

First, we restricted ourselves to the study of a Venus twin. While the performance for individual model parameters will not generalize well, the more general findings provide insights into biases inherent to retrievals (e.g., the dependence of the parameter posteriors on the forward model). However, as suggested in \citet{Robinson_SS_Retrievals}, Solar System planets provide an excellent benchmark for retrievals, because these atmospheres are known to be physical and estimates for the ground-truth values of the model parameters are available. Performing retrievals on different Solar System planets will help us generalize our predictions for the characterization performance.

Second, there are limitations inherent to our theoretical Venus-twin input spectrum. We assumed a fully mixed atmosphere (vertically constant abundances). While the same assumption was made in \pIII{}, the input spectra in \pV{} were based on variable abundance profiles. Nevertheless, the retrieval results from \pIIIaV{} are comparable, indicating that this simplification does not heavily impact retrievals at the spectral qualities considered. We also treated the clouds in a simplified manner. More realistic cloud models, which consider \pt{}-dependent cloud particle sizes and abundances or spatial variations in the cloud deck, could affect the MIR emission spectrum measurably. Further, we neglected both temporal and spatial variances in the atmospheric structure and composition. Real (exo)planet emission spectra can vary with time and depend on the viewing geometry \citep[e.g.,][]{Mettler2020,Mettler2022}. Retrieval studies on more realistic input spectra will provide more reliable estimates for \life{}'s performance and are foreseen in the future.

Third, the limitations above are also valid for the forward models used in the retrievals. However, only limited increases in the forward model complexity are possible, as they lead to a substantial rise in the retrievals' computational complexity. For example, we only retrieved for molecules present in the Venus-twin atmosphere. Including additional molecules could lead to false positive detections of gases and a mischaracterization of the atmosphere. However, a first robustness study for false positive detections in the appendix of \pIII{} justifies our approach. Another simplification is that we use a 1D forward model. While this is not problematic here, since the Venus-twin input is also calculated with a 1D model, it will be wrong for retrievals of real spectra and spectra from 3D models. However, in a recent study, \citet{Robinson_SS_Retrievals} compared the performance of their 1D retrieval suite (\texttt{rfast}) to results from a computationally expensive 3D retrieval \citep{Feng_Retrieval}. They concluded that for the \R{} and \SN{} we considered here, 1D retrievals suffice to obtain a first order understanding of how the spectral quality  affects the exoplanet characterization performance.

Fourth, we used \texttt{petitRADTRANS} to generate the Venus-twin input spectrum and as radiative transfer model in the retrieval. As discussed in \pV{} and \citet{Barstov2020retrievals}, systematic differences between the radiative transfer model used to generate the input spectrum and in the retrievals \citep[e.g., differences in the used line lists,][]{Alei+SPIE} can lead to biases in the posteriors. For retrievals on real exoplanet spectra, similar problems are unavoidable, since the radiative transfer model will never capture the full atmospheric physics and chemistry of the observed exoplanet. Thus, our results might be overly optimistic and it is indispensable to investigate the nature and magnitude of the resulting biases in future studies.

Last, important limitations are rooted in the \lifesim{} noise model. Currently, \lifesim{} models the dominant astrophysical noise terms but neglects systematic instrumental effects (\pII{}). Ideally, instrumental noise contributions will not dominate \life{}'s noise budget. Nevertheless, they will contribute to the observational noise by altering the relative distribution of noise across the wavelength range (Dannert et al., in prep.), which might affect the retrieval results. More accurate estimates will be possible once \life{}'s optical, thermal, and detector designs have matured and are accounted for by \lifesim{}. Also, we interpreted the \lifesim{} noise as uncertainty on the Venus-twin spectrum. Crucially, we did not randomize the values of individual spectral points. This decision might lead to overly optimistic results (see Sect. \ref{sec:input_spectra}). We expect the low \R{} and \SN{} cases to be more strongly affected by randomization. Further, the cloud inference and characterization capabilities could also be overly optimistic. For a detailed discussion on potential impacts of this simplification on the characterization performance, we refer to the appendix of \pIII{}.

%-------------------------------------------------------------------
\section{Conclusions and outlook}\label{sec:conclusion}

In this study, we ran retrievals for a cloudy Venus twin orbiting a G2V star at a distance of 10~pc. The goal was to investigate how the minimal \R{} and \SN{} requirements for \life{} defined in \pIII{} and verified in \pV{} are affected by clouds.

We approximated Venus' MIR thermal spectrum using the 1D radiative transfer model \texttt{petitRADTRANS} \citep[][]{Molliere:petitRADTRANS, Molliere:petitRADTRANS2,LIFE_V} assuming a third order polynomial \pt{} structure, vertically constant \ce{CO2}, \ce{H2O}, and \ce{CO} abundances, and a uniform, Mie scattering \ce{H2SO4}–\ce{H2O} cloud slab. The \lifesim{} tool (\pII) simulated \life{} observations of the Venus twin. Using an updated version of the retrieval suite from \pIIIaV{}, we ran retrievals for variable quality spectra (from the \mic{4-18.5}, \Rv{50}, \SNv{10} minimal \life{} requirements to \mic{3-20}, \Rv{100}, \SNv{20} spectra) assuming different (cloud-free and cloudy) forward models.

At the minimal \life{} requirements, we correctly retrieve the \pt{} structure above the cloud top (for pressures $\leq0.03\pm0.02$~bar, $T$ lies within $\leq\pm25$~K of the truth) and find high \ce{CO2} levels ($\geq 30\%$ in mass fraction). These findings allow us to discern Venus- from Earth-like exoplanets. Further, Venus' surface conditions are not constrainable via its MIR thermal emission, since the opaque atmospheric clouds block contributions from the lower atmospheric layers. The results for the planet radius \Rpl{}, equilibrium temperature \Teq{}, Bond albedo \Ab{}, and the cloud parameters depend strongly on the forward model choice. Overall, the cloud-free model yields the best estimates ($\Rpl{}=0.97\pm0.05~\mathrm{R}_\oplus$, $\Teq{}=233\pm5$~K, $\Ab{}=0.74\pm0.04$) and is favored by the Bayes factor analysis. This suggests that cloud presence cannot be inferred at the minimal \life{} requirements. For high quality spectra ($\SN{}\geq20$), the parameter constraints increase and the model preference shifts toward the correct opaque \ce{H2SO4} cloud model. While this suggests that retrieval based cloud inference is possible with \life{}, other approaches, such as followup UV/O/NIR observations with a {HWO-like} telescope, offer an alternative and synergistic approach.

Crucially, we find that our retrieval results for important planetary parameters (\Rpl{}, \Teq{}, \Ab{}) strongly depend on the chosen forward model. An incorrect forward model or an inadequate level of forward model complexity (e.g., too complex given the quality of the input spectrum) can heavily bias retrieval results. %This is not a major concern for our Venus-twin study, since the true atmospheric state and the ground truths for the model parameters are known. Thus, identifying the best preforming forward model is straightforward.
This is a major concern since, for observations of real exoplanets, the atmospheric state will be unknown. Furthermore, the work we presented here suggests that model selection via the Bayes factor will likely be hard and thus the risk of over- or misinterpretation of the available data is high. While sufficient quality MIR spectra of Earth- or Venus-like exoplanets will not be available in the near future, the James Webb Space Telescope will measure transmission spectra as well as thermal emission of many exoplanets in the upcoming years. Atmospheric retrievals will be used to analyze these spectra \citep{Cowan_2015, Greene_2016, Krissansen_Totton_2018, Nixon_2022}. Therefore, working toward a community-wide common approach for retrieval studies of exoplanet spectra is of great importance, as it would mitigate the risk of false characterization significantly and augment the comparability of different studies. Applied to empirical data from powerful future space missions, such as \life{}, the in-depth characterization of different types of terrestrial exoplanets seems within reach.

\begin{acknowledgements}
    This work has been carried out within the framework of the NCCR PlanetS supported by the Swiss National Science Foundation under grants 51NF40\_182901 and 51NF40\_205606. S.P.Q. and E.A. acknowledge the financial support from the SNSF. S.R. acknowledges support from the Natural Sciences and Engineering Research Council of Canada (NSERC) Discovery Grant, [2022-04588]. \emph{Author contributions.} B.S.K. carried out the analyses, created the figures, and wrote the manuscript. S.P.Q. initiated this project. S.P.Q. and E.A. provided regular guidance. All authors discussed the results and commented on the manuscript.
\end{acknowledgements}

%After discussing extensively how results from our Venus-Twin retrievals can be interpreted, we focus on the implications of our results for the \life{} requirements. Our Venus-twin has a low \Teq{}, which results in a roughly 40\% lower MIR emission than the Earth-twin discussed in \pIIIaV{}. While a Venus-twin at 10~pc would be detectable with \life{}, a thorough characterization is not with the current \life{} design. However, closer or larger Venus-like planets would be an excellent target for \life{}. As we discuss above, inferring cloud presence for terrestrial exoplanets via MIR thermal emission retrievals is challenging. At the minimal \life{} specifications, we find the cloud-free model to perform best, and thus no direct evidence for a cloud deck.

\bibliographystyle{aa}
\bibliography{bibliography}

\begin{thebibliography}{100}
\expandafter\ifx\csname natexlab\endcsname\relax\def\natexlab#1{#1}\fi

\bibitem[{{Alei} {et~al.}(2022{\natexlab{a}}){Alei}, {Konrad}, {Angerhausen},
  {Grenfell}, {Molli{\`e}re}, {Quanz}, {Rugheimer}, {Wunderlich}, \& {LIFE
  Collaboration}}]{LIFE_V}
{Alei}, E., {Konrad}, B.~S., {Angerhausen}, D., {et~al.} 2022{\natexlab{a}},
  \aap, 665, A106

\bibitem[{{Alei} {et~al.}(2022{\natexlab{b}}){Alei}, {Konrad}, {Molli{\`e}re},
  {Quanz}, {Angerhausen}, \& {Ranganathan}}]{Alei+SPIE}
{Alei}, E., {Konrad}, B.~S., {Molli{\`e}re}, P., {et~al.} 2022{\natexlab{b}},
  in SPIE Conference Series, Vol. 12180, Space Telescopes and Instrumentation
  2022: Optical, Infrared, and Millimeter Wave, 121803L

\bibitem[{{Anglada-Escud{\'e}} {et~al.}(2016){Anglada-Escud{\'e}}, {Amado},
  {Barnes}, {Berdi{\~n}as}, {Butler}, {Coleman}, {de La Cueva}, {Dreizler},
  {Endl}, {Giesers}, {Jeffers}, {Jenkins}, {Jones}, {Kiraga}, {K{\"u}rster},
  {L{\'o}pez-Gonz{\'a}lez}, {Marvin}, {Morales}, {Morin}, {Nelson}, {Ortiz},
  {Ofir}, {Paardekooper}, {Reiners}, {Rodr{\'\i}guez},
  {Rodr{\'\i}guez-L{\'o}pez}, {Sarmiento}, {Strachan}, {Tsapras}, {Tuomi}, \&
  {Zechmeister}}]{2016Natur.536..437A}
{Anglada-Escud{\'e}}, G., {Amado}, P.~J., {Barnes}, J., {et~al.} 2016, \nat,
  536, 437

\bibitem[{Arney \& Kane(2018)}]{arney2018venus}
Arney, G. \& Kane, S. 2018, Venus as an Analog for Hot Earths

\bibitem[{Baldridge {et~al.}(2009)Baldridge, Hook, Grove, \&
  Rivera}]{ECOSTRESS1}
Baldridge, A., Hook, S., Grove, C., \& Rivera, G. 2009, Remote Sensing of
  Environment, 113, 711

\bibitem[{{Barstow} {et~al.}(2016){Barstow}, {Aigrain}, {Irwin}, {Kendrew}, \&
  {Fletcher}}]{CloudAmb1}
{Barstow}, J.~K., {Aigrain}, S., {Irwin}, P.~G.~J., {Kendrew}, S., \&
  {Fletcher}, L.~N. 2016, \mnras, 458, 2657

\bibitem[{{Barstow} {et~al.}(2020){Barstow}, {Changeat}, {Garland}, {Line},
  {Rocchetto}, \& {Waldmann}}]{Barstov2020retrievals}
{Barstow}, J.~K., {Changeat}, Q., {Garland}, R., {et~al.} 2020, \mnras, 493,
  4884

\bibitem[{{Barstow} \& {Heng}(2020)}]{barstow_heng2020}
{Barstow}, J.~K. \& {Heng}, K. 2020, Space Science Reviews, 216, 82

\bibitem[{{Boukrouche} {et~al.}(2021){Boukrouche}, {Lichtenberg}, \&
  {Pierrehumbert}}]{Boukrouche2021}
{Boukrouche}, R., {Lichtenberg}, T., \& {Pierrehumbert}, R.~T. 2021, \apj, 919,
  130

\bibitem[{Brandt \& Spiegel(2014)}]{Brandt2014}
Brandt, T.~D. \& Spiegel, D.~S. 2014, Proceedings of the National Academy of
  Sciences, 111, 13278

\bibitem[{Bryson {et~al.}(2020)Bryson, Kunimoto, Kopparapu, Coughlin, Borucki,
  Koch, Aguirre, Allen, Barentsen, Batalha, Berger, Boss, Buchhave, Burke,
  Caldwell, Campbell, Catanzarite, Chandrasekaran, Chaplin, Christiansen,
  Christensen-Dalsgaard, Ciardi, Clarke, Cochran, Dotson, Doyle, Duarte,
  Dunham, Dupree, Endl, Fanson, Ford, Fujieh, III, Geary, Gilliland, Girouard,
  Gould, Haas, Henze, Holman, Howard, Howell, Huber, Hunter, Jenkins, Kjeldsen,
  Kolodziejczak, Larson, Latham, Li, Mathur, Meibom, Middour, Morris, Morton,
  Mullally, Mullally, Pletcher, Prsa, Quinn, Quintana, Ragozzine, Ramirez,
  Sanderfer, Sasselov, Seader, Shabram, Shporer, Smith, Steffen, Still, Torres,
  Troeltzsch, Twicken, Uddin, Cleve, Voss, Weiss, Welsh, Wohler, \&
  Zamudio}]{Bryson_2020}
Bryson, S., Kunimoto, M., Kopparapu, R.~K., {et~al.} 2020, \aj, 161, 36

\bibitem[{Buchner {et~al.}(2014)Buchner, Georgakakis, Nandra, Hsu, Rangel,
  Brightman, Merloni, Salvato, Donley, \& Kocevski}]{Buchner:PyMultinest}
Buchner, J., Georgakakis, A., Nandra, K., {et~al.} 2014, \aap, 564, A125

\bibitem[{Budyko(1969)}]{Budyko1969}
Budyko, M.~I. 1969, Tellus, 21, 611

\bibitem[{Burch {et~al.}(1969)Burch, Gryvnak, Patty, \& Bartky}]{Burch:69}
Burch, D.~E., Gryvnak, D.~A., Patty, R.~R., \& Bartky, C.~E. 1969, J. Opt. Soc.
  Am., 59, 267

\bibitem[{{Carri{\'o}n-Gonz{\'a}lez} {et~al.}(2020){Carri{\'o}n-Gonz{\'a}lez},
  {Garc{\'\i}a Mu{\~n}oz}, {Cabrera}, {Csizmadia}, {Santos}, \&
  {Rauer}}]{CG2020}
{Carri{\'o}n-Gonz{\'a}lez}, {\'O}., {Garc{\'\i}a Mu{\~n}oz}, A., {Cabrera}, J.,
  {et~al.} 2020, \aap, 640, A136

\bibitem[{{Chaverot} {et~al.}(2022){Chaverot}, {Turbet}, {Bolmont}, \&
  {Leconte}}]{Chaverot2022}
{Chaverot}, G., {Turbet}, M., {Bolmont}, E., \& {Leconte}, J. 2022, \aap, 658,
  A40

\bibitem[{Chen \& Kipping(2016)}]{Kipping:Forecaster}
Chen, J. \& Kipping, D. 2016, \apj, 834, 17

\bibitem[{Cowan {et~al.}(2015)Cowan, Greene, Angerhausen, Batalha, Clampin,
  Col{\'{o}}n, Crossfield, Fortney, Gaudi, Harrington, Iro, Lillie, Linsky,
  Lopez-Morales, Mandell, \& and}]{Cowan_2015}
Cowan, N.~B., Greene, T., Angerhausen, D., {et~al.} 2015, \pasp, 127, 311

\bibitem[{{Dannert} {et~al.}(2022){Dannert}, {Ottiger}, {Quanz}, {Laugier},
  {Fontanet}, {Gheorghe}, {Absil}, {Dandumont}, {Defr\`ere}, {Gasc\'on},
  {Glauser}, {Kammerer}, {Lichtenberg}, {Linz}, {Loicq}, \& {the LIFE
  collaboration}}]{DannertLifeSim}
{Dannert}, F.~A., {Ottiger}, M., {Quanz}, S.~P., {et~al.} 2022, \aap, 664, A22

\bibitem[{Deming {et~al.}(2018)Deming, Louie, \& Sheets}]{deming2018}
Deming, D., Louie, D., \& Sheets, H. 2018, PASP, 131, 013001

\bibitem[{{Dressing} \& {Charbonneau}(2015)}]{DressingCharbonneau2015}
{Dressing}, C.~D. \& {Charbonneau}, D. 2015, \apj, 807, 45

\bibitem[{{Ertel} {et~al.}(2020){Ertel}, {Defr{\`e}re}, {Hinz}, {Mennesson},
  {Kennedy}, {Danchi}, {Gelino}, {Hill}, {Hoffmann}, {Mazoyer}, {Rieke},
  {Shannon}, {Stapelfeldt}, {Spalding}, {Stone}, {Vaz}, {Weinberger},
  {Willems}, {Absil}, {Arbo}, {Bailey}, {Beichman}, {Bryden}, {Downey},
  {Durney}, {Esposito}, {Gaspar}, {Grenz}, {Haniff}, {Leisenring}, {Marion},
  {McMahon}, {Millan-Gabet}, {Montoya}, {Morzinski}, {Perera}, {Pinna}, {Pott},
  {Power}, {Puglisi}, {Roberge}, {Serabyn}, {Skemer}, {Su}, {Vaitheeswaran}, \&
  {Wyatt}}]{ertel2020}
{Ertel}, S., {Defr{\`e}re}, D., {Hinz}, P., {et~al.} 2020, \aj, 159, 177

\bibitem[{{Feinstein} {et~al.}(2022){Feinstein}, {Radica}, {Welbanks},
  {Murray}, {Ohno}, {Coulombe}, {Espinoza}, {Bean}, {Teske}, {Benneke}, {Line},
  {Rustamkulov}, {Saba}, {Tsiaras}, {Barstow}, {Fortney}, {Gao}, {Knutson},
  {MacDonald}, {Mikal-Evans}, {Rackham}, {Taylor}, {Parmentier}, {Batalha},
  {Berta-Thompson}, {Carter}, {Changeat}, {Dos Santos}, {Gibson}, {Goyal},
  {Kreidberg}, {L{\'o}pez-Morales}, {Lothringer}, {Miguel}, {Molaverdikhani},
  {Moran}, {Morello}, {Mukherjee}, {Sing}, {Stevenson}, {Wakeford}, {Ahrer},
  {Alam}, {Alderson}, {Allen}, {Batalha}, {Bell}, {Blecic}, {Brande},
  {Caceres}, {Casewell}, {Chubb}, {Crossfield}, {Crouzet}, {Cubillos}, {Decin},
  {D{\'e}sert}, {Harrington}, {Heng}, {Henning}, {Iro}, {Kempton}, {Kendrew},
  {Kirk}, {Krick}, {Lagage}, {Lendl}, {Mancini}, {Mansfield}, {May}, {Mayne},
  {Nikolov}, {Palle}, {Petit dit de la Roche}, {Piaulet}, {Powell}, {Redfield},
  {Rogers}, {Roman}, {Roy}, {Nixon}, {Schlawin}, {Tan}, {Tremblin}, {Turner},
  {Venot}, {Waalkes}, {Wheatley}, \& {Zhang}}]{Feinstein2022}
{Feinstein}, A.~D., {Radica}, M., {Welbanks}, L., {et~al.} 2022, arXiv
  e-prints, arXiv:2211.10493

\bibitem[{{Feng} {et~al.}(2018){Feng}, {Robinson}, {Fortney}, {Lupu}, {Marley},
  {Lewis}, {Macintosh}, \& {Line}}]{Feng_Retrieval}
{Feng}, Y.~K., {Robinson}, T.~D., {Fortney}, J.~J., {et~al.} 2018, \aj, 155,
  200

\bibitem[{Feroz {et~al.}(2009)Feroz, Hobson, \& Bridges}]{Feroz:Multinest}
Feroz, F., Hobson, M.~P., \& Bridges, M. 2009, \mnras, 398, 1601–1614

\bibitem[{{Foreman-Mackey} {et~al.}(2014){Foreman-Mackey}, {Hogg}, \&
  {Morton}}]{F&M2014}
{Foreman-Mackey}, D., {Hogg}, D.~W., \& {Morton}, T.~D. 2014, \apj, 795, 64

\bibitem[{{Gaudi} {et~al.}(2020){Gaudi}, {Seager}, {Mennesson}, {Kiessling},
  {Warfield}, {Cahoy}, {Clarke}, {Domagal-Goldman}, {Feinberg}, {Guyon},
  {Kasdin}, {Mawet}, {Plavchan}, {Robinson}, {Rogers}, {Scowen}, {Somerville},
  {Stapelfeldt}, {Stark}, {Stern}, {Turnbull}, {Amini}, {Kuan}, {Martin},
  {Morgan}, {Redding}, {Stahl}, {Webb}, {Alvarez-Salazar}, {Arnold}, {Arya},
  {Balasubramanian}, {Baysinger}, {Bell}, {Below}, {Benson}, {Blais}, {Booth},
  {Bourgeois}, {Bradford}, {Brewer}, {Brooks}, {Cady}, {Caldwell}, {Calvet},
  {Carr}, {Chan}, {Cormarkovic}, {Coste}, {Cox}, {Danner}, {Davis}, {Dewell},
  {Dorsett}, {Dunn}, {East}, {Effinger}, {Eng}, {Freebury}, {Garcia}, {Gaskin},
  {Greene}, {Hennessy}, {Hilgemann}, {Hood}, {Holota}, {Howe}, {Huang}, {Hull},
  {Hunt}, {Hurd}, {Johnson}, {Kissil}, {Knight}, {Kolenz}, {Kraus}, {Krist},
  {Li}, {Lisman}, {Mandic}, {Mann}, {Marchen}, {Marrese-Reading}, {McCready},
  {McGown}, {Missun}, {Miyaguchi}, {Moore}, {Nemati}, {Nikzad}, {Nissen},
  {Novicki}, {Perrine}, {Pineda}, {Polanco}, {Putnam}, {Qureshi}, {Richards},
  {Eldorado Riggs}, {Rodgers}, {Rud}, {Saini}, {Scalisi}, {Scharf}, {Schulz},
  {Serabyn}, {Sigrist}, {Sikkia}, {Singleton}, {Shaklan}, {Smith}, {Southerd},
  {Stahl}, {Steeves}, {Sturges}, {Sullivan}, {Tang}, {Taras}, {Tesch},
  {Therrell}, {Tseng}, {Valente}, {Van Buren}, {Villalvazo}, {Warwick}, {Webb},
  {Westerhoff}, {Wofford}, {Wu}, {Woo}, {Wood}, {Ziemer}, {Arney}, {Anderson},
  {Ma{\'\i}z-Apell{\'a}niz}, {Bartlett}, {Belikov}, {Bendek}, {Cenko},
  {Douglas}, {Dulz}, {Evans}, {Faramaz}, {Feng}, {Ferguson}, {Follette},
  {Ford}, {Garc{\'\i}a}, {Geha}, {Gelino}, {G{\"o}tberg}, {Hildebrand t}, {Hu},
  {Jahnke}, {Kennedy}, {Kreidberg}, {Isella}, {Lopez}, {Marchis}, {Macri},
  {Marley}, {Matzko}, {Mazoyer}, {McCandliss}, {Meshkat}, {Mordasini},
  {Morris}, {Nielsen}, {Newman}, {Petigura}, {Postman}, {Reines}, {Roberge},
  {Roederer}, {Ruane}, {Schwieterman}, {Sirbu}, {Spalding}, {Teplitz},
  {Tumlinson}, {Turner}, {Werk}, {Wofford}, {Wyatt}, {Young}, \&
  {Zellem}}]{2020arXiv200106683G}
{Gaudi}, B.~S., {Seager}, S., {Mennesson}, B., {et~al.} 2020, arXiv e-prints,
  arXiv:2001.06683

\bibitem[{Gilbert {et~al.}(2020)Gilbert, Barclay, Schlieder, Quintana, Hord,
  Kostov, Lopez, Rowe, Hoffman, Walkowicz, Silverstein, Rodriguez, Vanderburg,
  Suissa, Airapetian, Clement, Raymond, Mann, Kruse, Lissauer, Col{\'{o}}n,
  kumar Kopparapu, Kreidberg, Zieba, Collins, Quinn, Howell, Ziegler, Vrijmoet,
  Adams, Arney, Boyd, Brande, Burke, Cacciapuoti, Chance, Christiansen, Covone,
  Daylan, Dineen, Dressing, Essack, Fauchez, Galgano, Howe, Kaltenegger, Kane,
  Lam, Lee, Lewis, Logsdon, Mandell, Monsue, Mullally, Mullally, Paudel,
  Pidhorodetska, Plavchan, Reyes, Rinehart, Rojas-Ayala, Smith, Stassun,
  Tenenbaum, Vega, Villanueva, Wolf, Youngblood, Ricker, Vanderspek, Latham,
  Seager, Winn, Jenkins, Bakos, Brice{\~{n}}o, Ciardi, Cloutier, Conti,
  Couperus, Sora, Eisner, Everett, Gan, Hartman, Henry, Isopi, Jao, Jensen,
  Law, Mallia, Matson, Shappee, Wood, \& Winters}]{Gilbert_2020}
Gilbert, E.~A., Barclay, T., Schlieder, J.~E., {et~al.} 2020, \aj, 160, 116

\bibitem[{{Gillon} {et~al.}(2016){Gillon}, {Jehin}, {Lederer}, {Delrez}, {de
  Wit}, {Burdanov}, {Van Grootel}, {Burgasser}, {Triaud}, {Opitom}, {Demory},
  {Sahu}, {Bardalez Gagliuffi}, {Magain}, \& {Queloz}}]{2016Natur.533..221G}
{Gillon}, M., {Jehin}, E., {Lederer}, S.~M., {et~al.} 2016, \nat, 533, 221

\bibitem[{{Gillon} {et~al.}(2017){Gillon}, {Triaud}, {Demory}, {Jehin}, {Agol},
  {Deck}, {Lederer}, {de Wit}, {Burdanov}, {Ingalls}, {Bolmont}, {Leconte},
  {Raymond}, {Selsis}, {Turbet}, {Barkaoui}, {Burgasser}, {Burleigh}, {Carey},
  {Chaushev}, {Copperwheat}, {Delrez}, {Fernandes}, {Holdsworth}, {Kotze}, {Van
  Grootel}, {Almleaky}, {Benkhaldoun}, {Magain}, \&
  {Queloz}}]{2017Natur.542..456G}
{Gillon}, M., {Triaud}, A. H.~M.~J., {Demory}, B.-O., {et~al.} 2017, \nat, 542,
  456

\bibitem[{{Graham}(2021)}]{Graham2021}
{Graham}, R.~J. 2021, Astrobiology, 21, 1406

\bibitem[{{Graham} {et~al.}(2022){Graham}, {Lichtenberg}, \&
  {Pierrehumbert}}]{Graham2022}
{Graham}, R.~J., {Lichtenberg}, T., \& {Pierrehumbert}, R.~T. 2022, Journal of
  Geophysical Research: Planets, 127, e2022JE007456

\bibitem[{Greene {et~al.}(2016)Greene, Line, Montero, Fortney, Lustig-Yaeger,
  \& Luther}]{Greene_2016}
Greene, T.~P., Line, M.~R., Montero, C., {et~al.} 2016, \apj, 817, 17

\bibitem[{Hanel {et~al.}(1983)Hanel, Conrath, Kunde, Pearl, \&
  Pirraglia}]{AbSaturn}
Hanel, R., Conrath, B., Kunde, V., Pearl, J., \& Pirraglia, J. 1983, \icarus,
  53, 262

\bibitem[{{Hartmann} {et~al.}(2002){Hartmann}, {Boulet}, {Brodbeck}, {van
  Thanh}, {Fouchet}, \& {Drossart}}]{HB02}
{Hartmann}, J.~M., {Boulet}, C., {Brodbeck}, C., {et~al.} 2002, \jqsrt, 72, 117

\bibitem[{Harvey {et~al.}(1998)Harvey, Gallagher, \& Sengers}]{Harvey1998}
Harvey, A.~H., Gallagher, J.~S., \& Sengers, J. M. H.~L. 1998, Journal of
  Physical and Chemical Reference Data, 27, 761

\bibitem[{Haus {et~al.}(2016)Haus, Kappel, Tellmann, Arnold, Piccioni,
  Drossart, \& Häusler}]{AbVenus}
Haus, R., Kappel, D., Tellmann, S., {et~al.} 2016, \icarus, 272, 178

\bibitem[{{Hoffman} {et~al.}(2017){Hoffman}, {Abbot}, {Ashkenazy}, {Benn},
  {Brocks}, {Cohen}, {Cox}, {Creveling}, {Donnadieu}, {Erwin}, {Fairchild},
  {Ferreira}, {Goodman}, {Halverson}, {Jansen}, {Le Hir}, {Love}, {Macdonald},
  {Maloof}, {Partin}, {Ramstein}, {Rose}, {Rose}, {Sadler}, {Tziperman},
  {Voigt}, \& {Warren}}]{HoffmanSnowball2}
{Hoffman}, P.~F., {Abbot}, D.~S., {Ashkenazy}, Y., {et~al.} 2017, Science
  Advances, 3, e1600983

\bibitem[{{Hoffman} {et~al.}(1998){Hoffman}, {Kaufman}, {Halverson}, \&
  {Schrag}}]{SnowballHoffman}
{Hoffman}, P.~F., {Kaufman}, A.~J., {Halverson}, G.~P., \& {Schrag}, D.~P.
  1998, Science, 281, 1342

\bibitem[{Howett {et~al.}(2010)Howett, Spencer, Pearl, \& Segura}]{AbSatsat}
Howett, C., Spencer, J., Pearl, J., \& Segura, M. 2010, \icarus, 206, 573

\bibitem[{Jeffreys(1998)}]{Jeffreys:Theory_of_prob}
Jeffreys, H. 1998, The Theory of Probability, Oxford Classic Texts in the
  Physical Sciences (OUP Oxford), 432--441

\bibitem[{{Kammerer} \& {Quanz}(2018)}]{K&QLIFE}
{Kammerer}, J. \& {Quanz}, S.~P. 2018, \aap, 609, A4

\bibitem[{Karman {et~al.}(2019)Karman, Gordon, {van der Avoird}, Baranov,
  Boulet, Drouin, Groenenboom, Gustafsson, Hartmann, Kurucz, Rothman, Sun,
  Sung, Thalman, Tran, Wishnow, Wordsworth, Vigasin, Volkamer, \& {van der
  Zande}}]{KARMAN2019160}
Karman, T., Gordon, I.~E., {van der Avoird}, A., {et~al.} 2019, \icarus, 328,
  160

\bibitem[{{Kasting}(1988)}]{Kasting1988Icar...74..472K}
{Kasting}, J.~F. 1988, \icarus, 74, 472

\bibitem[{{Kasting} \& {Harman}(2021)}]{Kasting2021Natur.598..259K}
{Kasting}, J.~F. \& {Harman}, C.~E. 2021, \nat, 598, 259

\bibitem[{Kasting {et~al.}(1993)Kasting, Whitmire, \& Reynolds}]{HZ_Kasting93}
Kasting, J.~F., Whitmire, D.~P., \& Reynolds, R.~T. 1993, \icarus, 101, 108

\bibitem[{Kirschvink(1992)}]{SnowballKirschvink}
Kirschvink, J.~L. 1992, in The Proterozoic Biosphere: A Multidisciplinary
  Study, ed. J.~W. Schopf, C.~Klein, \& D.~Des~Maris (Cambridge University
  Press), 51--52

\bibitem[{{Kitzmann} {et~al.}(2011){Kitzmann}, {Patzer}, {von Paris}, {Godolt},
  \& {Rauer}}]{Kitzmann2011Emission}
{Kitzmann}, D., {Patzer}, A.~B.~C., {von Paris}, P., {Godolt}, M., \& {Rauer},
  H. 2011, \aap, 531, A62

\bibitem[{{Komacek} {et~al.}(2020){Komacek}, {Fauchez}, {Wolf}, \&
  {Abbot}}]{KomacekClouds}
{Komacek}, T.~D., {Fauchez}, T.~J., {Wolf}, E.~T., \& {Abbot}, D.~S. 2020,
  \apjl, 888, L20

\bibitem[{{Konrad} {et~al.}(2022){Konrad}, {Alei}, {Quanz}, {Angerhausen},
  {Carri\'on-Gonz\'alez}, {Fortney}, {Grenfell}, {Kitzmann}, {Molli\`ere},
  {Rugheimer}, {Wunderlich}, \& {the LIFE Collaboration}}]{konrad2021large}
{Konrad}, B.~S., {Alei}, E., {Quanz}, S.~P., {et~al.} 2022, \aap, 664, A23

\bibitem[{{Kopparapu} {et~al.}(2013){Kopparapu}, {Ramirez}, {Kasting}, {Eymet},
  {Robinson}, {Mahadevan}, {Terrien}, {Domagal-Goldman}, {Meadows}, \&
  {Deshpande}}]{HZ_kopparapu13}
{Kopparapu}, R.~K., {Ramirez}, R., {Kasting}, J.~F., {et~al.} 2013, \apj, 765,
  131

\bibitem[{Krasnopolsky(2015)}]{KRASNOPOLSKY2015327}
Krasnopolsky, V.~A. 2015, \icarus, 252, 327

\bibitem[{Krissansen-Totton {et~al.}(2018)Krissansen-Totton, Garland, Irwin, \&
  Catling}]{Krissansen_Totton_2018}
Krissansen-Totton, J., Garland, R., Irwin, P., \& Catling, D.~C. 2018, \aj,
  156, 114

\bibitem[{L{\'e}ger {et~al.}(2019)L{\'e}ger, Defr{\`e}re, Mu{\~n}oz, Godolt,
  Grenfell, Rauer, \& Tian}]{Leger2019}
L{\'e}ger, A., Defr{\`e}re, D., Mu{\~n}oz, A.~G., {et~al.} 2019, Astrobiology,
  19, 797, pMID: 30985192

\bibitem[{Li {et~al.}(2018)Li, Jiang, West, Gierasch, Perez-Hoyos,
  Sanchez-Lavega, Fletcher, Fortney, Knowles, Porco, Baines, Fry, Mallama,
  Achterberg, Simon, Nixon, Orton, Dyudina, Ewald, \& Schmude}]{AbJupiter}
Li, L., Jiang, X., West, R.~A., {et~al.} 2018, Nature Communications, 9, 3709

\bibitem[{Loftus {et~al.}(2019)Loftus, Wordsworth, \& Morley}]{Loftus_2019}
Loftus, K., Wordsworth, R.~D., \& Morley, C.~V. 2019, \apj, 887, 231

\bibitem[{{Lustig-Yaeger} {et~al.}(2019){Lustig-Yaeger}, {Meadows}, \&
  {Lincowski}}]{CloudAmb2}
{Lustig-Yaeger}, J., {Meadows}, V.~S., \& {Lincowski}, A.~P. 2019, \apjl, 887,
  L11

\bibitem[{Madhusudhan(2018)}]{Madhusudhan:Atmospheric_Retrieval}
Madhusudhan, N. 2018, Handbook of Exoplanets, 2153–2182

\bibitem[{Mallama(2017)}]{AbMercury}
Mallama, A. 2017, The Spherical Bolometric Albedo of Planet Mercury

\bibitem[{Matthews(2008)}]{AbMoon}
Matthews, G. 2008, \ao, 47, 4981

\bibitem[{Meerdink {et~al.}(2019)Meerdink, Hook, Roberts, \&
  Abbott}]{ECOSTRESS2}
Meerdink, S.~K., Hook, S.~J., Roberts, D.~A., \& Abbott, E.~A. 2019, Remote
  Sensing of Environment, 230, 111196

\bibitem[{{Mettler} {et~al.}(2020){Mettler}, {Quanz}, \&
  {Helled}}]{Mettler2020}
{Mettler}, J.-N., {Quanz}, S.~P., \& {Helled}, R. 2020, \aj, 160, 246

\bibitem[{{Mettler} {et~al.}(2022){Mettler}, {Quanz}, {Helled}, {Olson}, \&
  {Schwieterman}}]{Mettler2022}
{Mettler}, J.-N., {Quanz}, S.~P., {Helled}, R., {Olson}, S.~L., \&
  {Schwieterman}, E.~W. 2022, Earth as an Exoplanet: II. Earth's Time-Variable
  Thermal Emission and its Atmospheric Seasonality of Bio-Indicators

\bibitem[{Min {et~al.}(2005)Min, Hovenier, \& de~Koter}]{Min2005}
Min, M., Hovenier, J.~W., \& de~Koter, A. 2005, \aap, 432, 909–920

\bibitem[{{Molli{\`e}re} {et~al.}(2020){Molli{\`e}re}, {Stolker}, {Lacour},
  {Otten}, {Shangguan}, {Charnay}, {Molyarova}, {Nowak}, {Henning}, {Marleau},
  {Semenov}, {van Dishoeck}, {Eisenhauer}, {Garcia}, {Garcia Lopez}, {Girard},
  {Greenbaum}, {Hinkley}, {Kervella}, {Kreidberg}, {Maire}, {Nasedkin},
  {Pueyo}, {Snellen}, {Vigan}, {Wang}, {de Zeeuw}, \&
  {Zurlo}}]{Molliere:petitRADTRANS2}
{Molli{\`e}re}, P., {Stolker}, T., {Lacour}, S., {et~al.} 2020, \aap, 640, A131

\bibitem[{{Molli{\`e}re} {et~al.}(2019){Molli{\`e}re}, {Wardenier}, {van
  Boekel}, {Henning}, {Molaverdikhani}, \& {Snellen}}]{Molliere:petitRADTRANS}
{Molli{\`e}re}, P., {Wardenier}, J.~P., {van Boekel}, R., {et~al.} 2019, \aap,
  627, A67

\bibitem[{Mollière {et~al.}(2015)Mollière, Boekel, Dullemond, Henning, \&
  Mordasini}]{Molliere:Gravity_Abundance_Degeneracy}
Mollière, P., Boekel, R.~v., Dullemond, C., Henning, T., \& Mordasini, C.
  2015, \apj, 813, 47

\bibitem[{Mueller-Wodarg {et~al.}(2008)Mueller-Wodarg, Strobel, Moses, Waite,
  Crovisier, Yelle, Bougher, \& Roble}]{Mueller-Wodarg:SS_PT_Profiles}
Mueller-Wodarg, I. C.~F., Strobel, D.~F., Moses, J.~I., {et~al.} 2008, Neutral
  Atmospheres (New York, NY: Springer New York), 191--234

\bibitem[{{National Academies of Sciences, Engineering, and
  Medicine}(2021)}]{NAP26141}
{National Academies of Sciences, Engineering, and Medicine}. 2021, Pathways to
  Discovery in Astronomy and Astrophysics for the 2020s (Washington, DC: The
  National Academies Press)

\bibitem[{Nixon \& Madhusudhan(2022)}]{Nixon_2022}
Nixon, M.~C. \& Madhusudhan, N. 2022, \apj, 935, 73

\bibitem[{{Ortenzi} {et~al.}(2020){Ortenzi}, {Noack}, {Sohl}, {Guimond},
  {Grenfell}, {Dorn}, {Schmidt}, {Vulpius}, {Katyal}, {Kitzmann}, \&
  {Rauer}}]{2020NatSR..1010907O}
{Ortenzi}, G., {Noack}, L., {Sohl}, F., {et~al.} 2020, Scientific Reports, 10,
  10907

\bibitem[{Oschlisniok {et~al.}(2012)Oschlisniok, Häusler, Pätzold, Tyler,
  Bird, Tellmann, Remus, \& Andert}]{OSCHLISNIOK2012940}
Oschlisniok, J., Häusler, B., Pätzold, M., {et~al.} 2012, \icarus, 221, 940

\bibitem[{Pallé {et~al.}(2003)Pallé, Goode, Yurchyshyn, Qiu, Hickey,
  Montañés~Rodriguez, Chu, Kolbe, Brown, \& Koonin}]{AbEarth}
Pallé, E., Goode, P.~R., Yurchyshyn, V., {et~al.} 2003, Journal of Geophysical
  Research: Atmospheres, 108

\bibitem[{Palmer \& Williams(1975)}]{Palmer75H2SO4}
Palmer, K.~F. \& Williams, D. 1975, \ao, 14, 208

\bibitem[{{Petigura} {et~al.}(2013){Petigura}, {Howard}, \&
  {Marcy}}]{Petigura2013}
{Petigura}, E.~A., {Howard}, A.~W., \& {Marcy}, G.~W. 2013, Proceedings of the
  National Academy of Science, 110, 19273

\bibitem[{Pierrehumbert(2010)}]{pierrehumbert_2010}
Pierrehumbert, R.~T. 2010, Principles of Planetary Climate (Cambridge
  University Press)

\bibitem[{Pleskot \& Kieffer(1977)}]{AbMars}
Pleskot, L.~K. \& Kieffer, H.~H. 1977, \icarus, 30, 341

\bibitem[{{Quanz} {et~al.}(2021){Quanz}, {Absil}, {Angerhausen}, {Benz},
  {Bonfils}, {Berger}, {Brogi}, {Cabrera}, {Danchi}, {Defr{\`e}re}, {van
  Dishoeck}, {Ehrenreich}, {Ertel}, {Fortney}, {Gaudi}, {Girard}, {Glauser},
  {Grenfell}, {Ireland}, {Janson}, {Kammerer}, {Kitzmann}, {Kraus}, {Krause},
  {Labadie}, {Lacour}, {Lichtenberg}, {Line}, {Linz}, {Loicq}, {Mennesson},
  {Meyer}, {Miguel}, {Monnier}, {N'Diaye}, {Pall{\'e}}, {Queloz}, {Rauer},
  {Ribas}, {Rugheimer}, {Selsis}, {Serabyn}, {Snellen}, {Sozzetti},
  {Stapelfeldt}, {Triaud}, {Udry}, \&
  {Wyatt}}]{Quanz:exoplanets_and_atmospheric_characterization}
{Quanz}, S.~P., {Absil}, O., {Angerhausen}, D., {et~al.} 2021, Experimental
  Astronomy

\bibitem[{{Quanz} {et~al.}(2022){Quanz}, {Ottiger, M.}, {Fontanet, E.},
  {Kammerer, J.}, {Menti, F.}, {Dannert, F.}, {Gheorghe, A.}, {Absil, O.},
  {Airapetian, V. S.}, {Alei, E.}, {Allart, R.}, {Angerhausen, D.},
  {Blumenthal, S.}, {Buchhave, L. A.}, {Cabrera, J.}, {Carri\'on-Gonz\'alez,
  \'O.}, {Chauvin, G.}, {Danchi, W. C.}, {Dandumont, C.}, {Defr\'ere, D.},
  {Dorn, C.}, {Ehrenreich, D.}, {Ertel, S.}, {Fridlund, M.}, {Mu\~noz, A.
  Garc\'{\i}a}, {Gasc\'on, C.}, {Girard, J. H.}, {Glauser, A.}, {Grenfell, J.
  L.}, {Guidi, G.}, {Hagelberg, J.}, {Helled, R.}, {Ireland, M. J.}, {Janson,
  M.}, {Kopparapu, R. K.}, {Korth, J.}, {Kozakis, T.}, {Kraus, S.}, {L\'eger,
  A.}, {Leedj\"arv, L.}, {Lichtenberg, T.}, {Lillo-Box, J.}, {Linz, H.},
  {Liseau, R.}, {Loicq, J.}, {Mahendra, V.}, {Malbet, F.}, {Mathew, J.},
  {Mennesson, B.}, {Meyer, M. R.}, {Mishra, L.}, {Molaverdikhani, K.}, {Noack,
  L.}, {Oza, A. V.}, {Pall\'e, E.}, {Parviainen, H.}, {Quirrenbach, A.},
  {Rauer, H.}, {Ribas, I.}, {Rice, M.}, {Romagnolo, A.}, {Rugheimer, S.},
  {Schwieterman, E. W.}, {Serabyn, E.}, {Sharma, S.}, {Stassun, K. G.},
  {Szul\'agyi, J.}, {Wang, H. S.}, {Wunderlich, F.}, {Wyatt, M. C.}, \& {the
  LIFE Collaboration}}]{LIFE_I}
{Quanz}, S.~P., {Ottiger, M.}, {Fontanet, E.}, {et~al.} 2022, \aap, 664, A21

\bibitem[{Robinson \& Salvador(2022)}]{Robinson_SS_Retrievals}
Robinson, T.~D. \& Salvador, A. 2022, Exploring and Validating Exoplanet
  Atmospheric Retrievals with Solar System Analog Observations

\bibitem[{Rothman {et~al.}(2010)Rothman, Gordon, Barber, Dothe, Gamache,
  Goldman, Perevalov, Tashkun, \& Tennyson}]{ROTHMAN20102139}
Rothman, L., Gordon, I., Barber, R., {et~al.} 2010, \jqsrt, 111, 2139

\bibitem[{{Rugheimer} {et~al.}(2013){Rugheimer}, {Kaltenegger}, {Zsom},
  {Segura}, \& {Sasselov}}]{Rugheimer2013}
{Rugheimer}, S., {Kaltenegger}, L., {Zsom}, A., {Segura}, A., \& {Sasselov}, D.
  2013, Astrobiology, 13, 251

\bibitem[{Segelstein(1981)}]{siegel1981H2O}
Segelstein, D.~J. 1981, The complex refractive index of water

\bibitem[{Sellers(1969)}]{Sellers1969}
Sellers, W.~D. 1969, Journal of Applied Meteorology and Climatology, 8, 392

\bibitem[{Sergeev {et~al.}(2022)Sergeev, Lewis, Lambert, Mayne, Boutle,
  Manners, \& Kohary}]{Sergeev2022}
Sergeev, D.~E., Lewis, N.~T., Lambert, F.~H., {et~al.} 2022, Bistability of the
  atmospheric circulation on TRAPPIST-1e

\bibitem[{Skilling(2006)}]{Skilling:Nested_Sampling}
Skilling, J. 2006, Bayesian Anal., 1, 833

\bibitem[{{Sneep} \& {Ubachs}(2005)}]{2005JQSRT..92..293S}
{Sneep}, M. \& {Ubachs}, W. 2005, \jqsrt, 92, 293

\bibitem[{Spencer {et~al.}(1999)Spencer, Tamppari, Martin, \&
  Travis}]{AbEuropa}
Spencer, J.~R., Tamppari, L.~K., Martin, T.~Z., \& Travis, L.~D. 1999, Science,
  284, 1514 – 1516

\bibitem[{Taylor {et~al.}(2018)Taylor, Svedhem, \& Head}]{Taylor2018}
Taylor, F.~W., Svedhem, H., \& Head, J.~W. 2018, Space Science Reviews, 214, 35

\bibitem[{{The LUVOIR Team}(2019)}]{2017AAS...22940504P}
{The LUVOIR Team}. 2019, The LUVOIR Mission Concept Study Final Report

\bibitem[{{Titov} {et~al.}(2018){Titov}, {Ignatiev}, {McGouldrick}, {Wilquet},
  \& {Wilson}}]{Esposito1983}
{Titov}, D.~V., {Ignatiev}, N.~I., {McGouldrick}, K., {Wilquet}, V., \&
  {Wilson}, C.~F. 2018, \ssr, 214, 126

\bibitem[{Toon \& Ackerman(1981)}]{Toon:81}
Toon, O.~B. \& Ackerman, T.~P. 1981, \ao, 20, 3657

\bibitem[{{Turbet} {et~al.}(2021){Turbet}, {Bolmont}, {Chaverot}, {Ehrenreich},
  {Leconte}, \& {Marcq}}]{VenusHab?3}
{Turbet}, M., {Bolmont}, E., {Chaverot}, G., {et~al.} 2021, \nat, 598, 276

\bibitem[{{Vasquez} {et~al.}(2013){Vasquez}, {Schreier}, {Gimeno Garc{\'\i}a},
  {Kitzmann}, {Patzer}, {Rauer}, \& {Trautmann}}]{Vasquez2013}
{Vasquez}, M., {Schreier}, F., {Gimeno Garc{\'\i}a}, S., {et~al.} 2013, \aap,
  557, A46

\bibitem[{Villanueva {et~al.}(2018)Villanueva, Smith, Protopapa, Faggi, \&
  Mandell}]{Villanueva2018PSG}
Villanueva, G., Smith, M., Protopapa, S., Faggi, S., \& Mandell, A. 2018,
  \jqsrt, 217, 86–104

\bibitem[{{von Paris} {et~al.}(2013){von Paris}, {Hedelt}, {Selsis},
  {Schreier}, \& {Trautmann}}]{Paris2013}
{von Paris}, P., {Hedelt}, P., {Selsis}, F., {Schreier}, F., \& {Trautmann}, T.
  2013, \aap, 551, A120

\bibitem[{{Voyage 2050 Senior Committee}(2021)}]{ESAV2050}
{Voyage 2050 Senior Committee}. 2021, Voyage 2050 – Final Recommendations
  from the Voyage 2050 Senior Committee

\bibitem[{Way {et~al.}(2016)Way, Del~Genio, Kiang, Sohl, Grinspoon, Aleinov,
  Kelley, \& Clune}]{VenusHab?2}
Way, M.~J., Del~Genio, A.~D., Kiang, N.~Y., {et~al.} 2016, \grl, 43, 8376

\bibitem[{Wordsworth(2021)}]{Wordsworth_snowballs_inner_Hz}
Wordsworth, R. 2021, \apjl, 912, L14

\bibitem[{Yang {et~al.}(2014)Yang, Bou{\'{e}}, Fabrycky, \& Abbot}]{VenusHab?1}
Yang, J., Bou{\'{e}}, G., Fabrycky, D.~C., \& Abbot, D.~S. 2014, \apj, 787, L2

\end{thebibliography}

%-------------------------------------------------------------------
\begin{appendix}

%--------------------------------------------------------------------
\section{Cloud retrieval validation}\label{sec:validation}

\begin{figure*}
\centering
\includegraphics[width=0.93\textwidth]{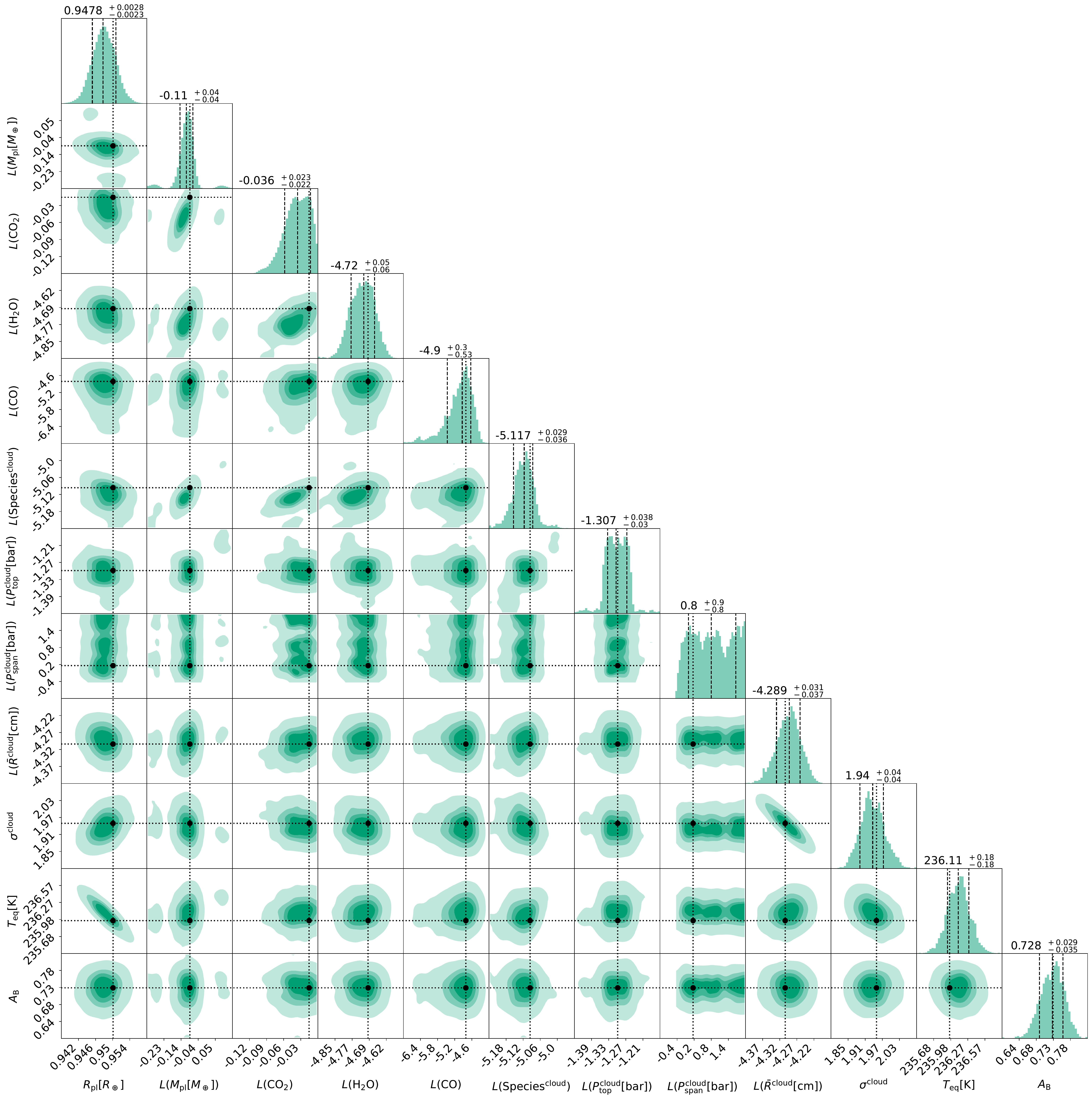}\\
\vspace{-0.937\textwidth}\hspace{0.441\textwidth}\includegraphics[width=0.5\textwidth]{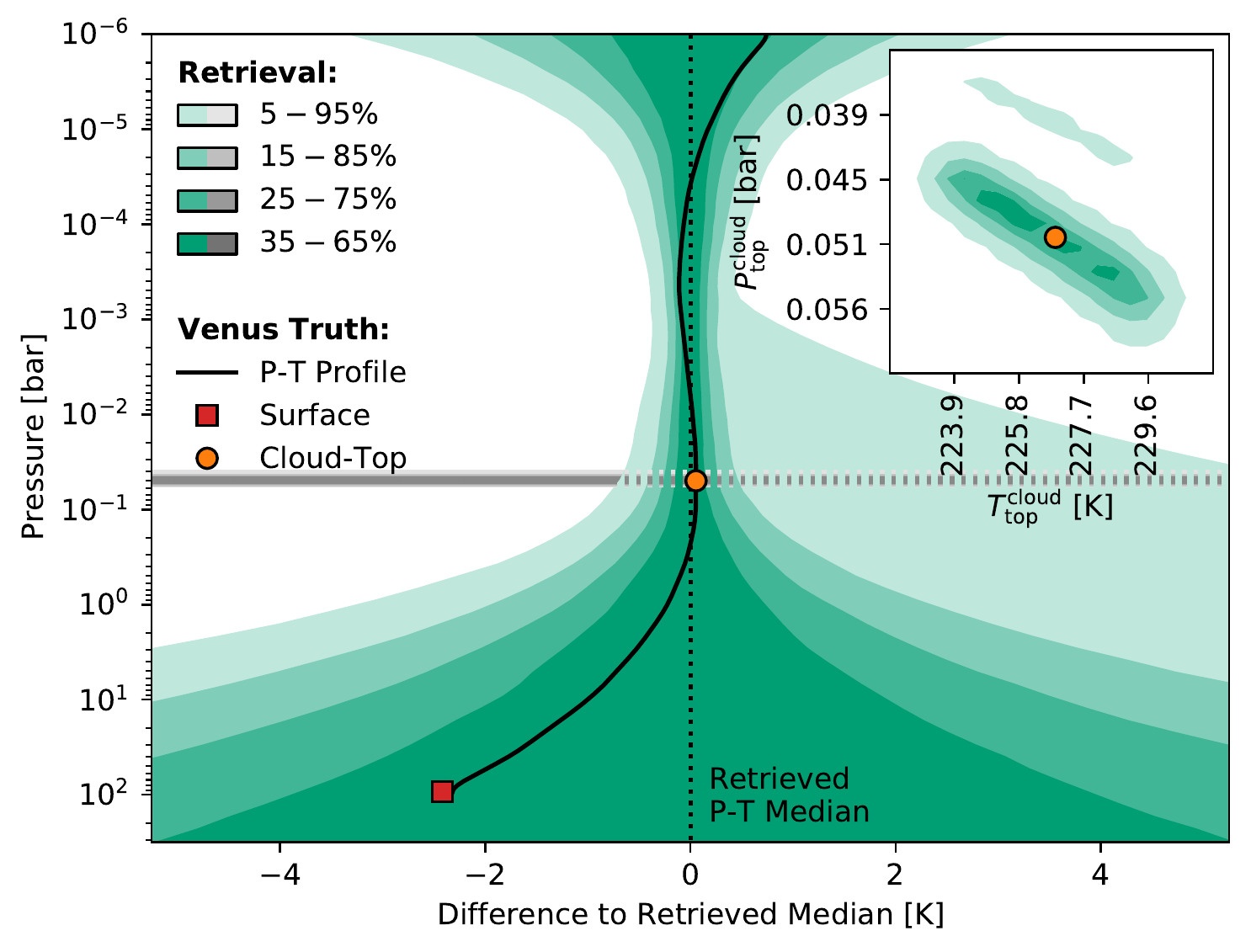}\\
\vspace{0.56\textwidth}\hspace{0.003\textwidth}\includegraphics[width=0.9371\textwidth]{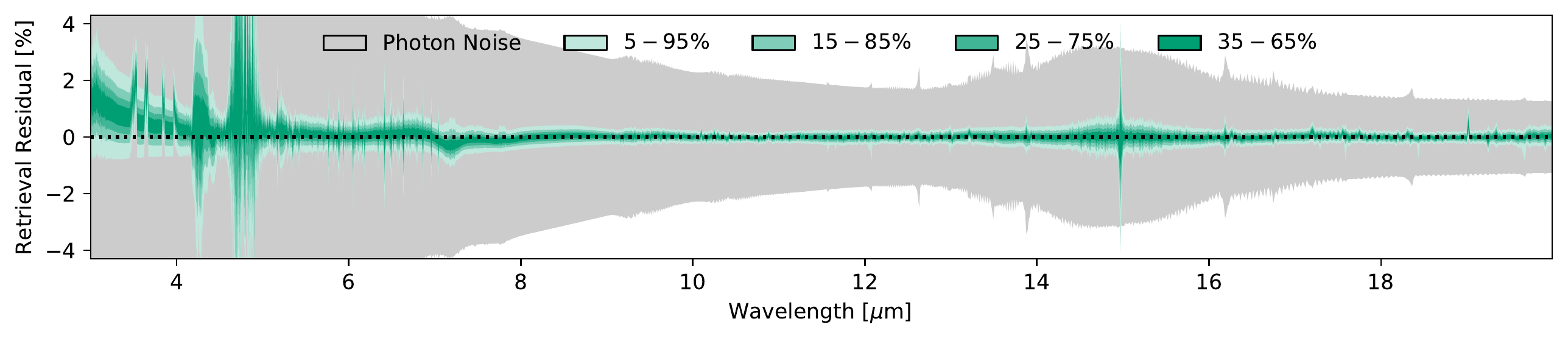}\\
\caption{{Results from the validation retrieval on the \mic{3-20}, \Rv{1000}, and photon noise \SNv{50} Venus-twin spectrum (opaque \ce{H2SO4} clouds) using the opaque \ce{H2SO4} cloud forward model. \emph{Top left}: Corner plot of the posterior distribution of the forward model parameters. Here, $L(\cdot)$ stands for \lgrt{\cdot}. We derived the equilibrium temperature \Teq{} and the Bond albedo \Ab{} from the other posteriors (see Appendix~\ref{sec:albedo_calc}). The dotted black lines indicate the true values. In the 1D posteriors, we show the 16th, 50th, and 84th percentiles as black dashed lines. \emph{Top right}: \pt{} profile residuals relative to the retrieved median \pt{} profile. Color-shaded regions indicate percentiles of the retrieved \pt{} profiles. The gray shaded regions indicate percentiles of the retrieved cloud-top pressure. The solid black line, the orange circular marker, and the red square marker represent the true Venus-twin \pt{} profile relative to the median retrieved \pt{} profile. In the inset figure, we plot the 2D \Ps{}-\Ts{} posterior. \emph{Bottom}: Residuals of the spectra corresponding to the retrieved posteriors relative to the Venus-twin input. Color-shaded areas represent different quantiles of the residuals. The gray area marks the 1$\sigma$ photon noise level.}}
\label{fig:valid}
\end{figure*}

{Here, we validate the updated retrieval routine and test if the newly added cloud model is constrainable. For this retrieval test, we considered a high resolution (\Rv{1000}), low noise (only photon noise, \SNv{50}), and \mic{3-20} cloudy Venus-twin MIR spectrum. For this validation retrieval, we used the opaque \ce{H2SO4} forward model and the priors specified in Table~\ref{table:True_Values}. We summarize the results in Fig.~\ref{fig:valid}.}

{The top left panel of Fig.~\ref{fig:valid}, shows the corner plot of the posterior distribution, excluding the \pt{} parameters $a_i$ for clarity (the full corner plot is in Appendix~\ref{app:corners}). Additionally, we derived the equilibrium temperature \Teq{} and the Bond albedo \Ab{} from the posteriors following the method outlined in Appendix~\ref{sec:albedo_calc}. We see that most posteriors are well constrained and centered on the true values. We especially emphasize the strong constraints on both \Rpl{} and \Ab{}. The ability to constrain \Rpl{} and \Ab{} simultaneously is unique to studies of the MIR thermal emission spectrum. Comparable constraints are not inferrable from reflected light spectra, due to a strong degeneracy between \Rpl{} an \Ab{} \citep{Feng_Retrieval, CG2020}. }

{Further, we find that the atmosphere is \ce{CO2} dominated (abundance $\geq$85\%) and manage to detect both \ce{H2O} and \ce{CO}. Most cloud parameters are well retrieved, which indicates that there are spectral features from the clouds. Only \Pcs{} is not well constrained. The posterior is best described by a step function, where any $\Pcs{}\geq0.5$~bar is possible. This lower limit on \Pcs{} corresponds to the thickness at which the cloud layer becomes opaque. Thicker cloud layers are not excluded since the spectrum contains no signatures from the lower atmospheric layers. For thinner layers, there are contributions from the lower atmospheric layers, which result in a bad fit to the input spectrum.}

%First, the retrieved \Rpl{} value is underestimated. This offset in the posterior likely occurs due to differences between the forward model in the retrieval and the model used to generate the input spectrum. For the input spectrum, we assumed \Ps{} to reach 93~bar. In contrast, for the opaque \ce{H2SO4} retrieval model, we fixed \Ps{} to an arbitrary $10^4$~bar (see Sect.~\ref{sec:ret-models}). Since both models divide the atmosphere into 100 discrete layers between \Ps{} and $10^{-6}$~bar, the atmospheric layers in the two models will have different pressures, temperatures, and thicknesses. These differences could lead to shifts in the \Rpl{} posterior. Such a shift in the posterior is linked to the limited resolution of our atmospheric model. This effect is observable here, because first we consider a high \SN{} and \R{} spectrum and second the atmospheric forward model uses a fixed pressure grid. Since we derived the \Teq{} posterior from the \Rpl{} posterior (see Appendix~\ref{sec:albedo_calc}), a similar shift to higher \Teq{} is observable. We further used \Rpl{} to derive \Ab{}. However, the uncertainties on the planet-star separation and the stellar luminosity we assumed to derive \Ab{} resulted in large uncertainties on \Ab{} (significantly larger than the shift caused by \Rpl{}). Therefore, the shift is not observable for \Ab{}.

%\input{Tables/Validation}
{The model parameters are generally well retrieved. However, there are interesting features in some posteriors. First, the rectangular shape of the \Pct{} posterior is linked to the fixed atmospheric pressure grid of the opaque \ce{H2SO4} cloud forward model. All \Pct{} values in the rectangular peak fall within the same atmospheric layer and thus result in the same cloud-top. Their posterior probability is therefore equal. Thus, at the high \R{} and \SN{} considered here, we are sensitive to the discrete nature of our atmosphere model. Cloud-tops in lower or higher layers are very unlikely and are linked to the secondary peaks seen in both the 2D and 1D posteriors of other parameters (e.g., \Mpl{}). }

{Second, the strength of the \Mpl{} constraint opposes the results from \pIIIaV{}, where \Mpl{} was less strongly constrained relative to the prior due to a known degeneracy between the mass (surface gravity) and the abundances of the atmospheric gases \citep[\Mpl{} and abundances simultaneously over- or undersetimated; see also, e.g.,][]{Molliere:Gravity_Abundance_Degeneracy, Feng_Retrieval, Madhusudhan:Atmospheric_Retrieval, Quanz:exoplanets_and_atmospheric_characterization}. This degeneracy is visible for \ce{CO2} and \ce{H2O}, but not for \ce{CO} (due to overall weaker abundance constraints). However, the \ce{CO2} posterior converges toward the upper edge of the prior range and is more strongly constrained (tail to high abundances is cut off). This breaks the degeneracy and thus also \Mpl{} is strongly constrained. The secondary peaks in the \Mpl{} posterior correspond to the secondary peaks in the \Pct{} posterior (cloud top in different layers).}

{The top right panel of Fig.~\ref{fig:valid} shows the constraints on the atmospheric \pt{} structure relative to the retrieved median (we plot the relative retrieved \pt{} profile, since deviances from the truth are more visible). The absolute \pt{} plot is given in Appendix~\ref{app:corners}. We visualize percentiles of the retrieved profiles as green-shaded regions and compare them to the true profile (red square, orange circle, solid black line). Further, we plot the percentiles of the retrieved cloud-top as gray areas. The inset plot shows the 2D histogram of the retrieved \Pct{} and \Tct{} (cloud-top temperature; calculated from the retrieved \pt{} profiles).}

{Both the \pt{} profile above the cloud-top and the cloud-top are accurately retrieved. The retrieved \pt{} structure above the cloud-top is centered on the true profile, and roughly $90\%$ of the profiles lie within $\leq\pm2$~K of the truth. Similarly, roughly $90\%$ of the retrieved \Pct{} and \Tct{} values are distributed symmetrically around the true cloud-top and lie within $\leq\pm0.01$~bar and $\leq\pm3$~K of the truth. The $5-95\%$ \pt{} envelope shows an asymmetry toward positive relative temperatures. In the \Pct{}-\Tct{} posterior, the $5-95\%$ envelope exhibits a second isolated region (same \Tct{}, lower \Pct{}). These two outliers correspond to the cases discussed above where the cloud-top falls within a different atmospheric layer.} %Since \Tct{} is roughly the same for both layers, the retrieved \pt{} profile is warmer at lower pressures. This explains the tail to positive relative temperatures.

{The uncertainty on the retrieved \pt{} profile grows with decreasing pressure. This increase is due to a lack of spectral features from these optically thin, low-pressure layers in Venus' MIR spectrum at the \R{} and \SN{} considered here. Similarly, the uncertainty on the retrieved \pt{} structure increases below the cloud-top, because the optically thick cloud deck blocks spectral contributions from these atmospheric regions. This lack of spectral features can be seen in the emission contribution function provided in Appendix~\ref{app:corners}. Constraints on the high- and low-pressure regions are obtained by extrapolating the \pt{} model.}

{In the bottom panel of Fig.~\ref{fig:valid}, we show the residual of the emission spectra corresponding to the retrieved parameter posteriors relative to our Venus-twin input spectrum. We indicate percentiles of the residuals as green-shaded areas and the $1\sigma$ photon noise level as a gray area. Overall, the retrieval output is well centered on the input spectrum and the deviations from the truth mostly lie below $1\%$. The observed deviations grow larger for wavelengths where the flux is low and the photon noise is large (e.g., below \mic{6}, in the \ce{CO2} band between \mic{14-16}). However, in general, the assumed forward model accurately reproduces the input spectrum, despite the points mentioned above. This demonstrates that our cloudy Venus-twin model is suited for our retrieval study.}

\clearpage
\FloatBarrier

\section{Equilibrium temperature and Bond albedo}\label{sec:albedo_calc}

The planetary equilibrium temperature \Teq{} and the Bond albedo \Ab{} are two parameters, which were of particular interest to us. These parameters provide important information about the energy budget of the planet. However, both parameters were not directly determined by our retrieval framework. Therefore, we required a method to derive their values from the retrieved posterior distributions of the model parameters.

% daniel.kitzmann: Why do you restrict the emission to this wavelength range? The planet's emission is definitely not a black body, thus using a temperature derived from assuming the black body traces the continuum there probably overestimates the planet's emission. In practice, what you actually want to know here is not a temperature but the emitted thermal flux.
%In case of brown dwarf retrievals, for example,  I get this information in a post-process step. That means, I calculate all posterior spectra over a wide wavelength range and then integrate the flux over the entire range. This yields the total emitted flux. Via the Stefan-Boltzmann law one can then compute the effective temperature of a black body with the same flux, but this step is actually not necessary in your case. 
%All you want here, is the total outgoing thermal emission flux of the planet. Since you know how much flux the planet receives, you can use the difference between the incident flux and the outgoing thermal emission to infer the amount of scattered light and, therefore, the albedo.

In a first step, we generated a set of MIR spectra that was representative of the retrieved parameter posteriors. Next, we determined the planet's \Teq{} by fitting the black body emission, $F_{BB}\left(\lambda\right)$, expected for a spherical planet with radius \Rpl{} to the continuum (between \mic{8} and \mic{11}) of these MIR spectra:
\begin{equation}
    F_{BB}\left(\lambda\right)=\frac{2\pi hc^2}{\lambda^5}{e^{1-\frac{hc}{\lambda k_B\Teq{}}}}\cdot\frac{\Rpl{}^2}{d^2}.
\end{equation}
Here, $d$ is the distance to the observer (10~pc in our case), $\lambda$ the wavelength, $h$ the Planck constant, $c$ the speed of light, and $k_B$ the Boltzmann constant. This provided us with a probability distribution for the planetary equilibrium temperature \Teq{}. From the \Teq{} distribution, we then calculated the probability distribution for \Ab{} using the following relation:
\begin{equation}\label{eq:bond_albedo}
    \Ab = 1 - 16\,\pi\sigma \frac{a^2_P\Teq^4}{L_*}.
\end{equation}
Here, $\sigma$ is the Stefan–Boltzmann constant, $a_P$ is the semi-major axis of the planetary orbit around the host star, and $L_*$ is the luminosity of the host star. For the \Ab{} calculation, we assumed to know both $a_P$ (Venus-like: $a_P=0.72$~AU) and $L_*$ (Sun-like: $L_*=1\,L_\odot$) with an uncertainty of $\pm$5\%. For each sample from the \Teq{} distribution, we drew a random $a_P$ and $L_*$ value from two uncorrelated normal distributions and used these random values to calculate \Ab{}. This yielded a probability distribution for the planetary Bond albedo \Ab{}.

This approach for determining \Teq{} and \Ab{} is advantageous, because correlations between the posteriors of the retrieved planetary parameters are taken into account when deriving the \Teq{} and \Ab{} distributions. Furthermore, the resulting correlations between the distributions of \Teq{}, \Ab{}, and the retrieved parameter posteriors are correct and can be analyzed.

\section{Additional plots for retrieval results}\label{app:corners}

\begin{figure*}
   \centering
    \includegraphics[width=0.99\textwidth]{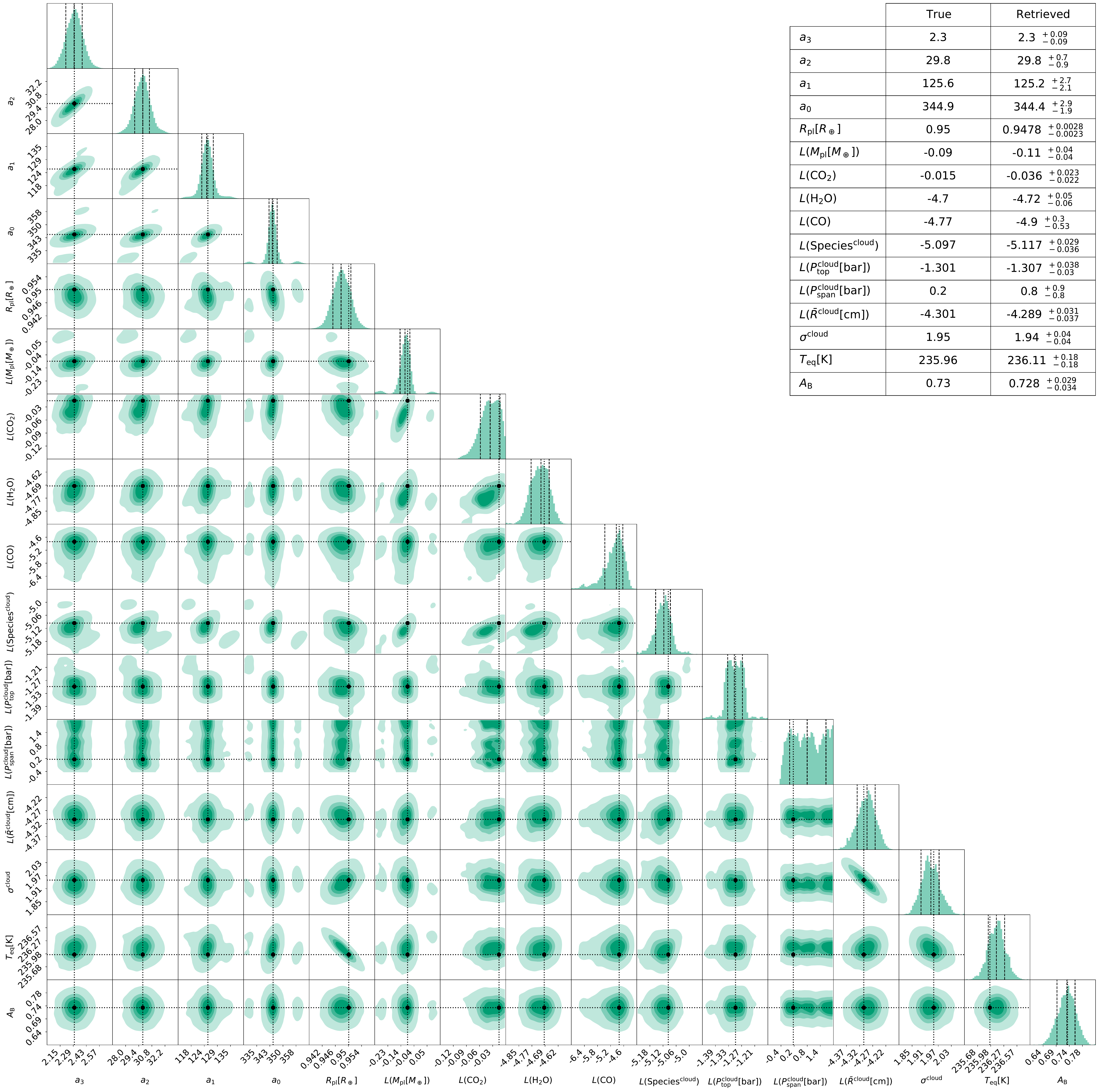}\\
    \vspace{-0.9956\textwidth}\hspace{0.05\textwidth}\includegraphics[width=0.38\textwidth]{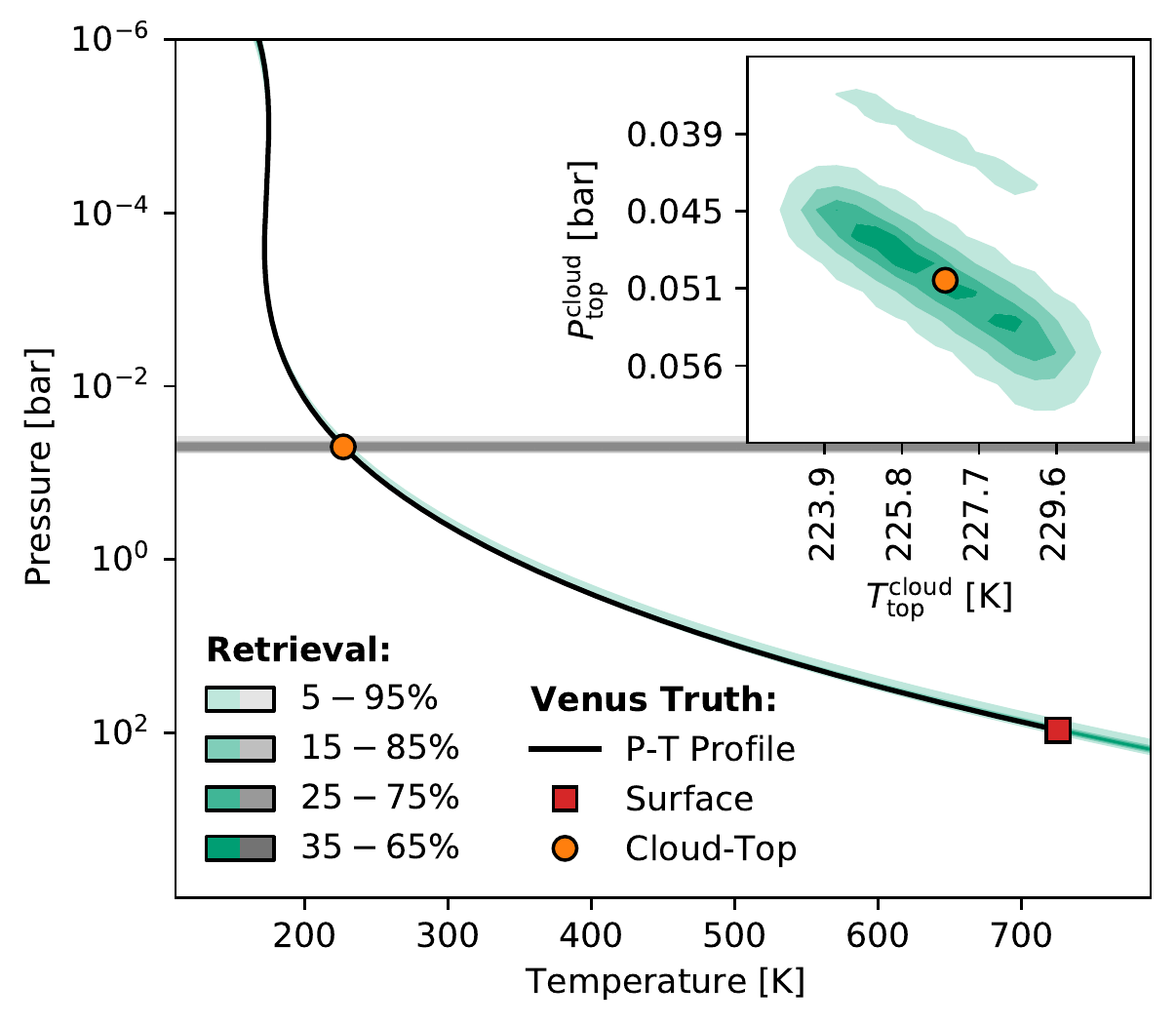}\\
    \vspace{0.06\textwidth}\hspace{0.675\textwidth}\includegraphics[width=0.325\textwidth]{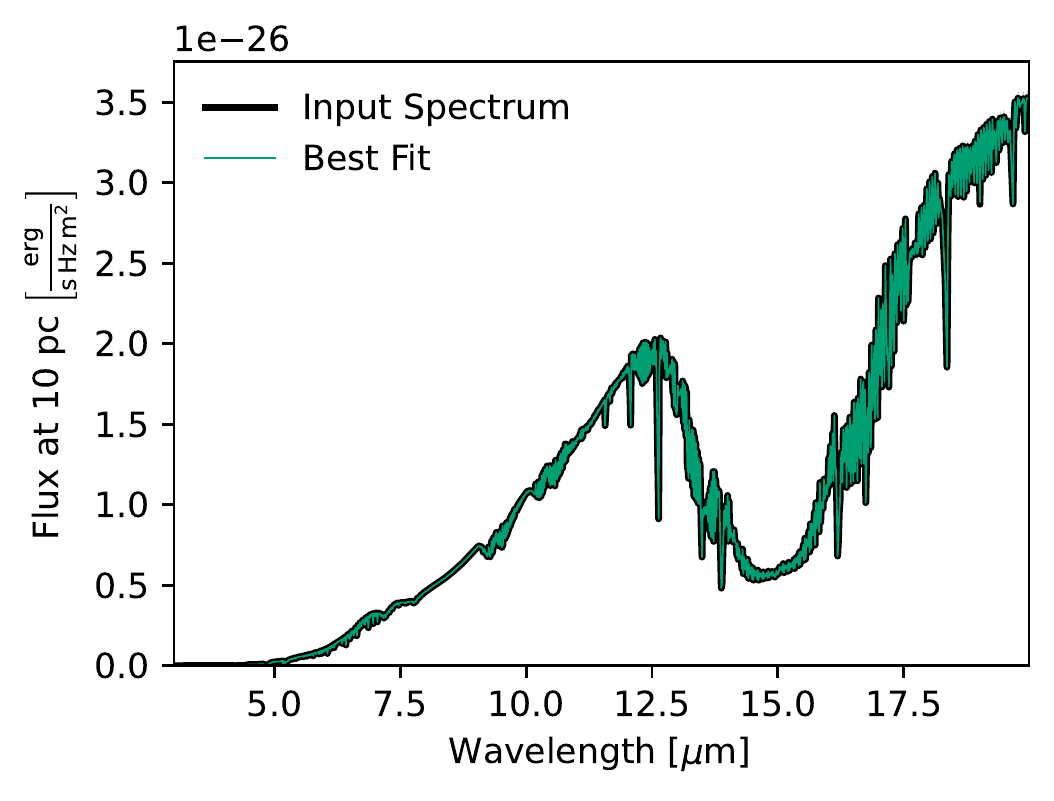}\\
    \vspace{-0.005\textwidth}\hspace{0.81\textwidth}\includegraphics[width=0.19\textwidth]{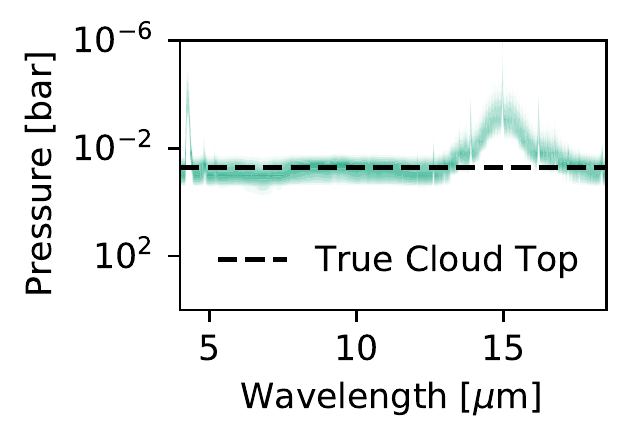}
    \vspace{0.21\textwidth}
    \caption{Results from the validation retrieval on the \mic{3-20}, \Rv{1000}, and photon noise \SNv{50} Venus-twin spectrum (opaque \ce{H2SO4} clouds) using the opaque \ce{H2SO4} cloud forward model (see Sect.~\ref{sec:validation}). \emph{Lower left half}: Corner plot of the posterior distribution of the forward model parameters. Here, $L(\cdot)$ stands for \lgrt{\cdot}. We derived the equilibrium temperature \Teq{} and the Bond albedo \Ab{} from the other posteriors (see Appendix~\ref{sec:albedo_calc}). The dotted black lines indicate the true values. In the 1D posteriors, we show the 16th, 50th, and 84th percentiles as black dashed lines. \emph{Top center}: \pt{} profiles corresponding to the retrieved \pt{} parameters. Color-shaded regions indicate percentiles of the retrieved \pt{} profiles. The gray shaded regions indicate percentiles of the retrieved cloud-top pressure. The solid black line, the orange circular marker, and the red square marker represent the true Venus-twin \pt{} profile. In the inset figure, we plot the 2D \Ps{}-\Ts{} posterior. \emph{Top right}: A table summarizing the true values of the forward model parameters and the 16th, 50th, and 84th percentiles of the parameter posteriors. \emph{Center right}: Comparison of the retrievals best fit to the Venus-twin input spectrum. The thin colored line represents the best fit, the thick black line the Venus-twin input spectrum. The uncertainties on the best fit are too small to be visible. Below the spectrum plot, we show the mean wavelength-dependent contribution of the atmospheric layers to the emission spectrum corresponding to the retrieved parameter posteriors. Darker colors indicate higher contributions. The dashed black line indicates the position of the cloud-deck assumed to simulate the input spectrum (opaque \ce{H2SO4} clouds).}
    \label{fig:corner_full_valid}
\end{figure*}

\begin{figure*}
   \centering
    \includegraphics[width=0.99\textwidth]{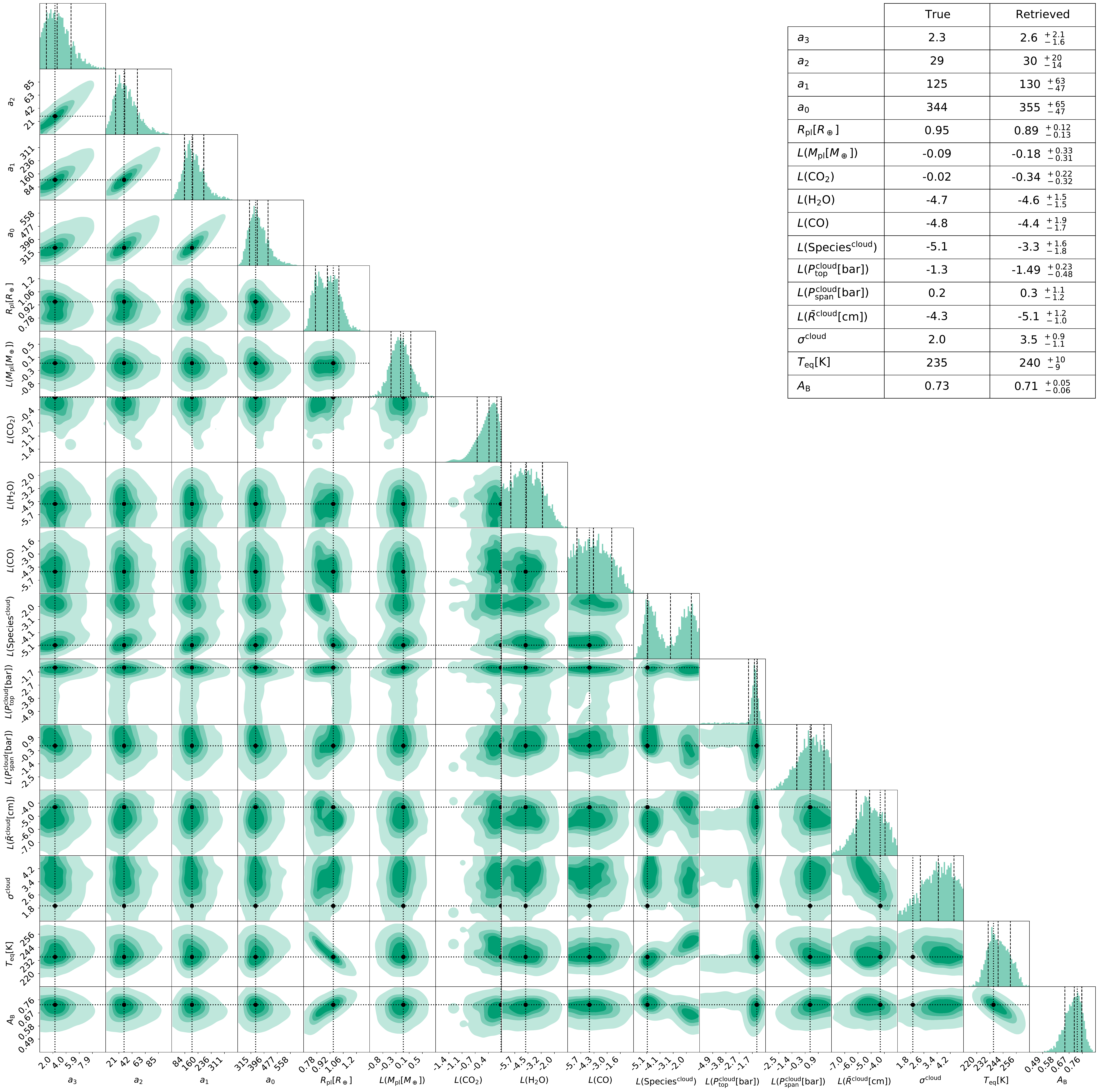}\\
    \vspace{-0.9956\textwidth}\hspace{0.05\textwidth}\includegraphics[width=0.38\textwidth]{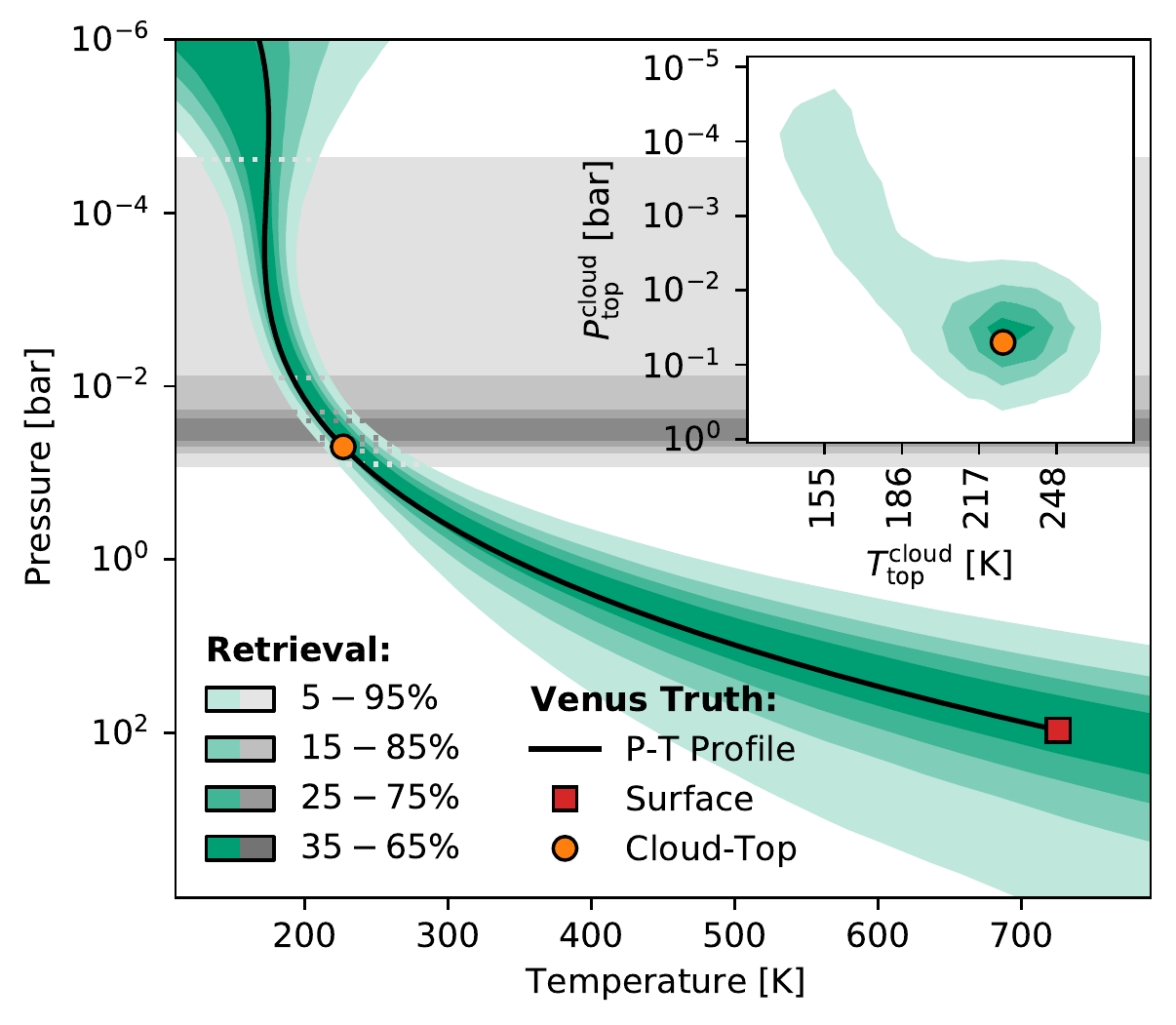}\\
    \vspace{0.06\textwidth}\hspace{0.675\textwidth}\includegraphics[width=0.325\textwidth]{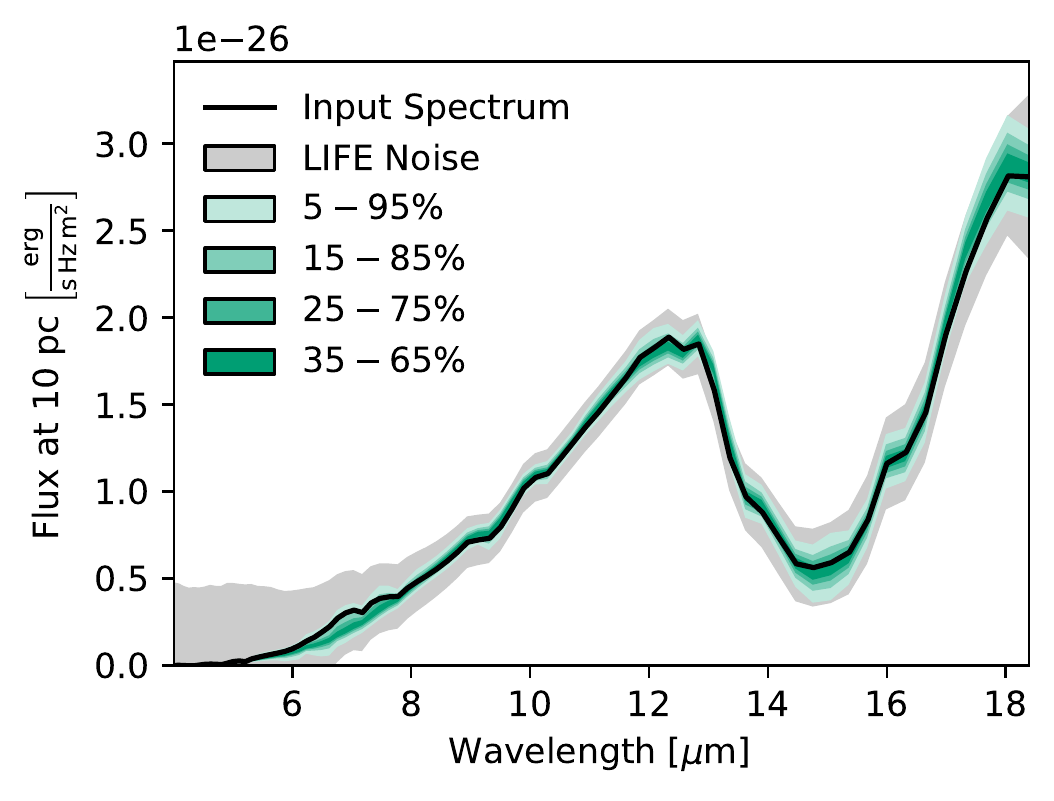}\\
    \vspace{-0.005\textwidth}\hspace{0.81\textwidth}\includegraphics[width=0.19\textwidth]{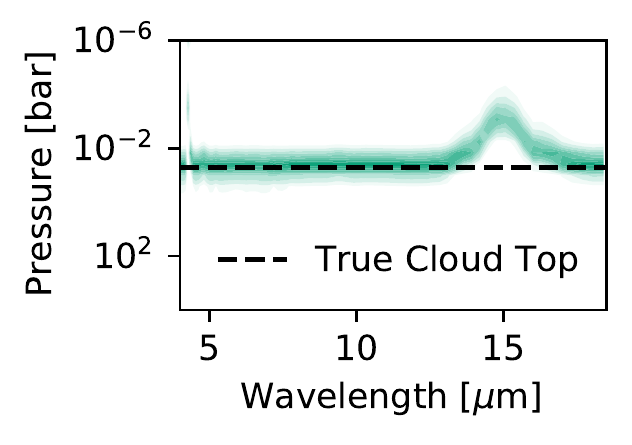}
    \vspace{0.21\textwidth}
    \caption{Results from the retrieval on the \mic{4-18.5}, \Rv{50}, and \lifesim{} noise \SNv{10} Venus-twin spectrum (opaque \ce{H2SO4} clouds) using the opaque \ce{H2SO4} cloud forward model (see Sect.~\ref{sec:LIFEsim_papIII_ret_res}). \emph{Lower left half}: Corner plot of the posterior distribution of the forward model parameters. Here, $L(\cdot)$ stands for \lgrt{\cdot}. We derived the equilibrium temperature \Teq{} and the Bond albedo \Ab{} from the other posteriors (see Appendix~\ref{sec:albedo_calc}). The dotted black lines indicate the true values. In the 1D posteriors, we show the 16th, 50th, and 84th percentiles as black dashed lines. \emph{Top center}: \pt{} profiles corresponding to the retrieved \pt{} parameters. Color-shaded regions indicate percentiles of the retrieved \pt{} profiles. The gray shaded regions indicate percentiles of the retrieved cloud-top pressure. The solid black line, the orange circular marker, and the red square marker represent the true Venus-twin \pt{} profile. In the inset figure, we plot the 2D \Ps{}-\Ts{} posterior (if retrieved; otherwise \Pct{}-\Tct{}). \emph{Top right}: A table summarizing the true values of the forward model parameters and the 16th, 50th, and 84th percentiles of the parameter posteriors. \emph{Center right}: Spectra corresponding to the retrieved posteriors in comparison to the Venus-twin input. Color-shaded areas represent different quantiles of the retrieved spectra. The solid black line represents the Venus-twin input spectrum. The gray area marks the 1$\sigma$ \lifesim{} noise level. Below the spectrum plot, we show the mean wavelength-dependent contribution of the atmospheric layers to the emission spectrum corresponding to the retrieved parameter posteriors. Darker colors indicate higher contributions. The dashed black line indicates the position of the cloud-deck assumed to simulate the input spectrum (opaque \ce{H2SO4} clouds).}
    \label{fig:corner_O_H2SO4}
\end{figure*}

\begin{figure*}
   \centering
    \includegraphics[width=0.99\textwidth]{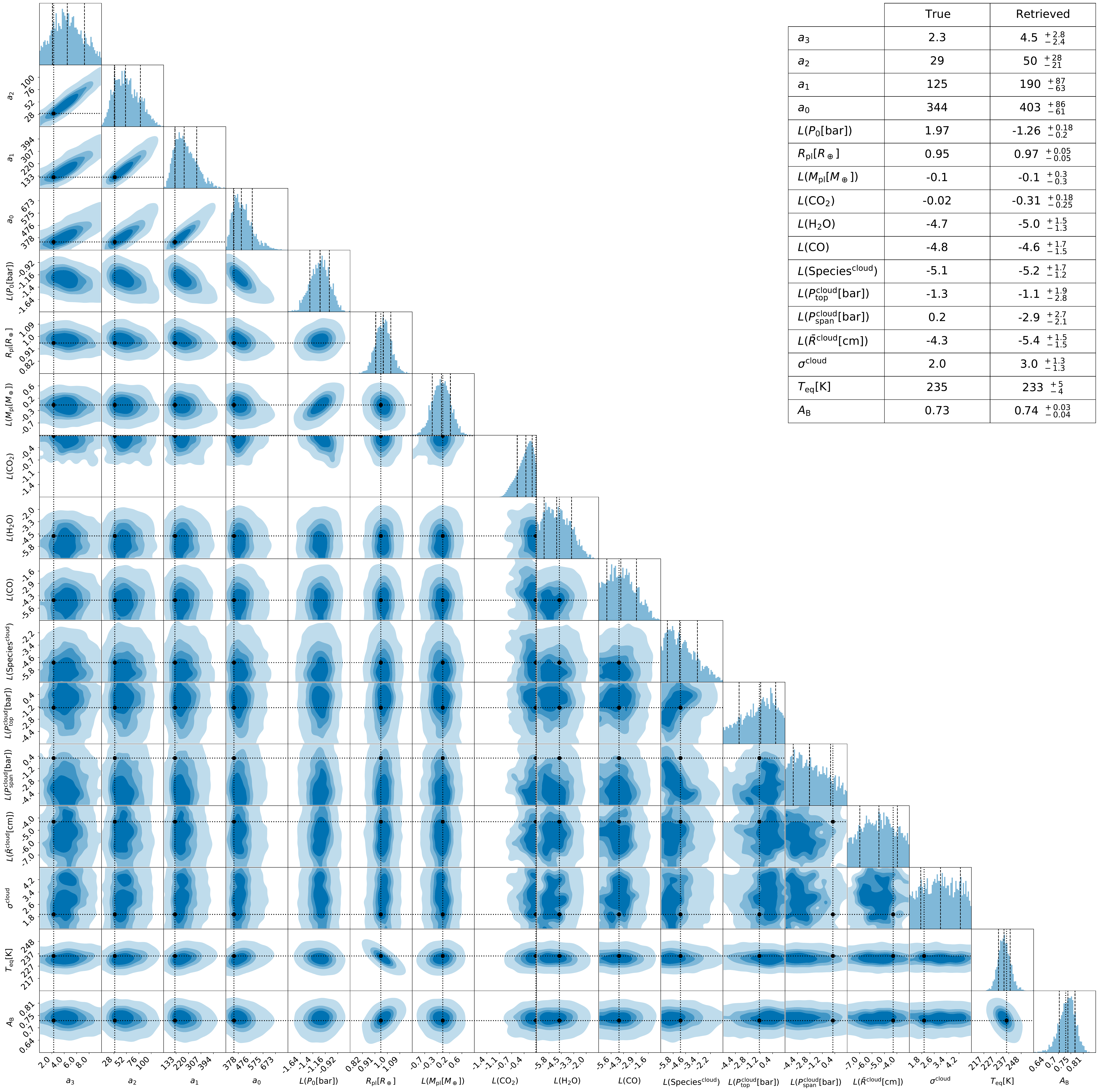}\\
    \vspace{-0.9956\textwidth}\hspace{0.05\textwidth}\includegraphics[width=0.38\textwidth]{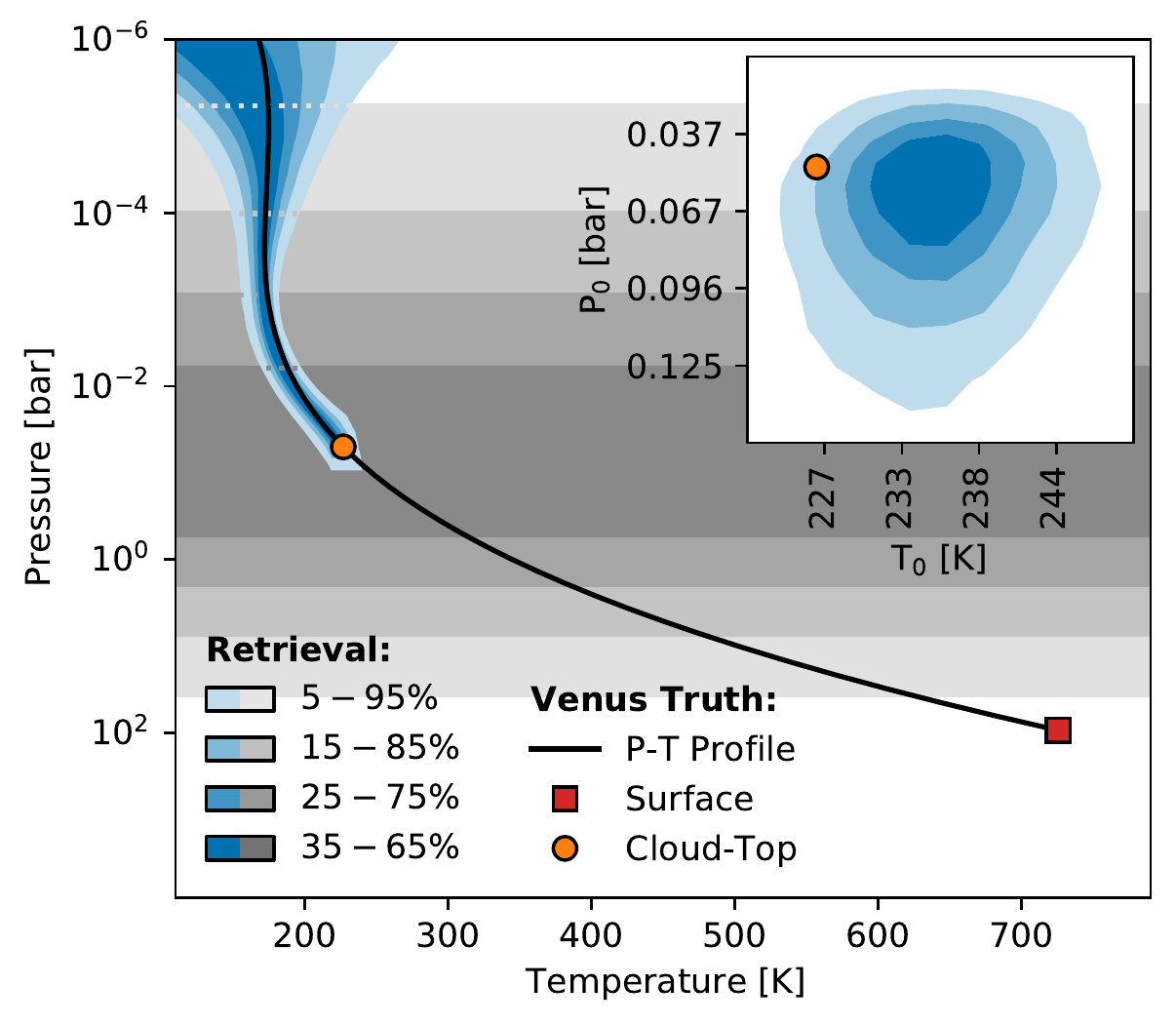}\\
    \vspace{0.06\textwidth}\hspace{0.675\textwidth}\includegraphics[width=0.325\textwidth]{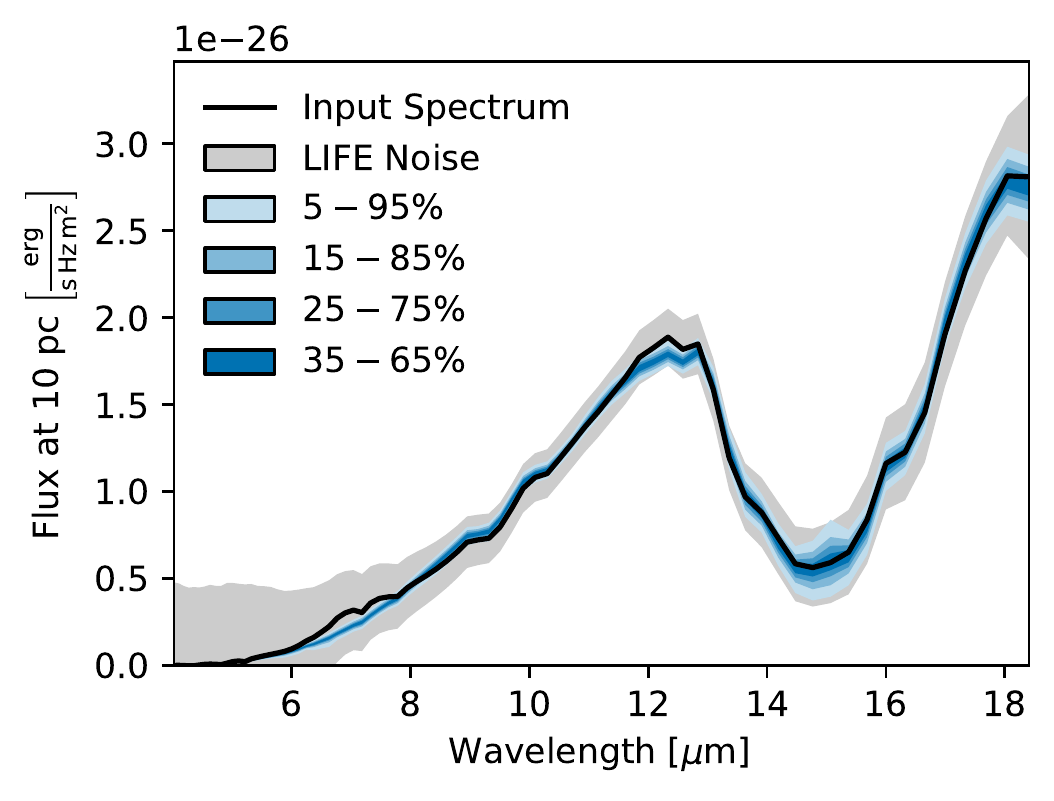}\\
    \vspace{-0.005\textwidth}\hspace{0.81\textwidth}\includegraphics[width=0.19\textwidth]{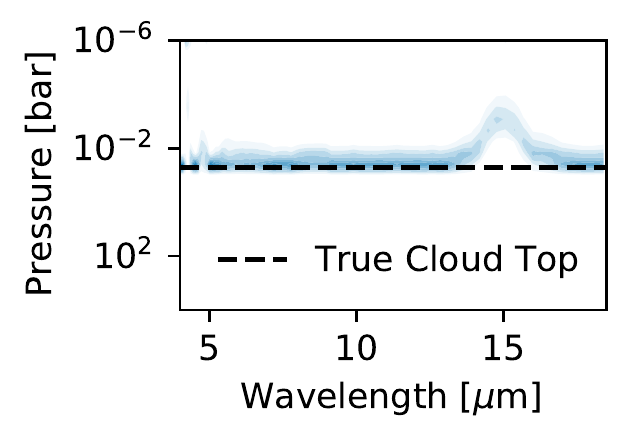}
    \vspace{0.21\textwidth}
    \caption{Same as Fig.~\ref{fig:corner_O_H2SO4}, but for the retrieval on the \mic{4-18.5}, \Rv{50}, and \lifesim{} noise \SNv{10} Venus-twin spectrum (opaque \ce{H2SO4} clouds) using the transparent \ce{H2SO4} cloud forward model (see Sect.~\ref{sec:LIFEsim_papIII_ret_res}).}
    \label{fig:corner_T_H2SO4}
\end{figure*}

\begin{figure*}
   \centering
    \includegraphics[width=0.99\textwidth]{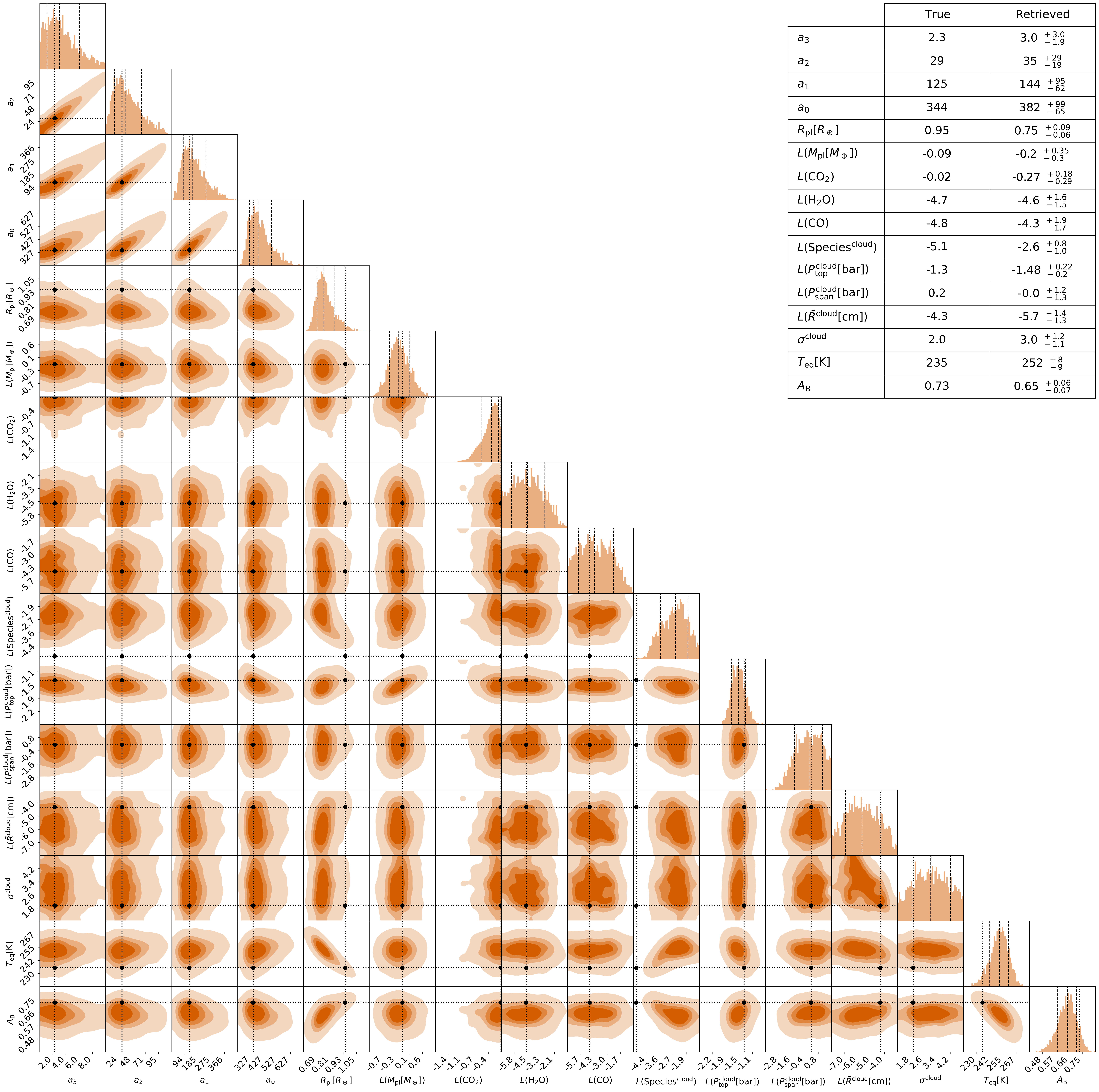}\\         \vspace{-0.9956\textwidth}\hspace{0.05\textwidth}\includegraphics[width=0.38\textwidth]{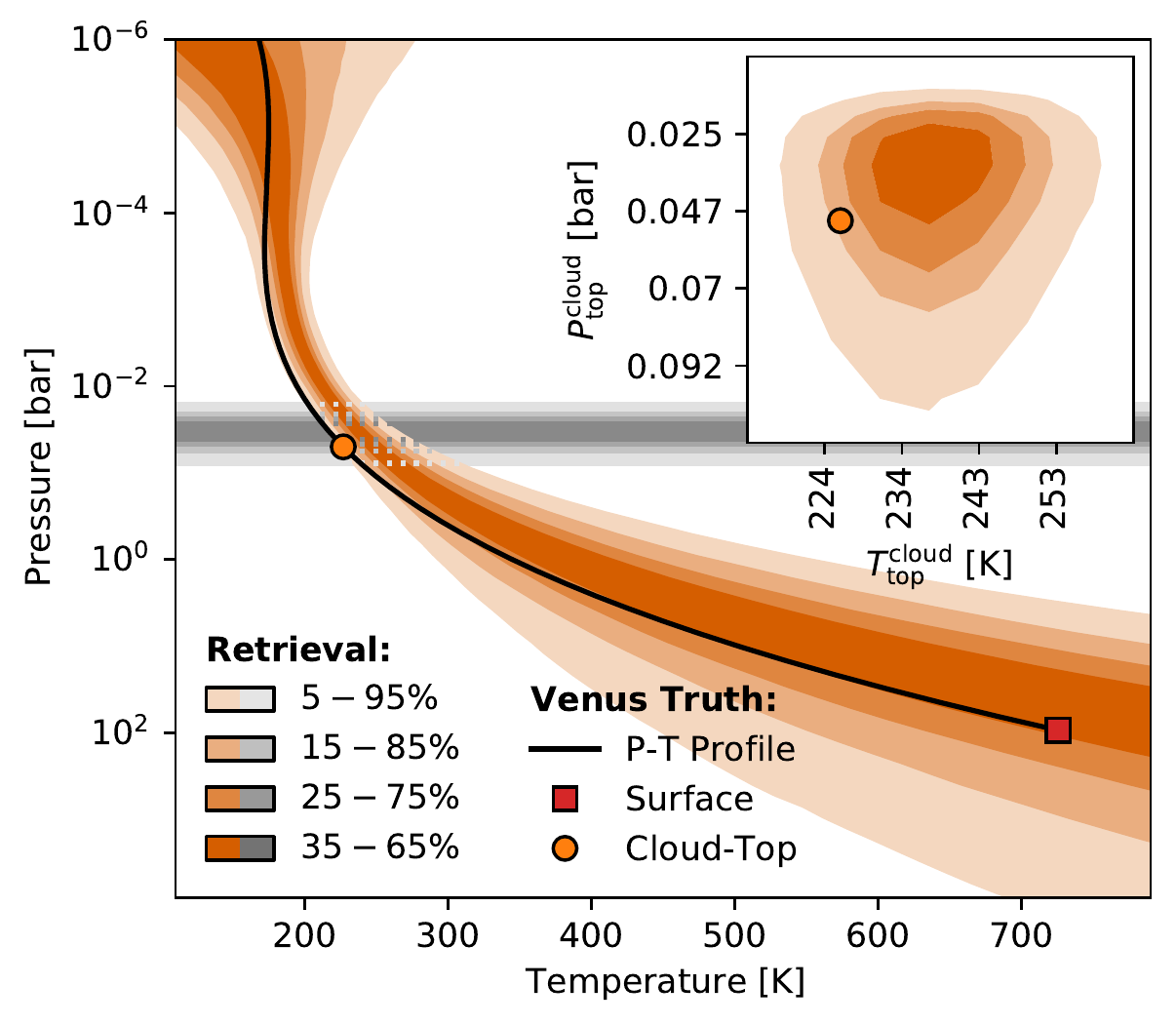}\\
    \vspace{0.06\textwidth}\hspace{0.675\textwidth}\includegraphics[width=0.325\textwidth]{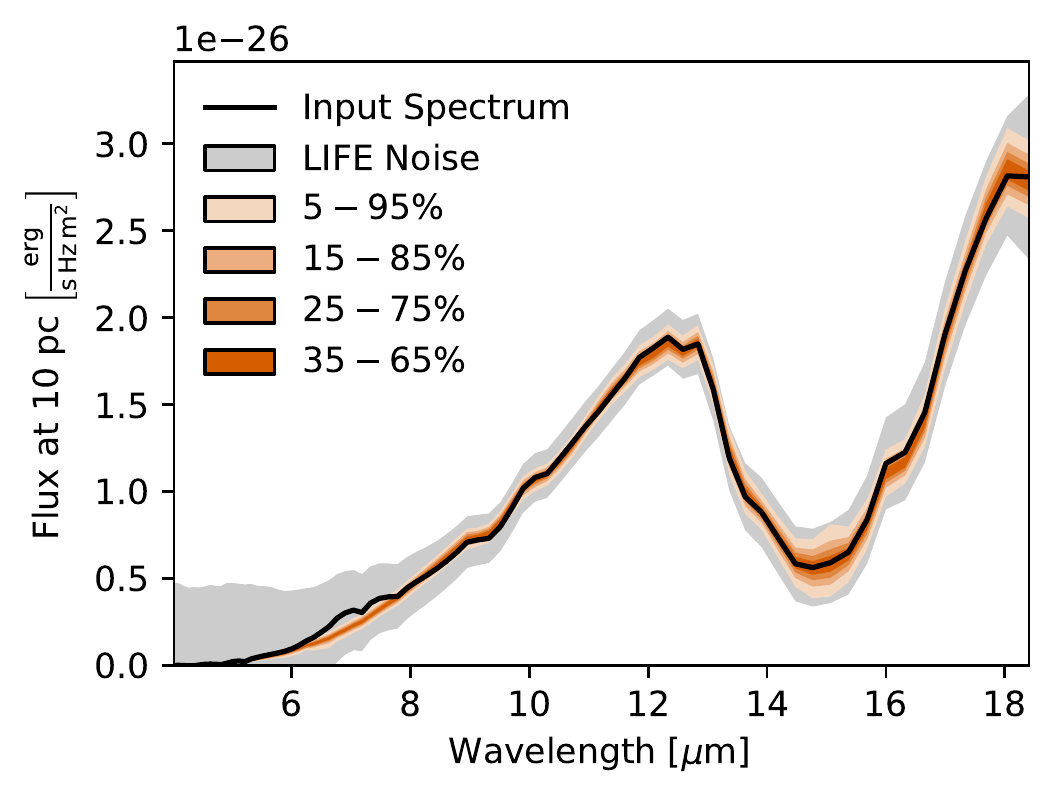}\\
    \vspace{-0.005\textwidth}\hspace{0.81\textwidth}\includegraphics[width=0.19\textwidth]{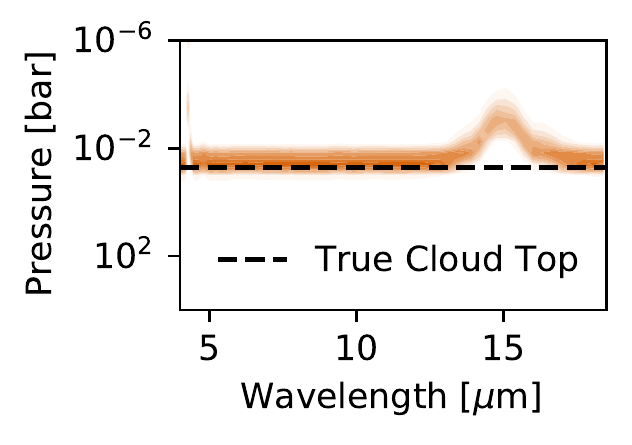}
    \vspace{0.21\textwidth}
    \caption{Same as Fig.~\ref{fig:corner_O_H2SO4}, but for the retrieval on the \mic{4-18.5}, \Rv{50}, and \lifesim{} noise \SNv{10} Venus-twin spectrum (opaque \ce{H2SO4} clouds) using the opaque \ce{H2O} cloud forward model (see Sect.~\ref{sec:LIFEsim_papIII_ret_res}).}
    \label{fig:corner_O_H2O}
\end{figure*}

\begin{figure*}
   \centering
    \includegraphics[width=0.99\textwidth]{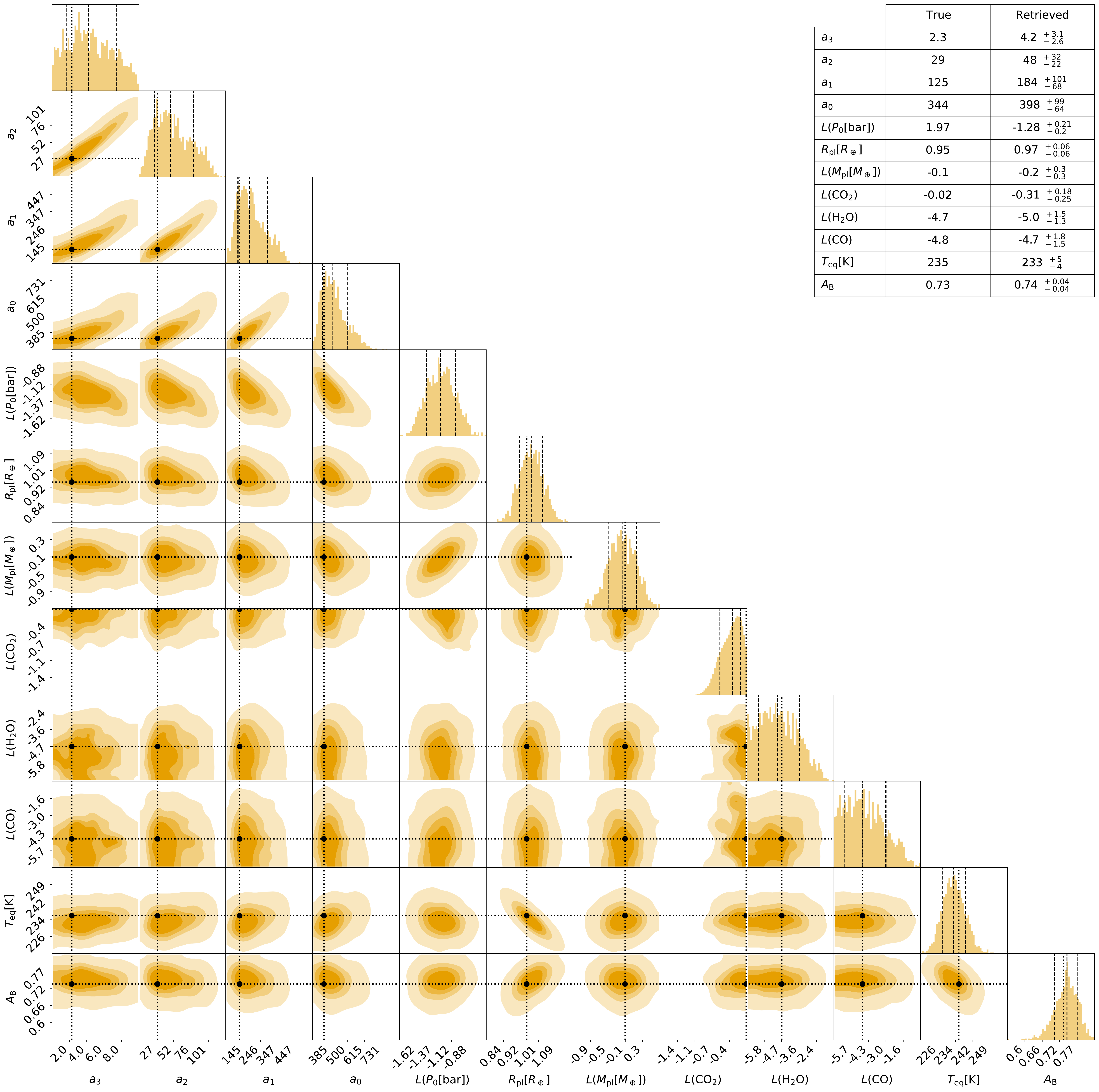}\\         
    \vspace{-0.9956\textwidth}\hspace{0.05\textwidth}\includegraphics[width=0.38\textwidth]{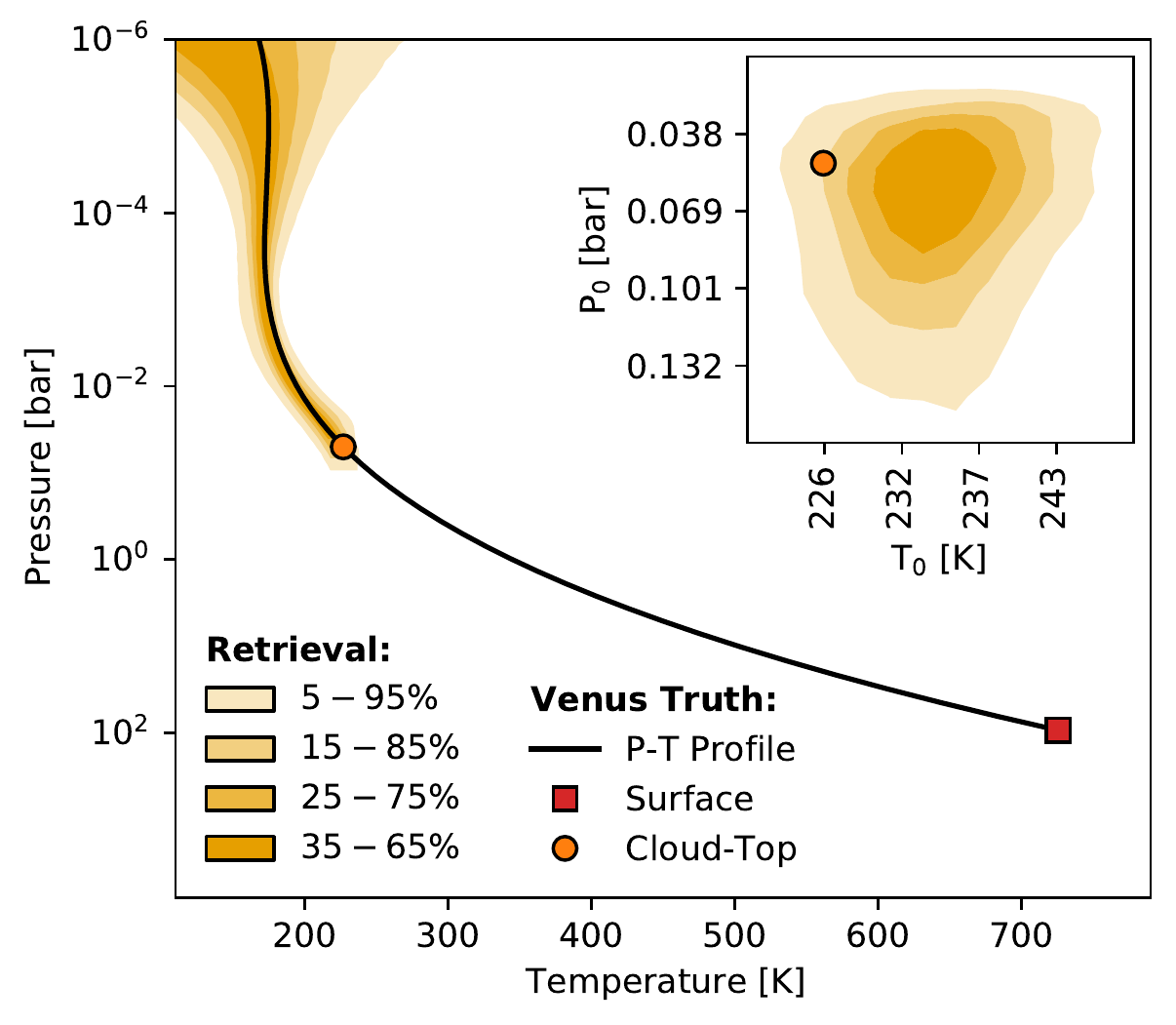}\\
    \vspace{0.03\textwidth}\hspace{0.675\textwidth}\includegraphics[width=0.325\textwidth]{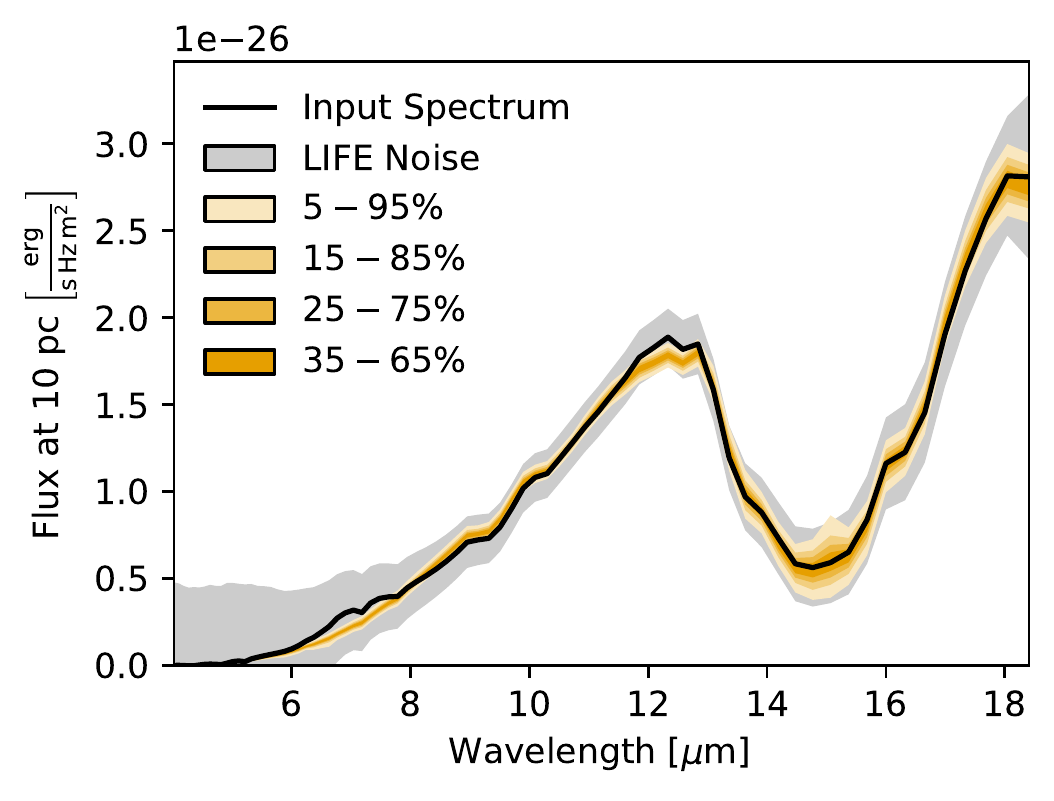}\\
    \vspace{-0.005\textwidth}\hspace{0.81\textwidth}\includegraphics[width=0.19\textwidth]{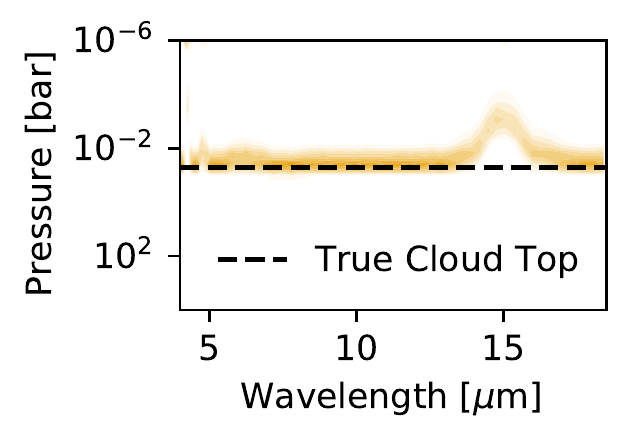}
    \vspace{0.24\textwidth}
    \caption{Same as Fig.~\ref{fig:corner_O_H2SO4}, but for the retrieval on the \mic{4-18.5}, \Rv{50}, and \lifesim{} noise \SNv{10} Venus-twin spectrum (opaque \ce{H2SO4} clouds) using the cloud-free forward model (see Sect.~\ref{sec:LIFEsim_papIII_ret_res}).}
    \label{fig:corner_CF}
\end{figure*}

Additional plots for the results from the validation retrieval (see Sect.~\ref{sec:validation}) and the retrievals of the \mic{4-18.5}, \Rv{50}, and \SNv{10} \lifesim{} spectrum (see Sect.~\ref{sec:LIFEsim_papIII_ret_res}). We plot the retrieval results from the validation retrieval in Fig.~\ref{fig:corner_full_valid}. For the \mic{4-18.5}, \Rv{50}, and \SNv{10} \lifesim{} spectrum, we plot the results for the opaque \ce{H2SO4} forward model in Fig.~\ref{fig:corner_O_H2SO4}, for the transparent \ce{H2SO4} forward model in Fig.~\ref{fig:corner_T_H2SO4}, for the opaque \ce{H2O} forward model in Fig.~\ref{fig:corner_O_H2O}, and for the cloud-free forward model in Fig.~\ref{fig:corner_CF}.
%The table in the top right corner of each figure lists the true values for each model parameter, along with the numeric values corresponding to the 16th, 50th, 84th percentiles of retrieved posteriors.

\clearpage

\section{Supplementary results from retrievals on higher quality input spectra}\label{app:extended_retrievals}

Here we plot the posteriors for all parameters found in the retrieval analysis for higher quality input spectra discussed in Sect.~\ref{sec:high_LIFEsim_res}. We include the \Teq{} and \Ab{} distributions, which we calculated from the posteriors following the method introduced in Appendix~\ref{sec:albedo_calc}. Further, we add the results for the intermediate \SNv{15} noise levels. In Figs.~\ref{fig:full_better_input_post_3-20_1} and \ref{fig:full_better_input_post_3-20_2}, we provide the retrieval results for the \mic{3-20} input spectra. In Figs.~\ref{fig:full_better_input_post_4-185_1} and \ref{fig:full_better_input_post_4-185_2}, we display the results obtained for the \mic{4-18.5} input spectra.

%{Cool! We should however figure out the order of the quality for the intermediate cases (for the appendix plots) to make sure it is at increasing quality (or if it should be somehow rearranged for the SNR 15 cases)}

\begin{figure*}
   \centering
    \includegraphics[width=0.0129\textwidth]{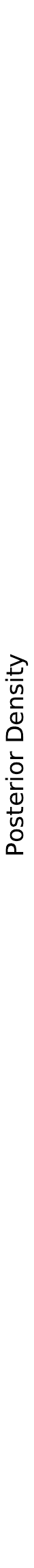}
    \includegraphics[width=0.193\textwidth]{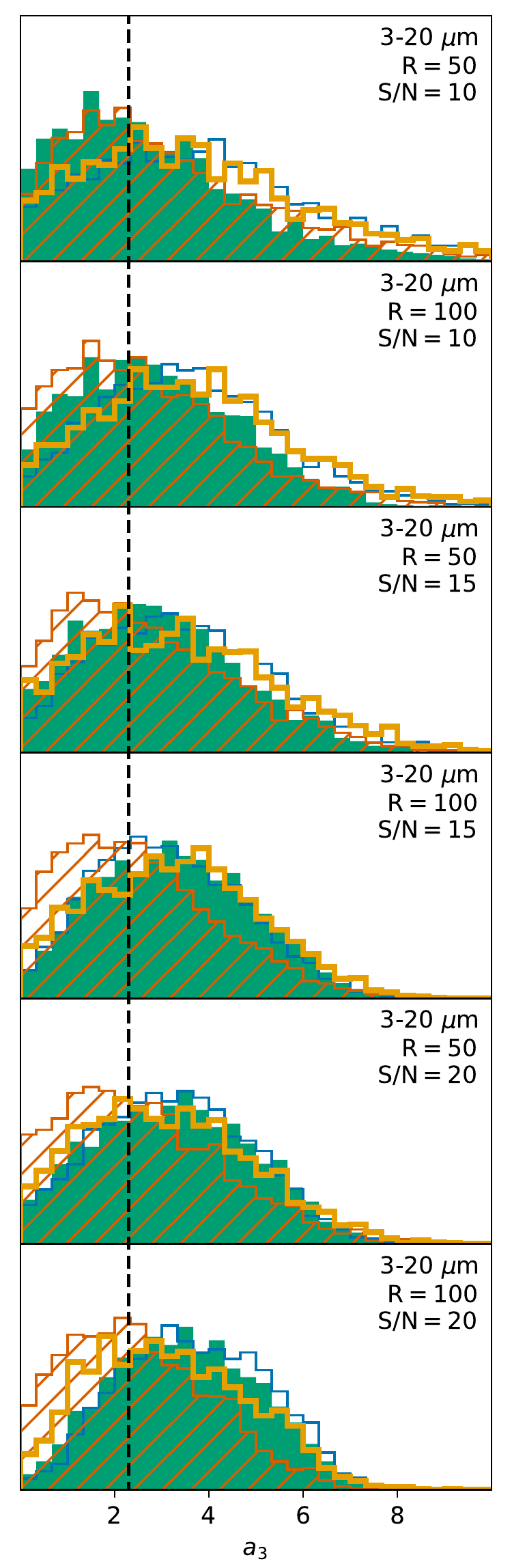}
    \includegraphics[width=0.193\textwidth]{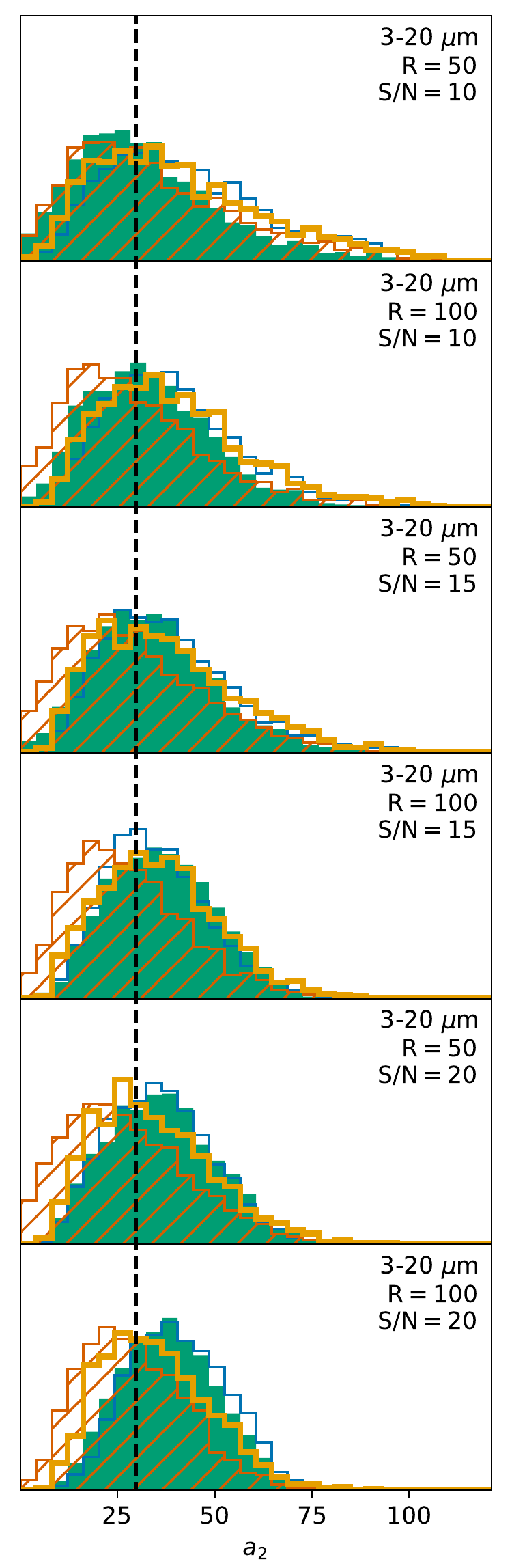}
    \includegraphics[width=0.193\textwidth]{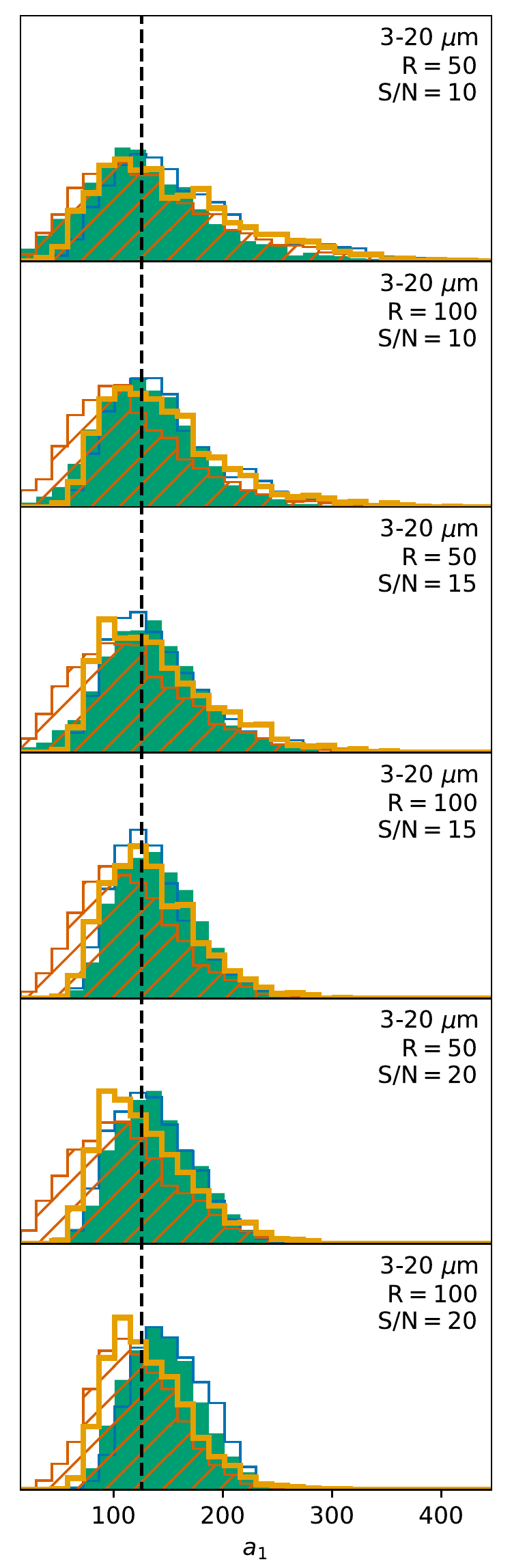}
    \includegraphics[width=0.193\textwidth]{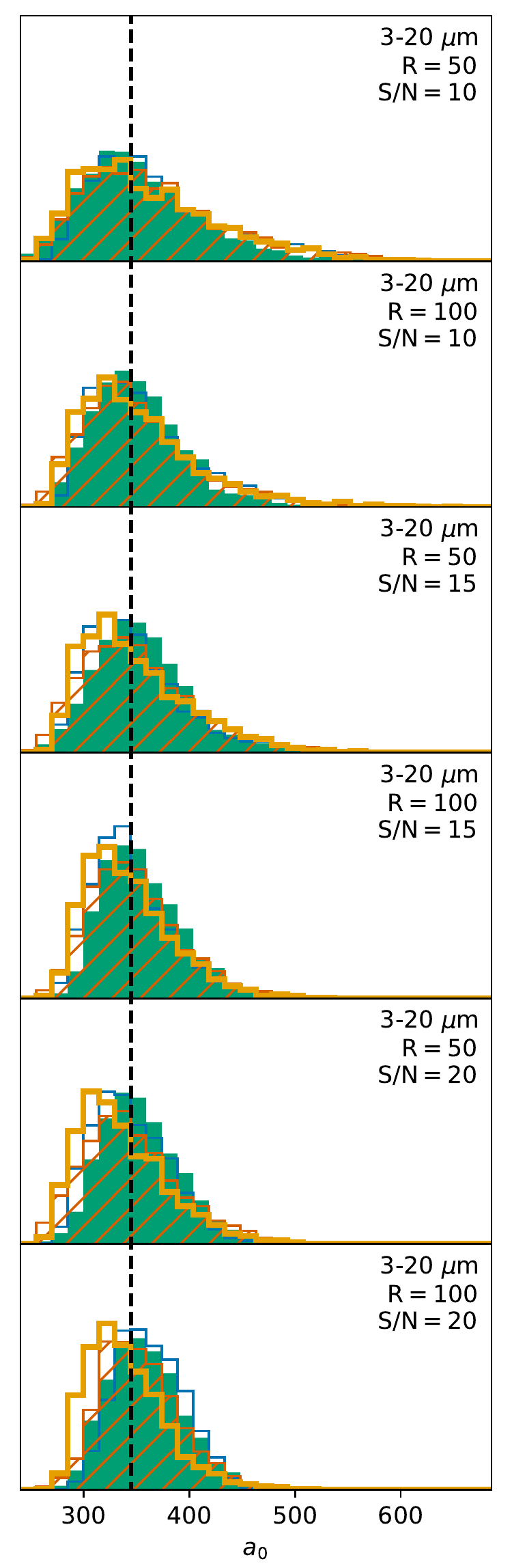}\\
    
    \includegraphics[width=0.0129\textwidth]{Figures/Extended_Retrievals(3-20)/Posterior_ylabel.pdf}
    \includegraphics[width=0.193\textwidth]{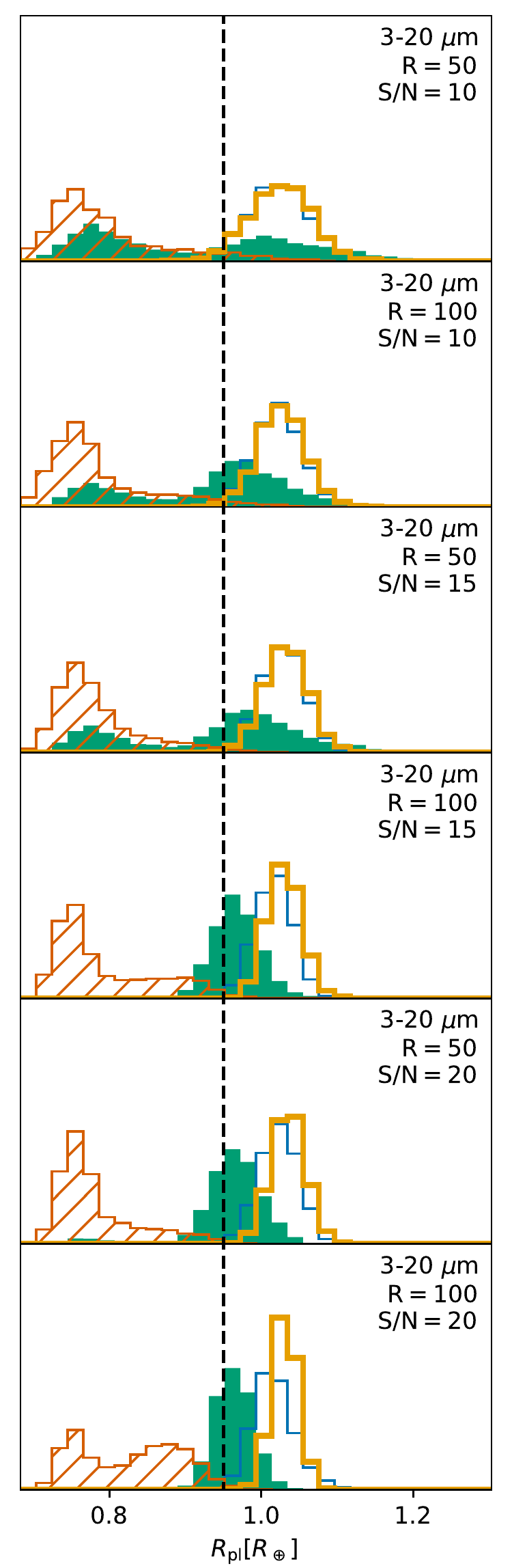}
    \includegraphics[width=0.193\textwidth]{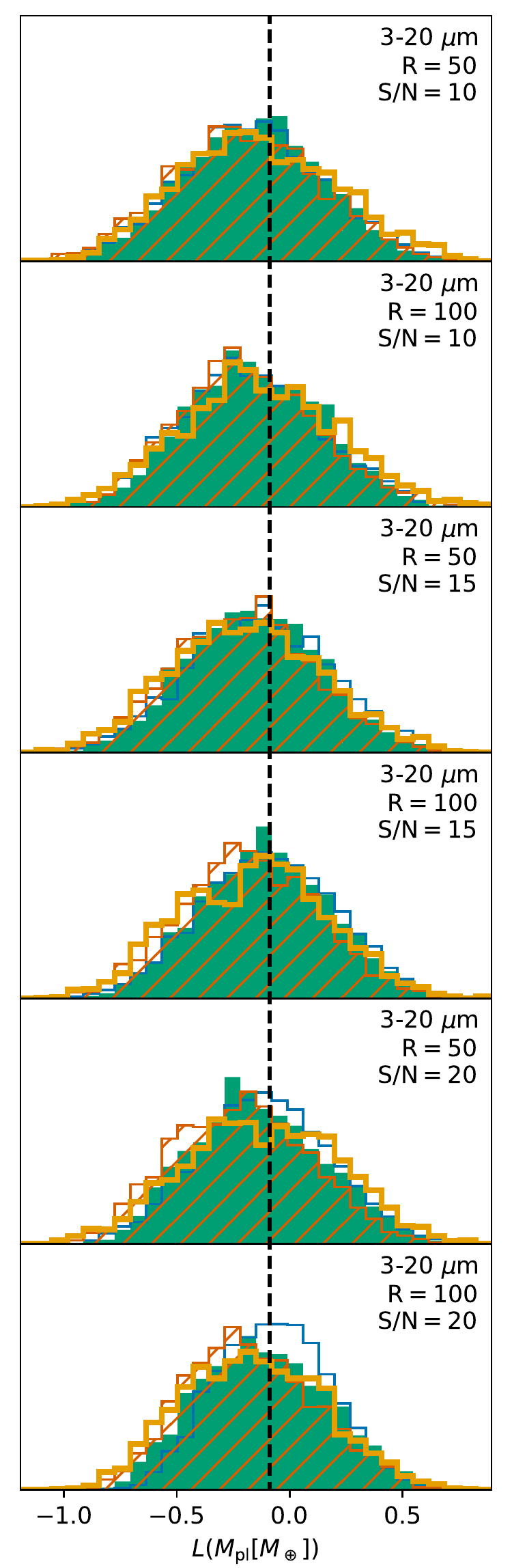}
    \includegraphics[width=0.193\textwidth]{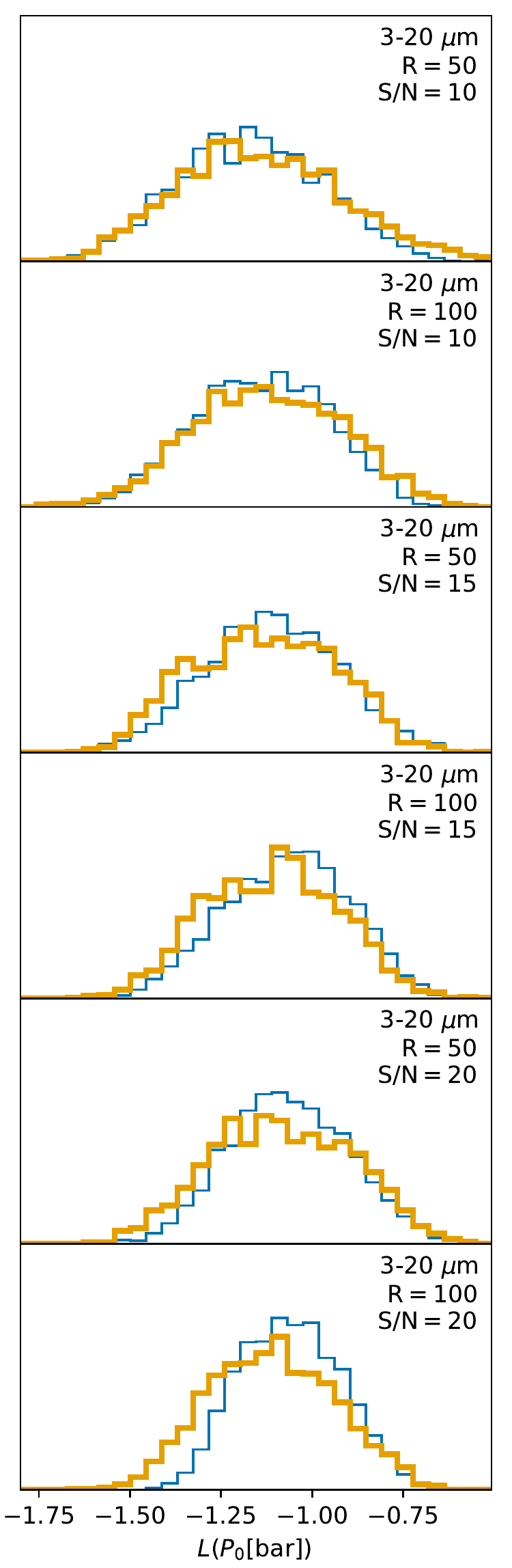}
    \includegraphics[width=0.193\textwidth]{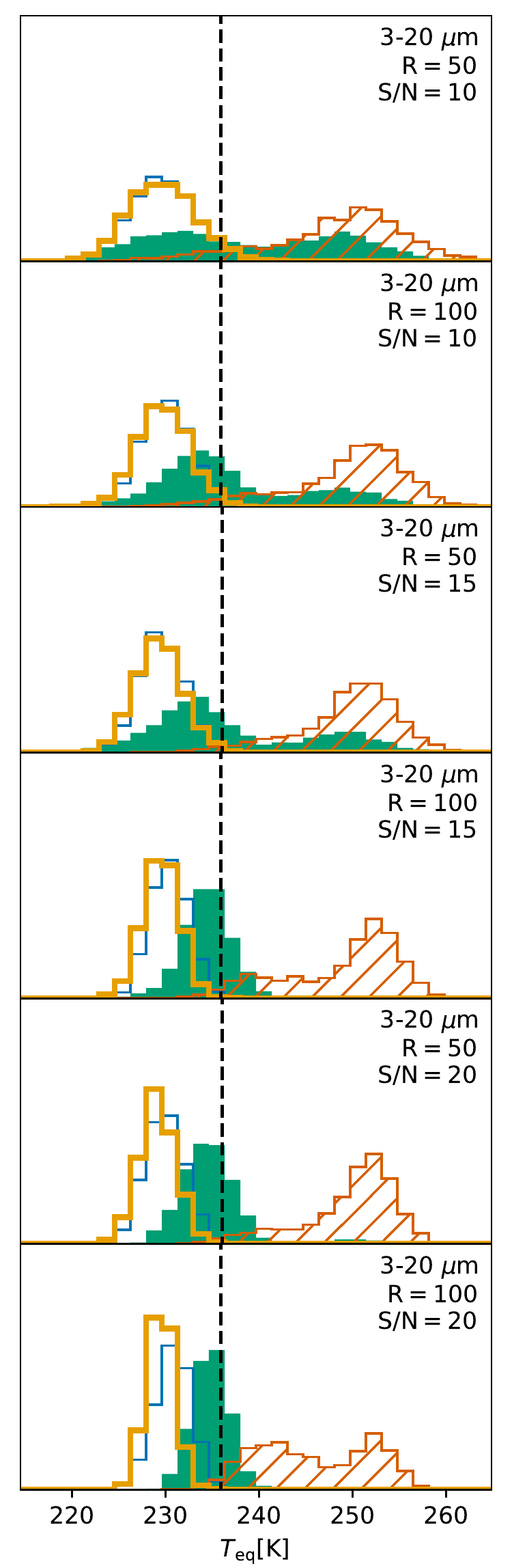}
    \includegraphics[width=0.193\textwidth]{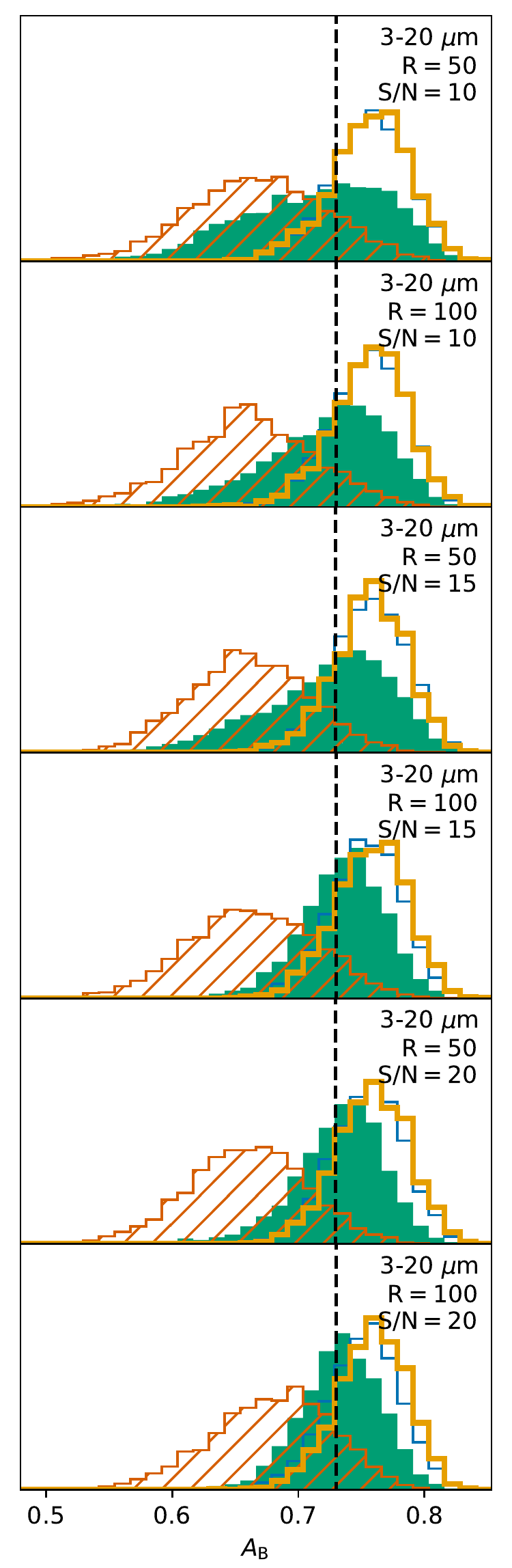}\\

    \includegraphics[width=0.523\textwidth]{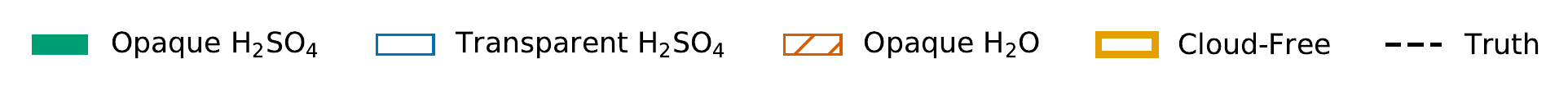}
    \caption{Model parameter posteriors for increased quality \mic{3-20} Venus-twin spectra (opaque \ce{H2SO4} clouds). Here, $L(\cdot)$ abbreviates $\lgrt{\cdot}$. Dashed black lines indicate the true values (see Table~\ref{table:True_Values}). Solid green area $-$ opaque \ce{H2SO4} clouds; thin-blue outlined area $-$ transparent \ce{H2SO4} clouds; orange-hatched area $-$ opaque \ce{H2O} clouds; thick-yellow outlined area $-$ cloud-free. Rows (top to bottom) represent different input qualities: \Rv{50}, \SNv{10}; \Rv{100}, \SNv{10}; \Rv{50}, \SNv{15}; \Rv{100}, \SNv{15}; \Rv{50}, \SNv{20}; \Rv{100}, \SNv{20}. Continuation in Fig.~\ref{fig:full_better_input_post_3-20_2}.}
          \label{fig:full_better_input_post_3-20_1}%
\end{figure*}

\begin{figure*}
   \centering
    \includegraphics[width=0.0129\textwidth]{Figures/Extended_Retrievals(3-20)/Posterior_ylabel.pdf}
    \includegraphics[width=0.193\textwidth]{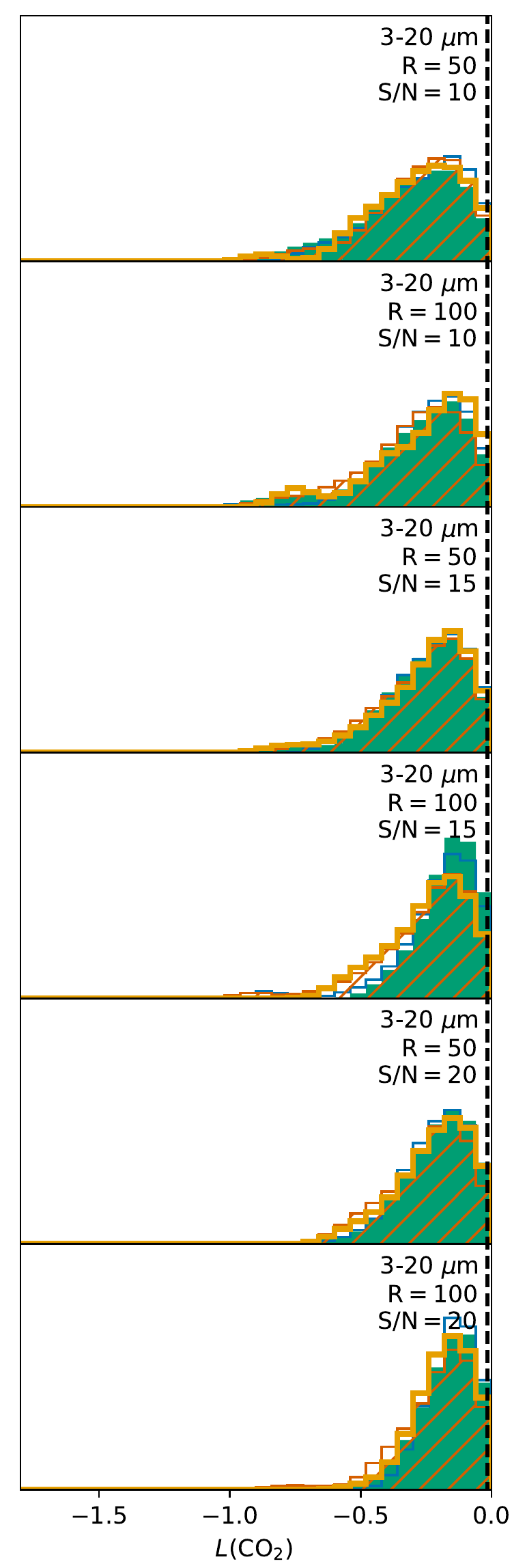}
    \includegraphics[width=0.193\textwidth]{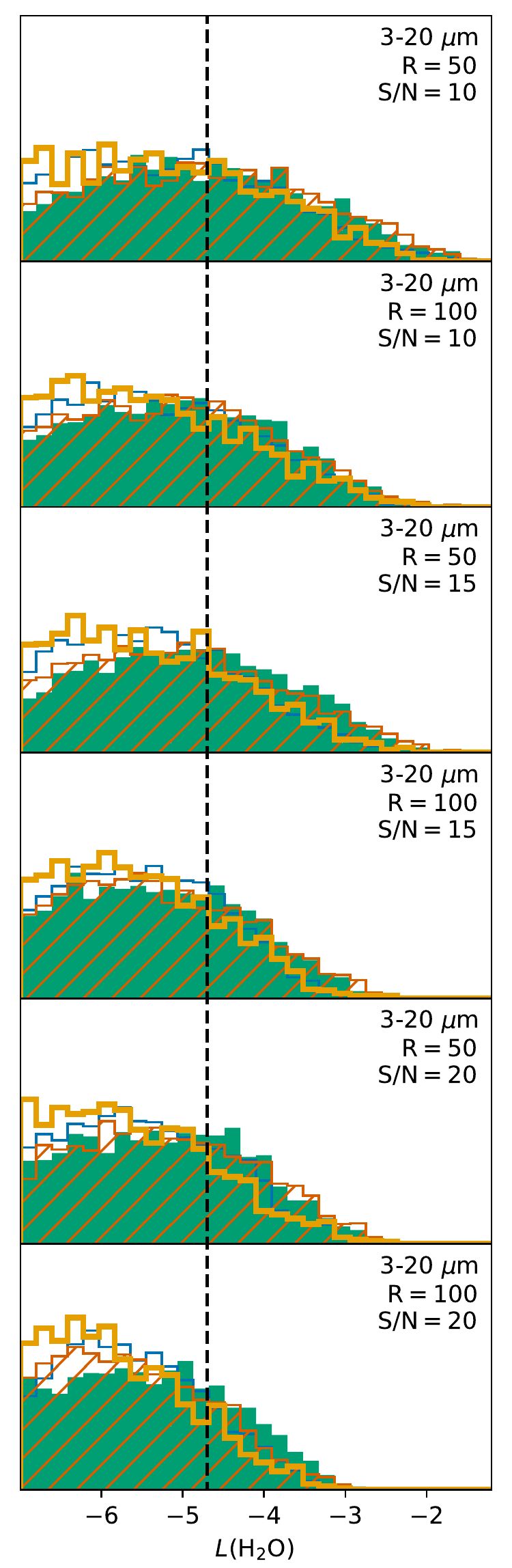}
    \includegraphics[width=0.193\textwidth]{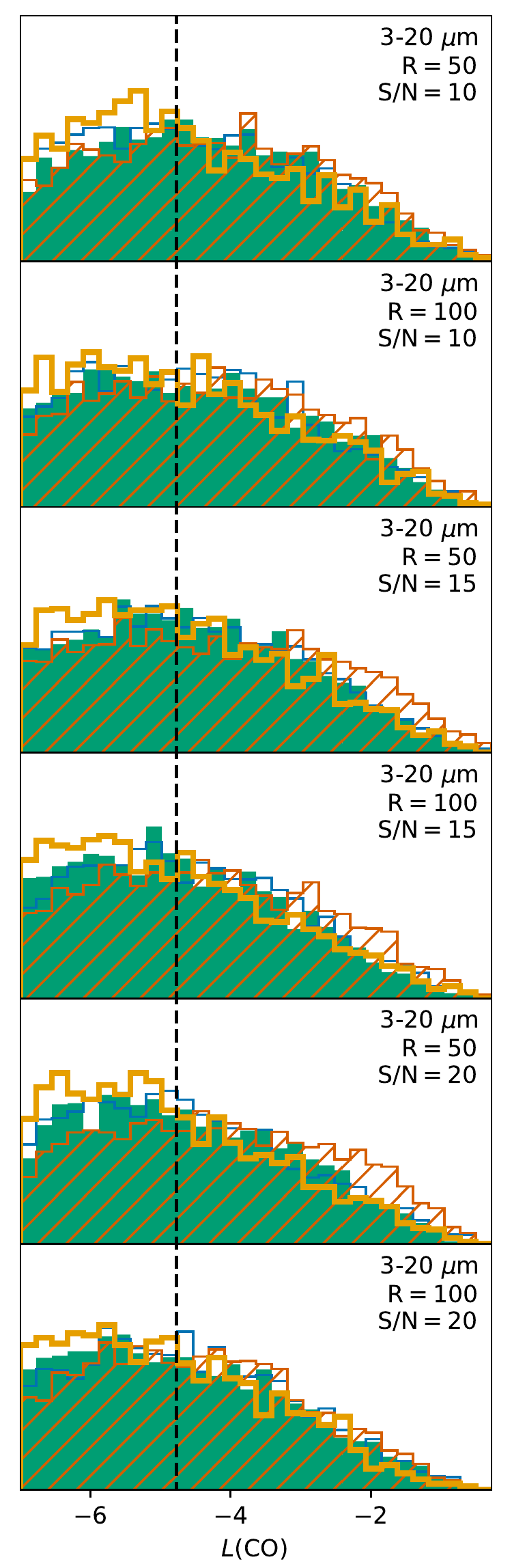}
    \includegraphics[width=0.193\textwidth]{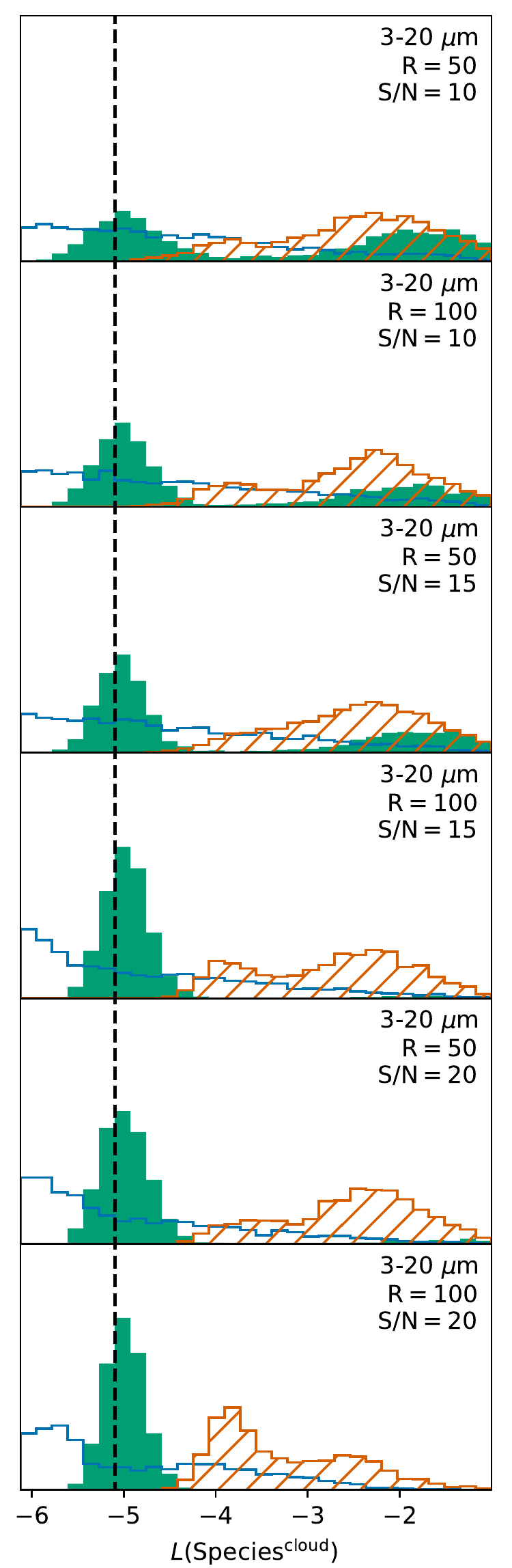}\\
    
    \includegraphics[width=0.0129\textwidth]{Figures/Extended_Retrievals(3-20)/Posterior_ylabel.pdf}
    \includegraphics[width=0.193\textwidth]{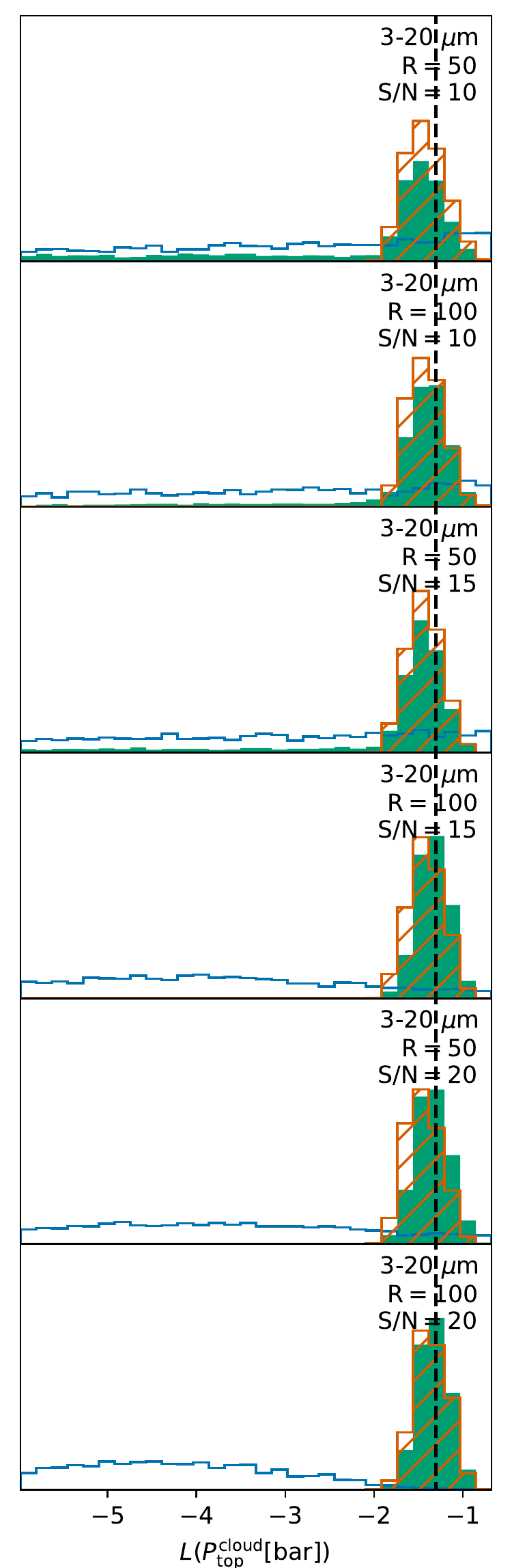}
    \includegraphics[width=0.193\textwidth]{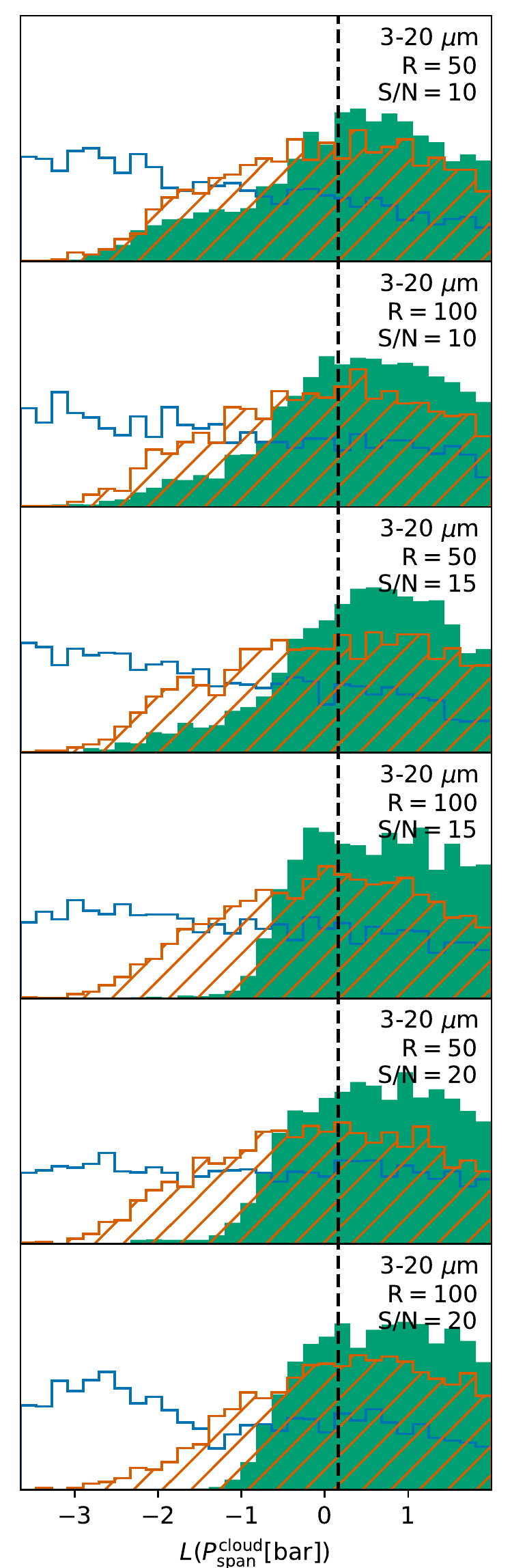}
    \includegraphics[width=0.193\textwidth]{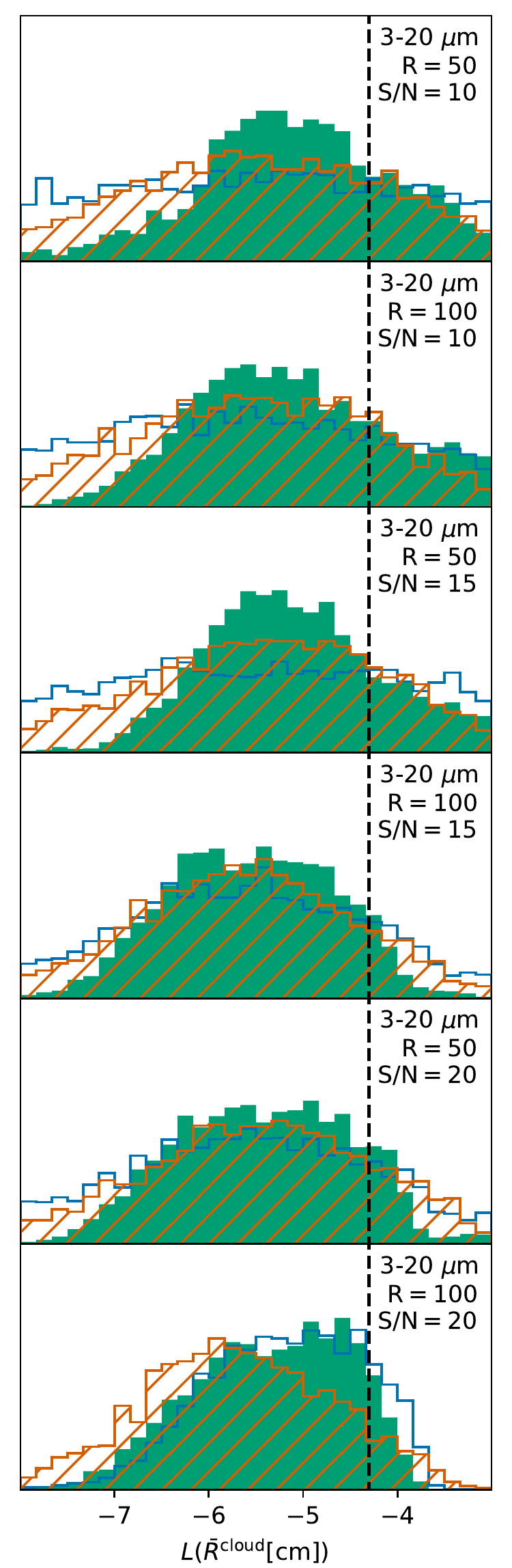}
    \includegraphics[width=0.193\textwidth]{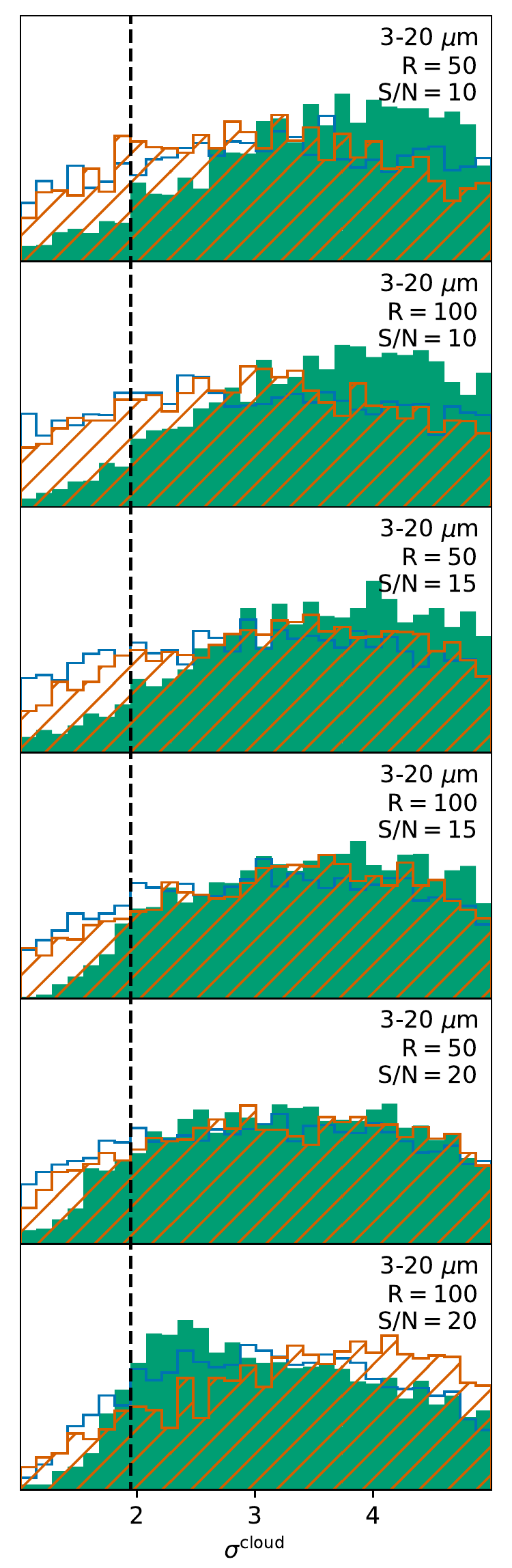}\\

    \includegraphics[width=0.523\textwidth]{Figures/Extended_Retrievals(3-20)/A_B_Posterior_Legend.pdf}
    \caption{Continuation of Fig.~\ref{fig:full_better_input_post_3-20_1}. For $\mathrm{Species^{cloud}}$, the true value is the \ce{H2SO4} mass fraction in the Venus-twin model.}
          \label{fig:full_better_input_post_3-20_2}%
\end{figure*}

\begin{figure*}
   \centering
    \includegraphics[width=0.0129\textwidth]{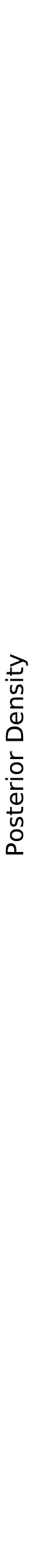}
    \includegraphics[width=0.193\textwidth]{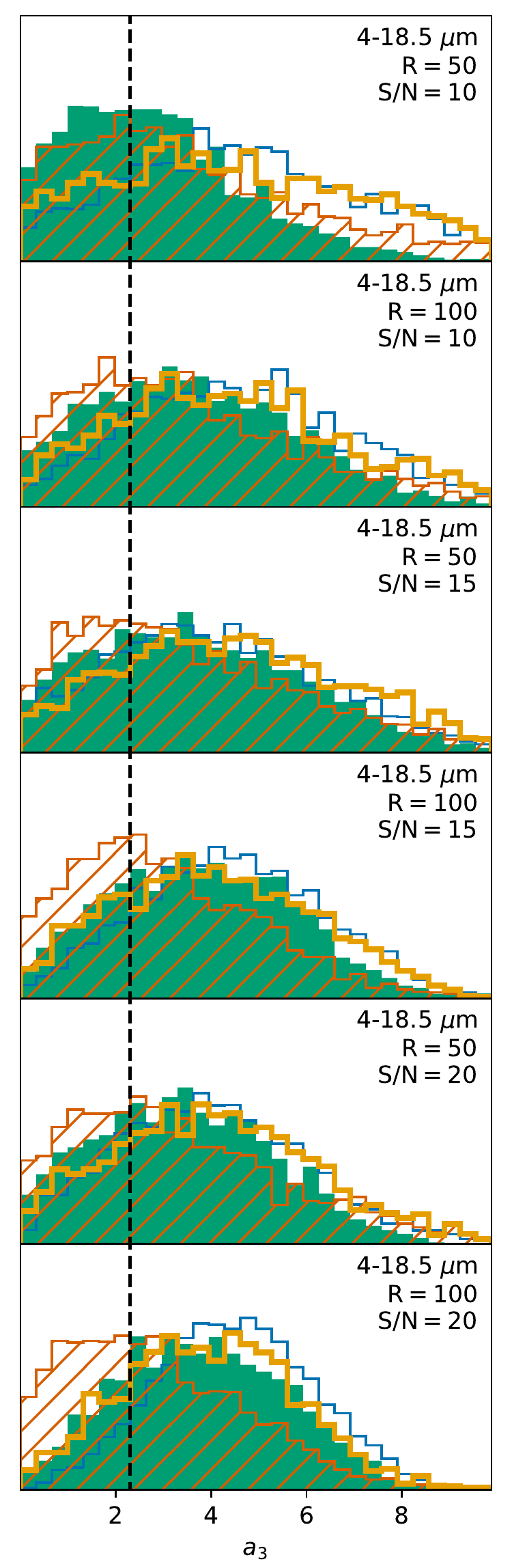}
    \includegraphics[width=0.193\textwidth]{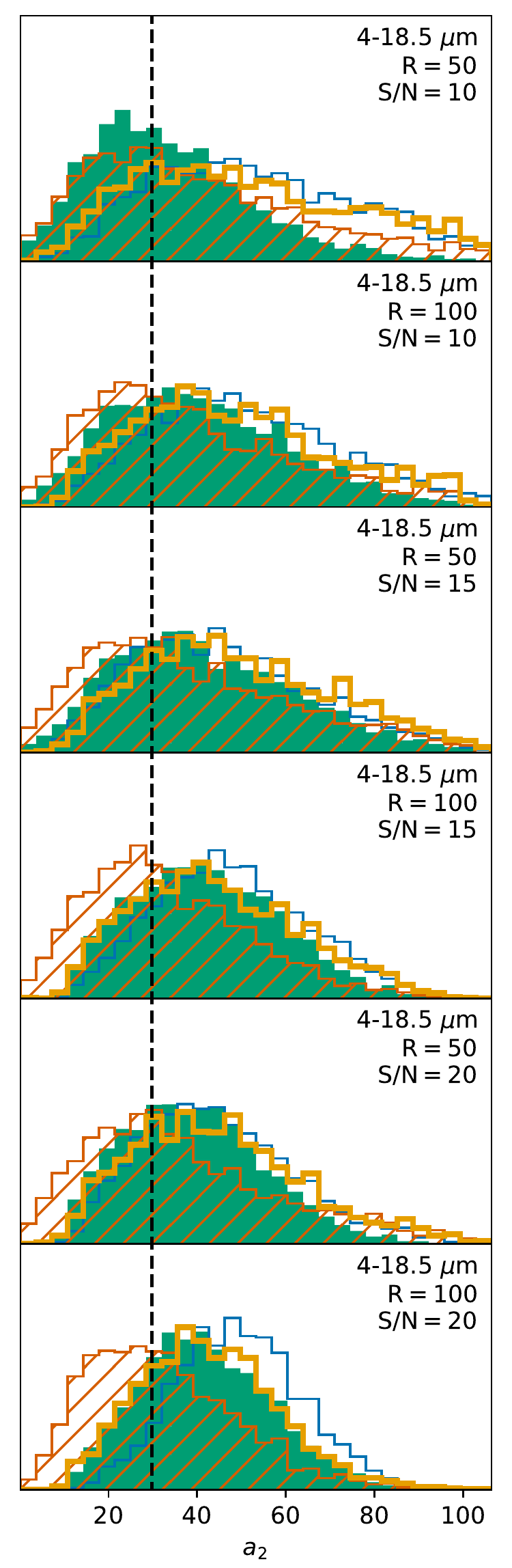}
    \includegraphics[width=0.193\textwidth]{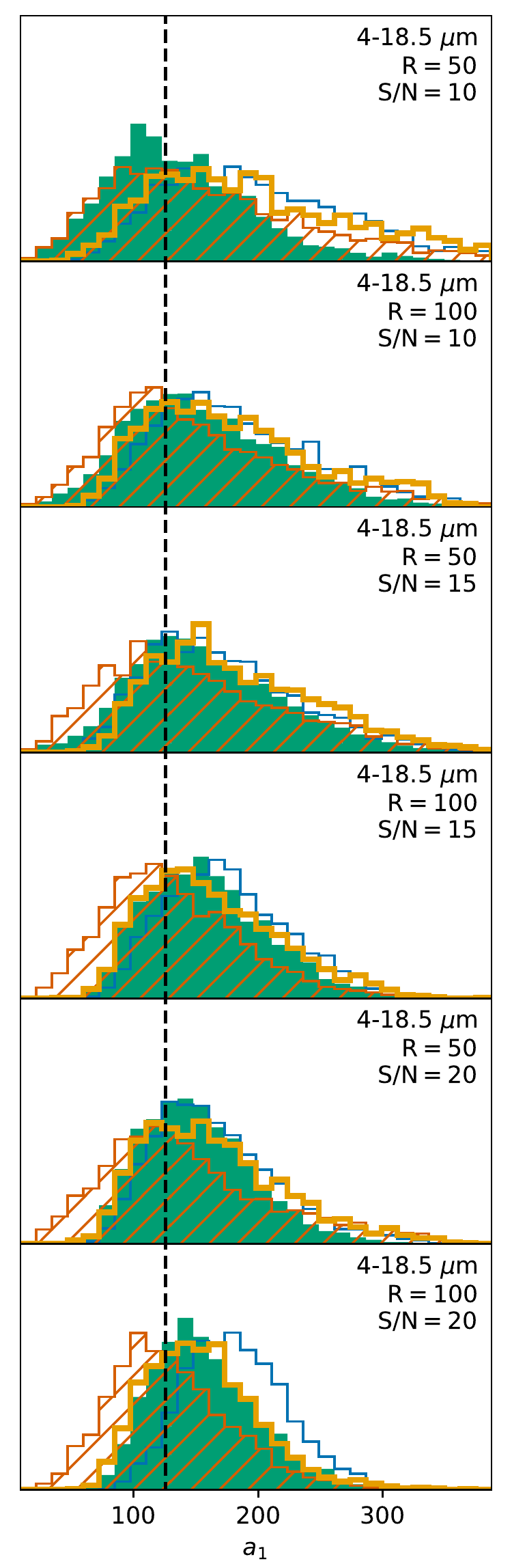}
    \includegraphics[width=0.193\textwidth]{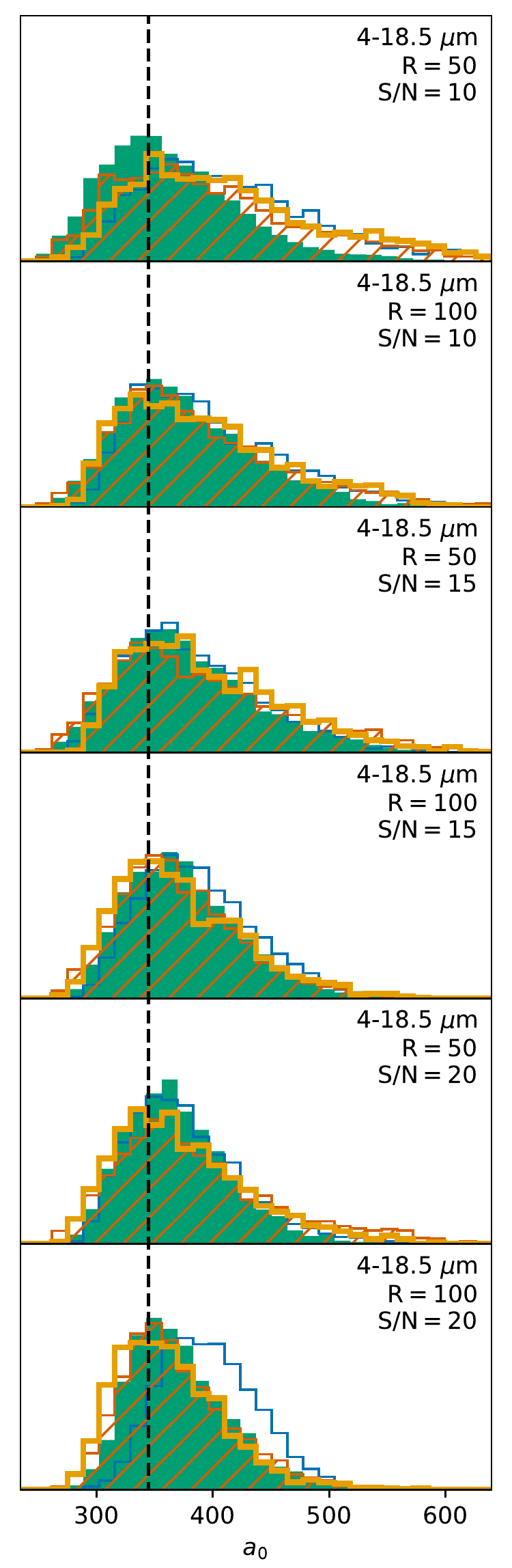}\\
    
    \includegraphics[width=0.0129\textwidth]{Figures/Extended_Retrievals(4-185)/Posterior_ylabel.pdf}
    \includegraphics[width=0.193\textwidth]{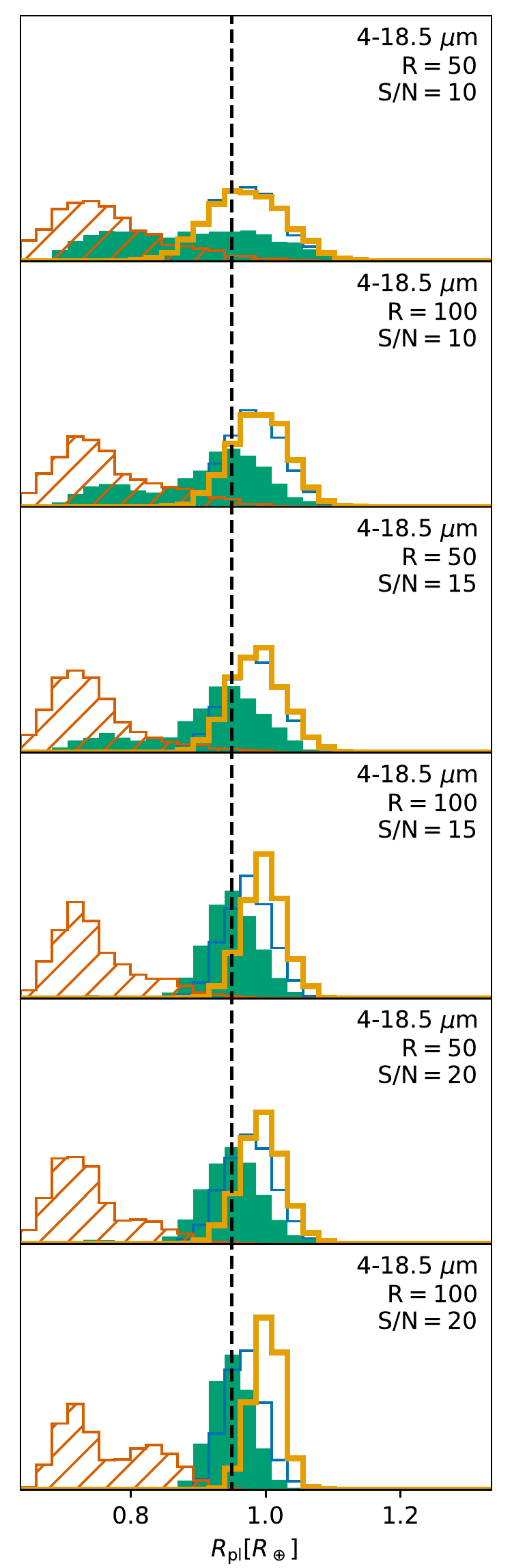}
    \includegraphics[width=0.193\textwidth]{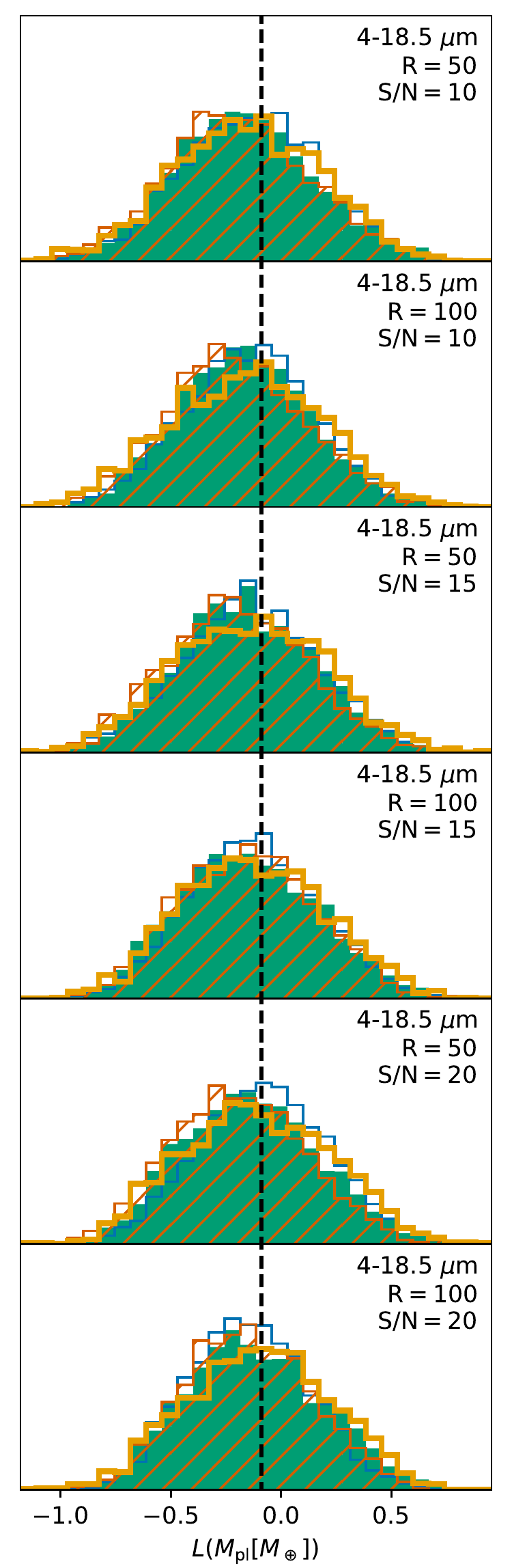}
    \includegraphics[width=0.193\textwidth]{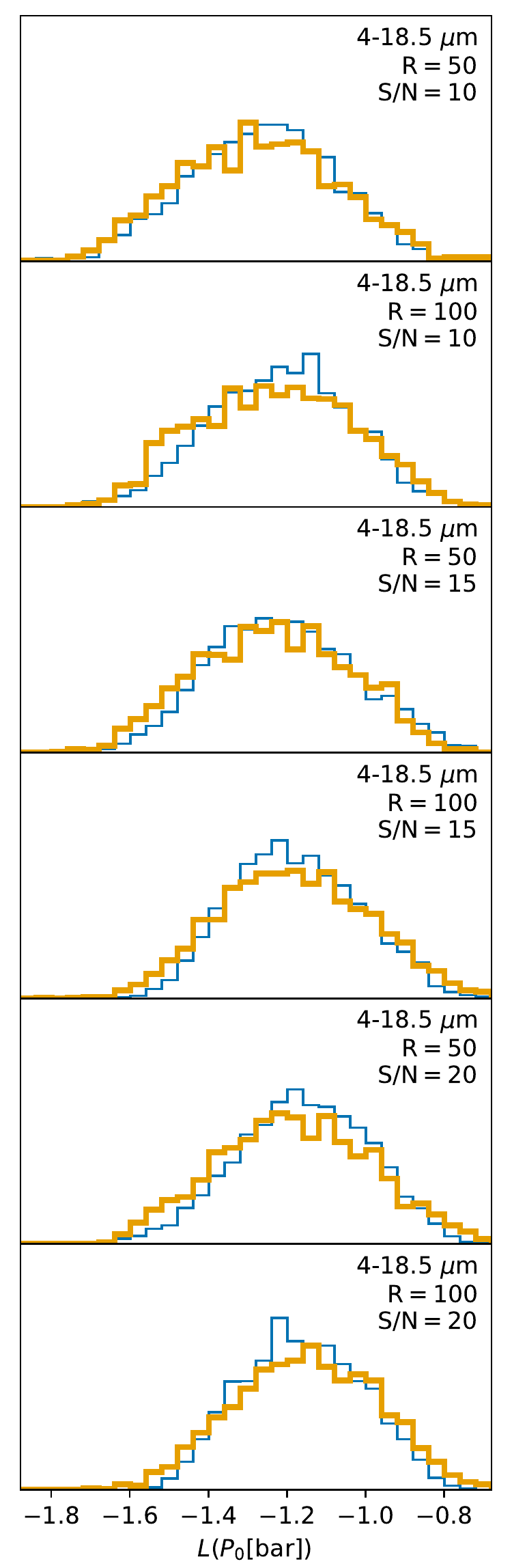}
    \includegraphics[width=0.193\textwidth]{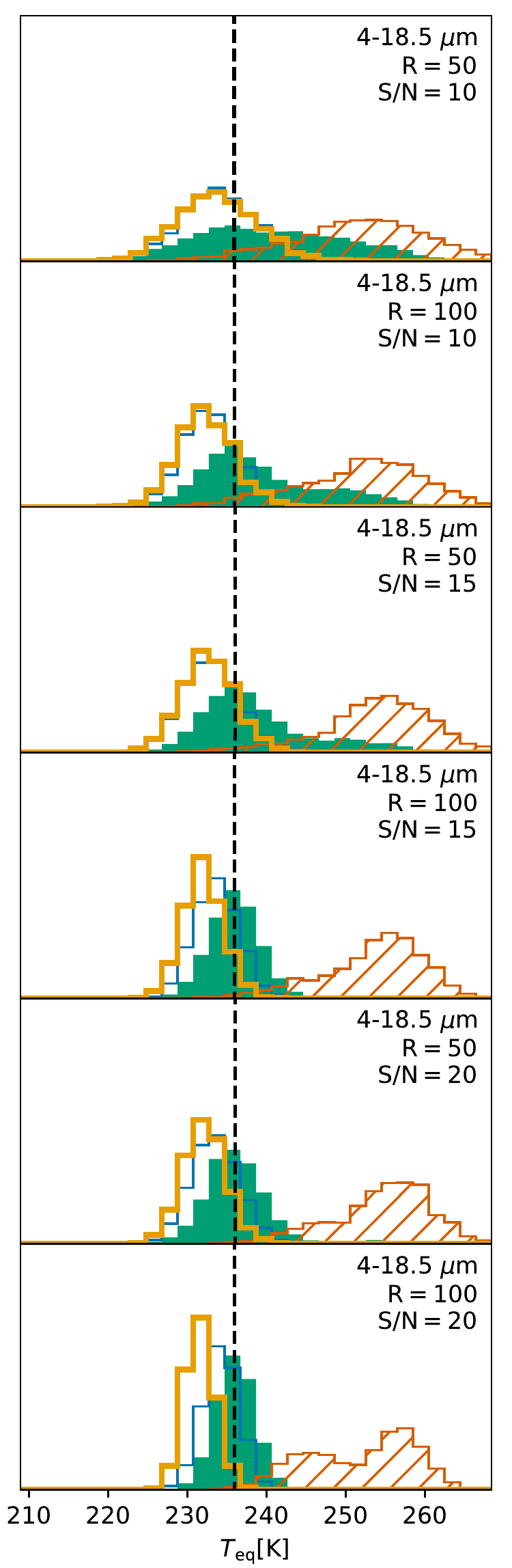}
    \includegraphics[width=0.193\textwidth]{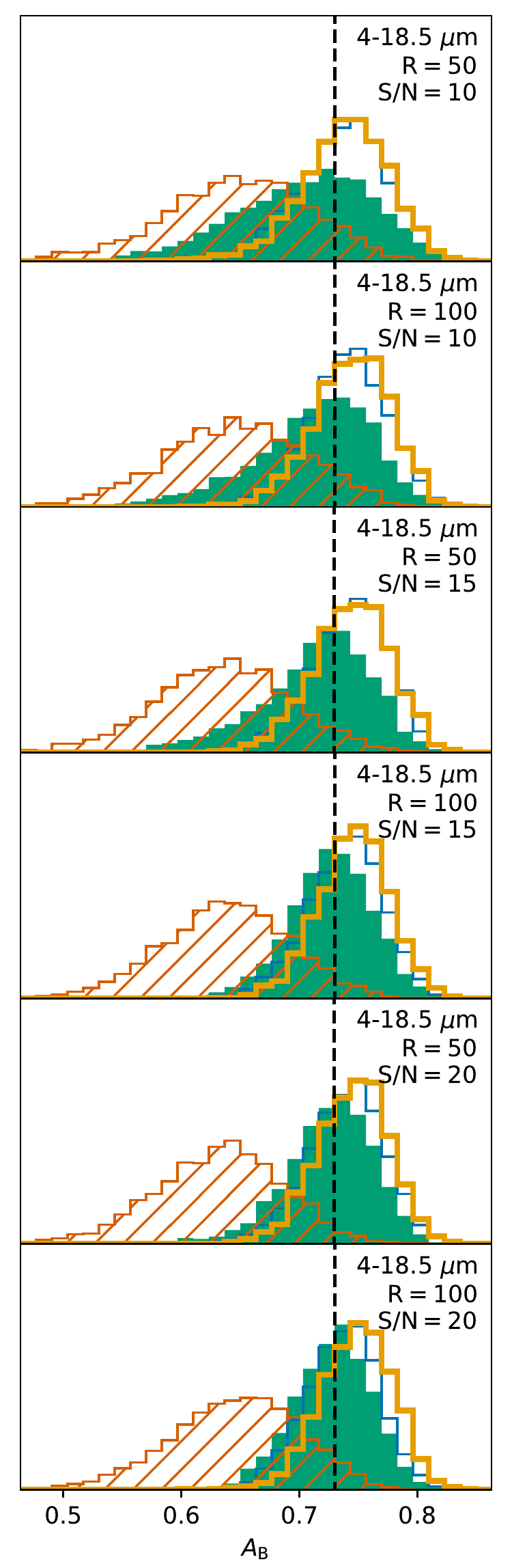}\\

    \includegraphics[width=0.523\textwidth]{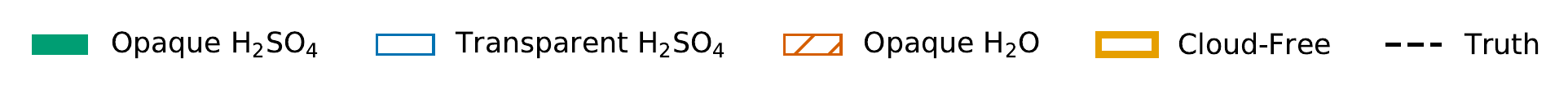}
    \caption{Model parameter posteriors for increased quality \mic{4-18.5} Venus-twin spectra (opaque \ce{H2SO4} clouds). Here, $L(\cdot)$ abbreviates $\lgrt{\cdot}$. Dashed black lines indicate the true values (see Table~\ref{table:True_Values}). Solid green area $-$ opaque \ce{H2SO4} clouds; thin-blue outlined area $-$ transparent \ce{H2SO4} clouds; orange-hatched area $-$ opaque \ce{H2O} clouds; thick-yellow outlined area $-$ cloud-free. Rows (top to bottom) represent different input qualities: \Rv{50}, \SNv{10}; \Rv{100}, \SNv{10}; \Rv{50}, \SNv{15}; \Rv{100}, \SNv{15}; \Rv{50}, \SNv{20}; \Rv{100}, \SNv{20}. Continuation in Fig.~\ref{fig:full_better_input_post_4-185_2}.}
          \label{fig:full_better_input_post_4-185_1}%
\end{figure*}

\begin{figure*}
   \centering
    \includegraphics[width=0.0129\textwidth]{Figures/Extended_Retrievals(4-185)/Posterior_ylabel.pdf}
    \includegraphics[width=0.193\textwidth]{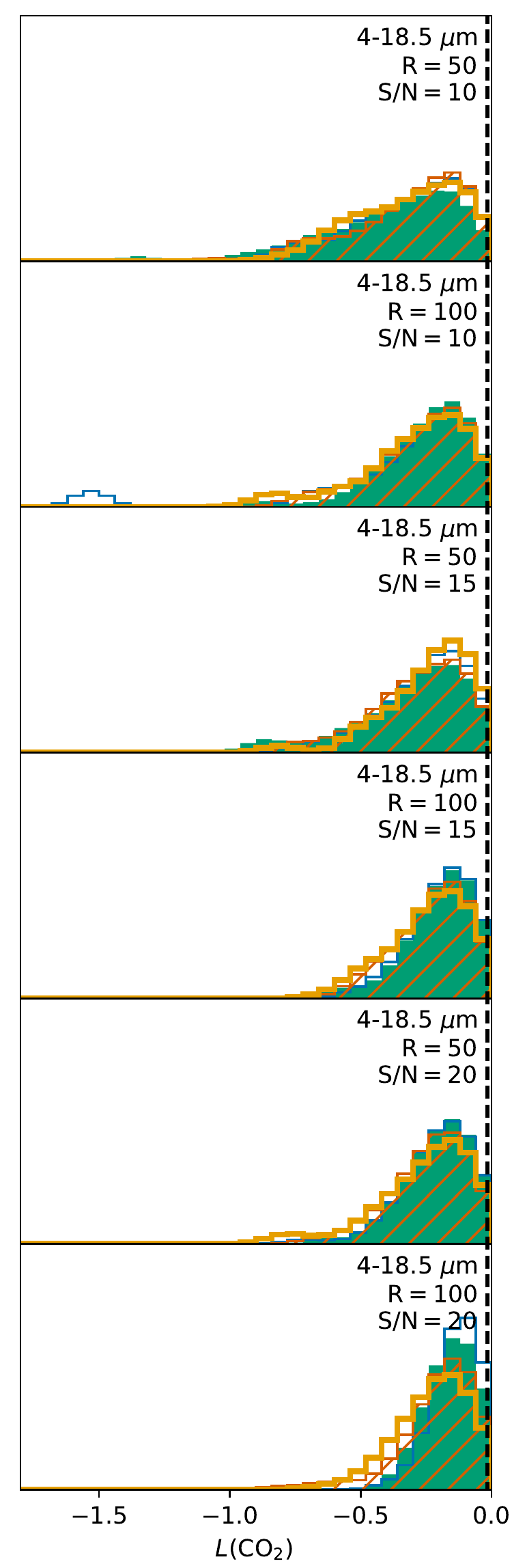}
    \includegraphics[width=0.193\textwidth]{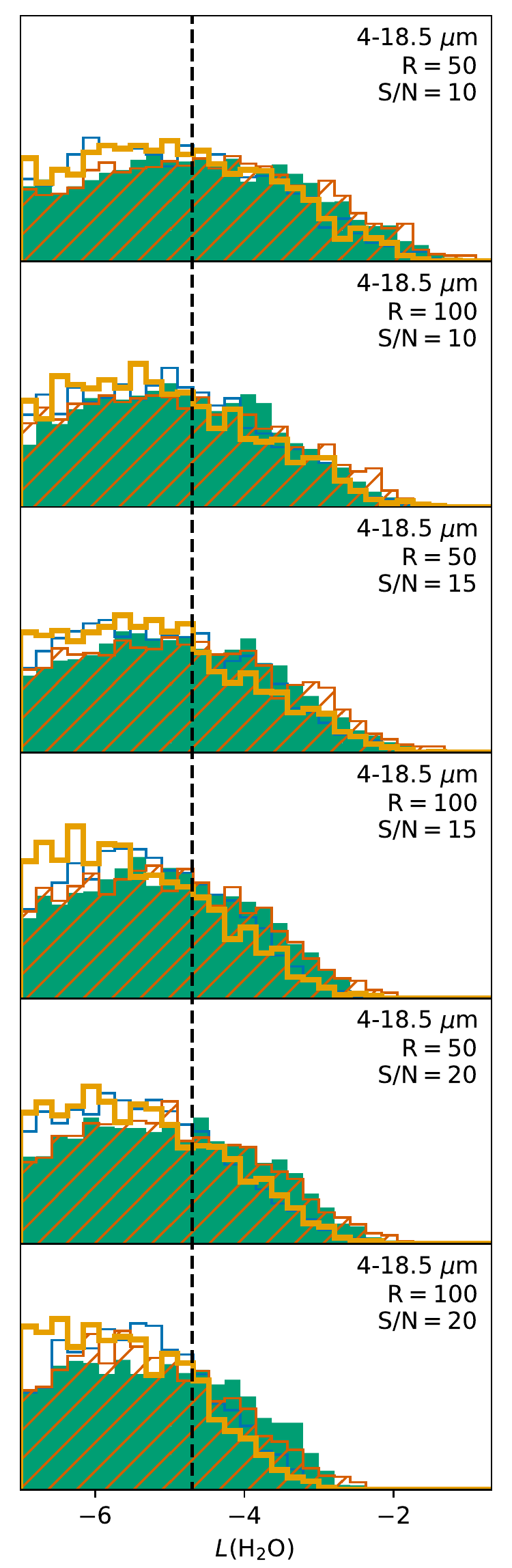}
    \includegraphics[width=0.193\textwidth]{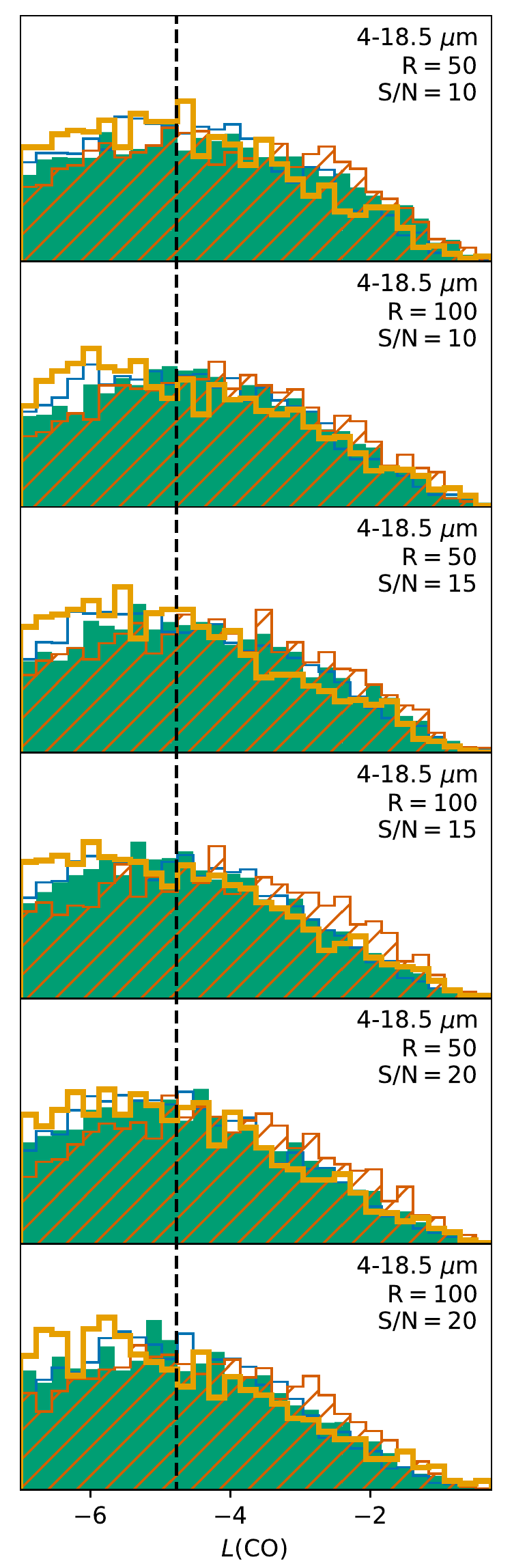}
    \includegraphics[width=0.193\textwidth]{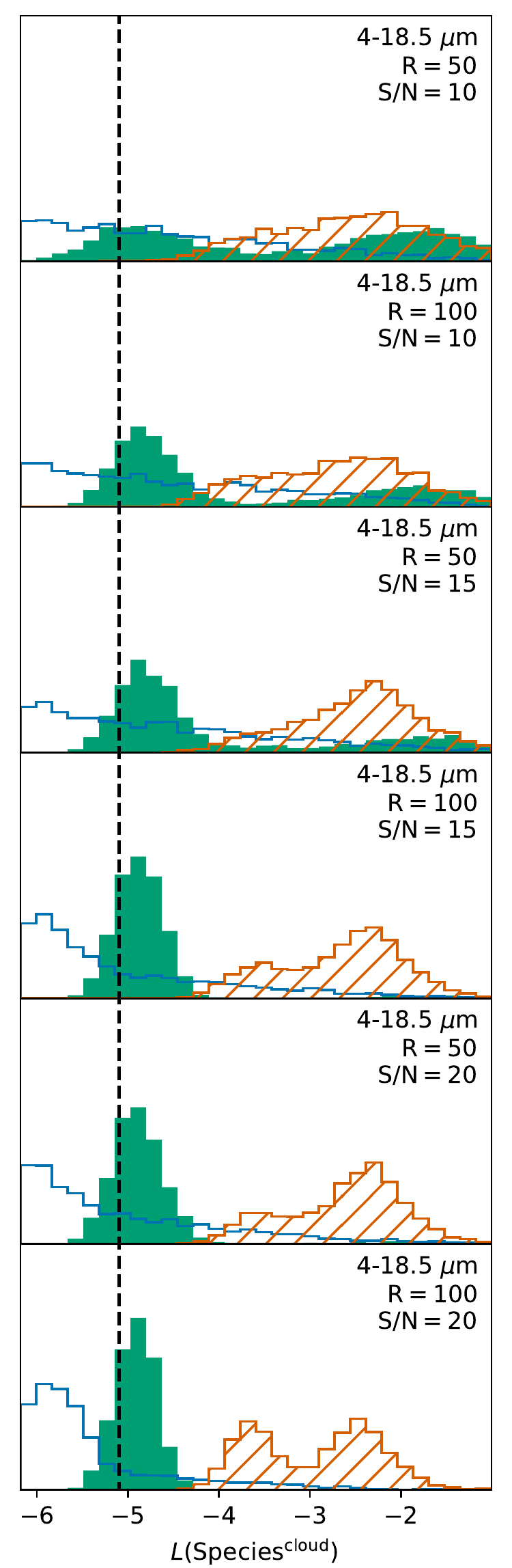}\\
    
    \includegraphics[width=0.0129\textwidth]{Figures/Extended_Retrievals(4-185)/Posterior_ylabel.pdf}
    \includegraphics[width=0.193\textwidth]{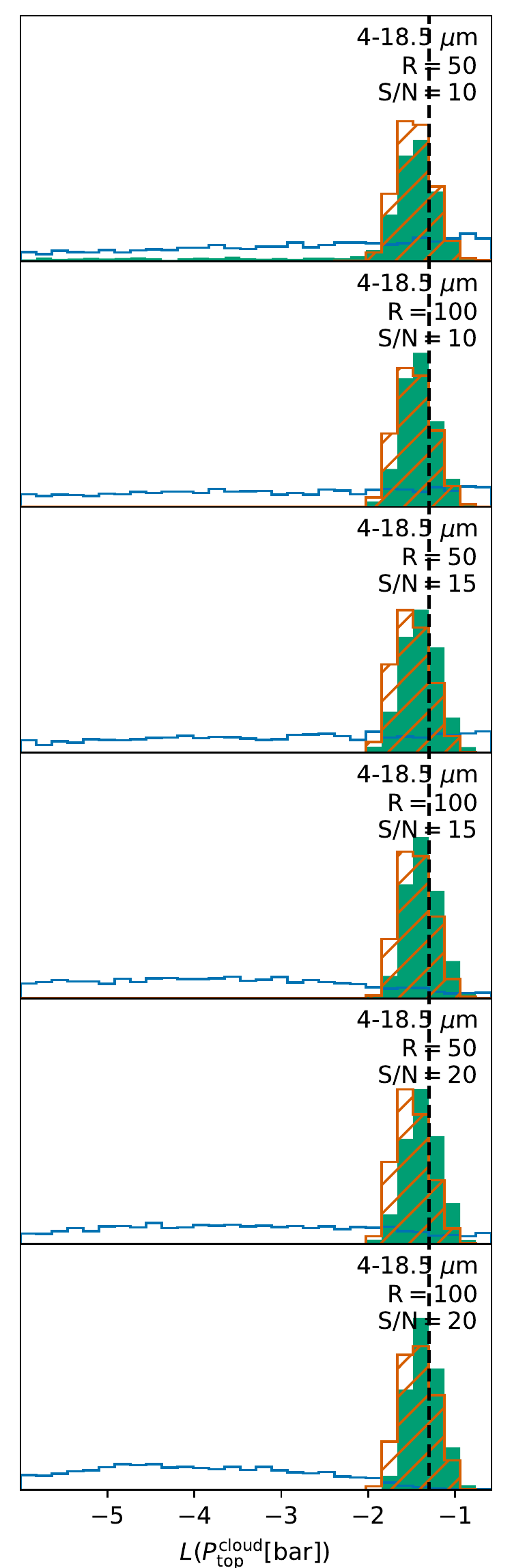}
    \includegraphics[width=0.193\textwidth]{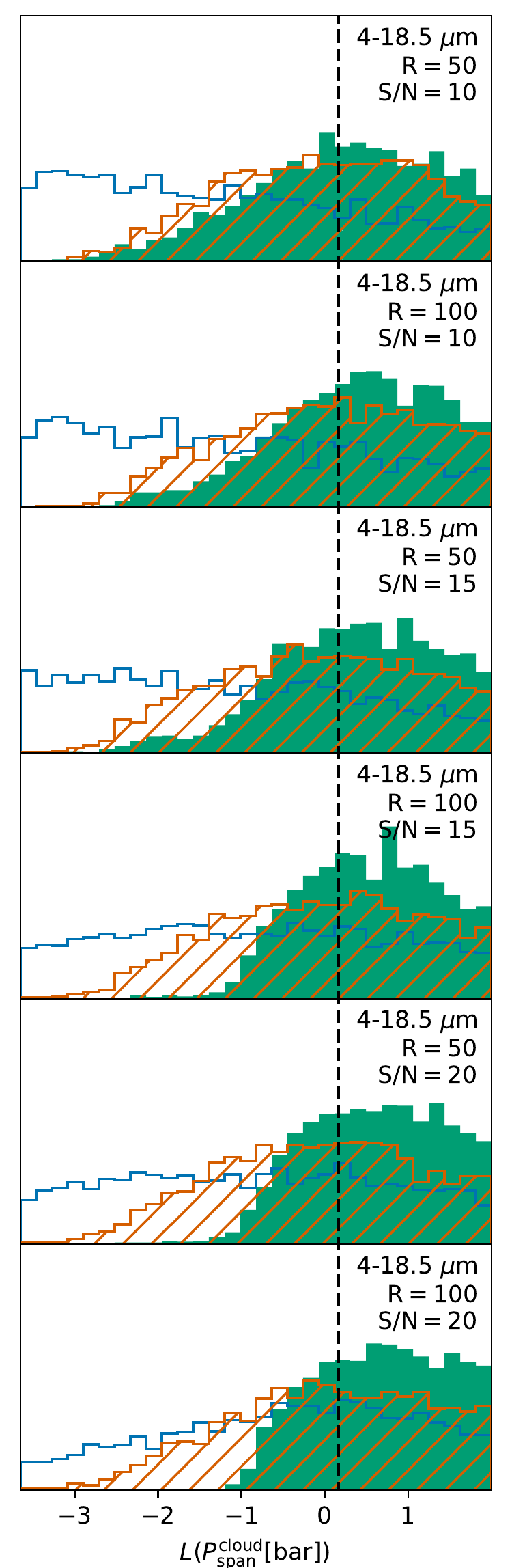}
    \includegraphics[width=0.193\textwidth]{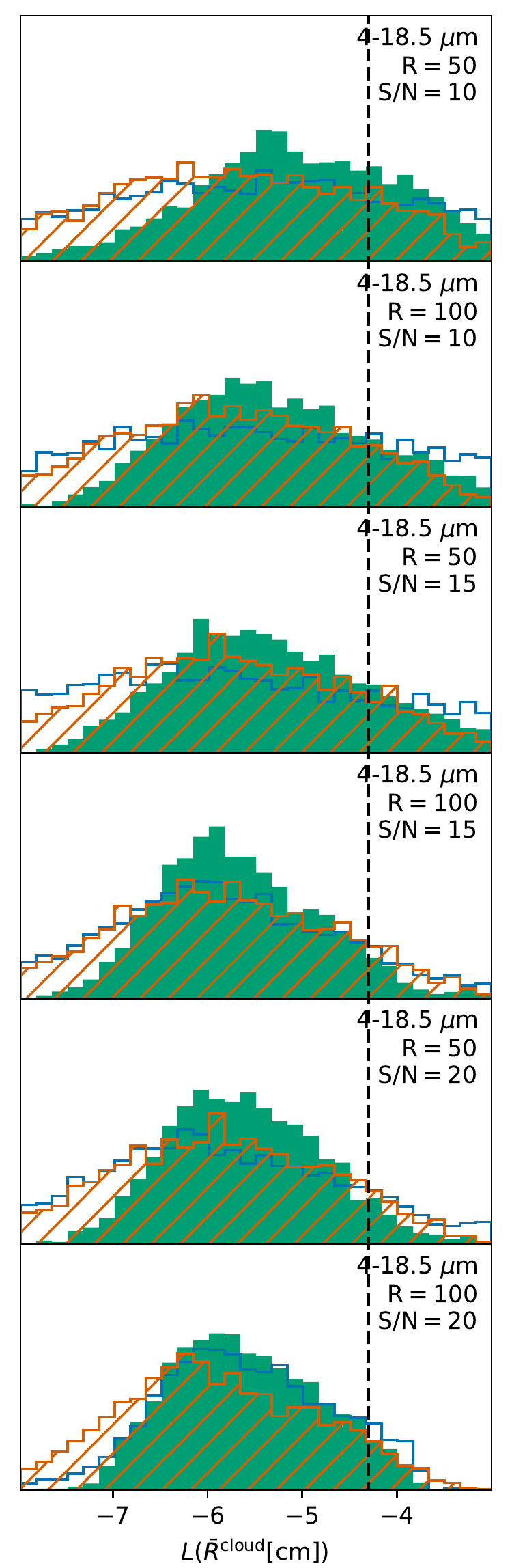}
    \includegraphics[width=0.193\textwidth]{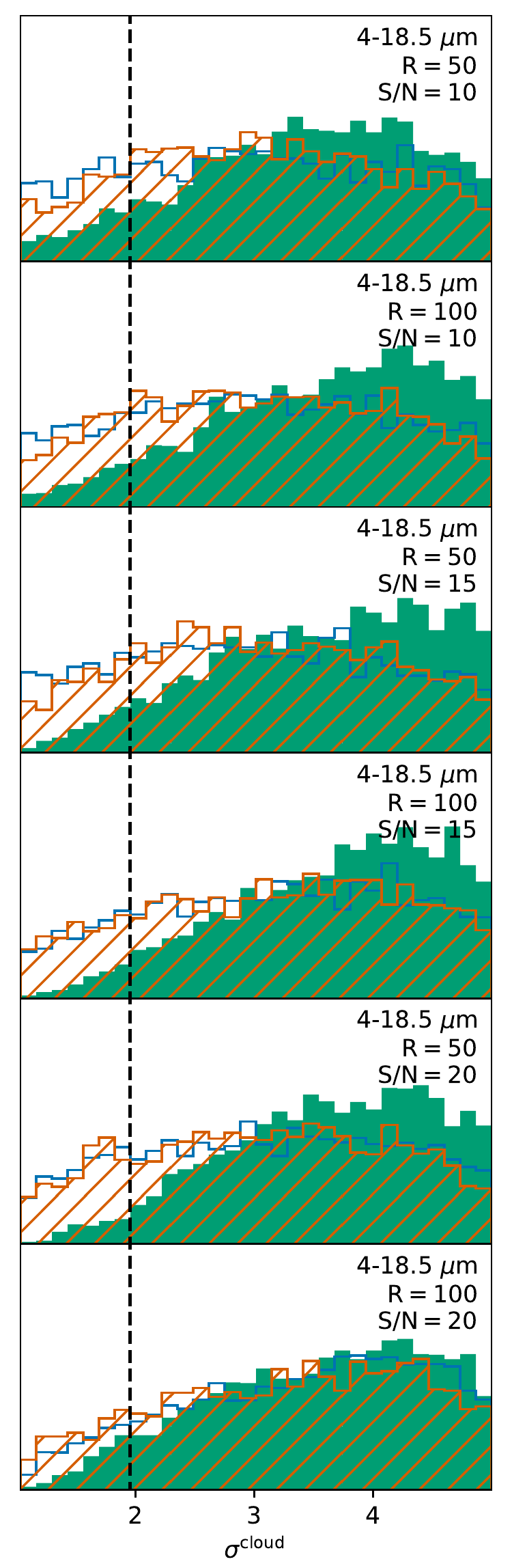}\\

    \includegraphics[width=0.523\textwidth]{Figures/Extended_Retrievals(4-185)/A_B_Posterior_Legend.pdf}
    \caption{Continuation of Fig.~\ref{fig:full_better_input_post_4-185_1}. For $\mathrm{Species^{cloud}}$, the true value is the \ce{H2SO4} mass fraction in the Venus-twin model.}
          \label{fig:full_better_input_post_4-185_2}%
\end{figure*}

\clearpage
\FloatBarrier

\section{Cloudy versus icy scenario - Comparison of absolute fluxes}\label{app:abs_flux_comp}

\begin{figure}
   \centering
    \includegraphics[width=0.48\textwidth]{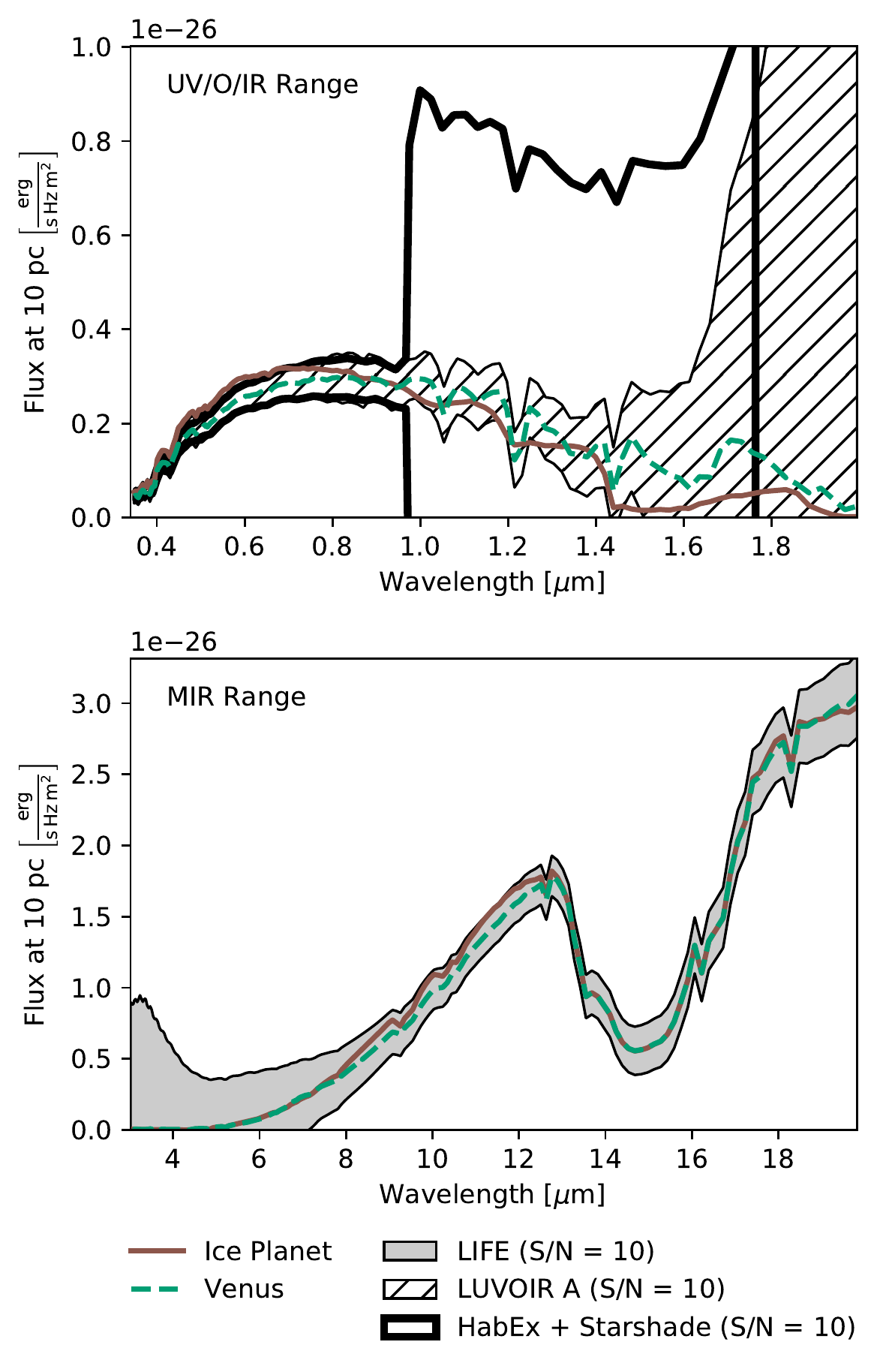}
    \caption{Flux of a cloudy Venus-twin exoplanet (opaque \ce{H2SO4} clouds; dashed green line) compared to an icy world with a thin \ce{CO2}-dominated atmosphere (solid brown line). In the top panel, we plot the UV/O/NIR wavelength range. The hatched area represents the expected wavelength-dependent \SNv{10} noise for the LUVOIR A mission concept \citep{2017AAS...22940504P}. The thick, solid black line represents the expected \SNv{10} noise for the HabEx + Starshade mission concept \citep{2020arXiv200106683G}. In the bottom panel, we plot the MIR wavelength range. The gray shaded region indicates the 1$\sigma$ \lifesim{} noise level at \SNv{10}.}
    \label{fig:Clouds_VS_Ice_absolute}
\end{figure}

In Fig.~\ref{fig:Clouds_VS_Ice_absolute}, we compare the absolute fluxes of the cloudy and icy exoplanet scenarios discussed in Sect.~\ref{sec:life_model_interpretation}. In the UV/O/NIR, we show the noise expected for observations of the cloudy Venus-twin exoplanet with the HabEx~+~Starshade \citep{2020arXiv200106683G} or LUVOIR A \citep{2017AAS...22940504P} instruments \citep[\SNv{10} at \mic{0.6}, calculated with the NASA-GSFC Planetary Spectrum Generator;][]{Villanueva2018PSG}. In the MIR, we show the wavelength-dependent expected \lifesim{} noise (\SNv{10} at \mic{11.2}).

The increase in the HabEx noise at roughly \mic{1} is caused by the HabEx~+~Starshade instrument design. Above \mic{1}, the throughput of the instrument is strongly decreased. Thus, the \R{} of the spectrum above \mic{1} is decreased to increase the \SN{} of the measured spectrum. However, the Venus-twin spectrum plotted in Fig.~\ref{fig:Clouds_VS_Ice_absolute} is constant in \R{} over the full wavelength range considered. Further, the integration time is also constant over the full spectrum. Thus, the HabEx~+~Starshade noise above ~\mic{1} is significantly larger than at shorter wavelengths.

\clearpage

\end{appendix}
\end{document}